\documentclass[11pt, oneside]{article}   	
\usepackage[T1]{fontenc}
\usepackage{lmodern}
\usepackage{geometry}                		
\usepackage{graphicx}				
\usepackage{amsmath}		
\usepackage{amssymb,amsfonts,amsthm}
\usepackage{mathtools,cool}
\usepackage{extarrows}
\usepackage{bbold}
\usepackage{braket}

\title{\textbf{Extended Massive Ambitwistor String II}}
\author{Christian Kunz \\ \small{\textit{E-mail:} \href{mailto:kunz.christian.321@gmail.com}{kunz.christian.321@gmail.com}}}
\usepackage{plain}
\usepackage[dvipsnames]{xcolor}
\usepackage{hyperref}
\hypersetup{
    colorlinks = true,
    linkcolor = {Fuchsia},
    citecolor = {blue},
    urlcolor={blue},
}
\usepackage{cleveref}

\newcommand{\ud}{\mathrm{d}}
\numberwithin{equation}{section}
\allowdisplaybreaks
\begin{document}
  \maketitle
  \tableofcontents

\begin{abstract}
  This article continues previous work done in \cite{Kunz:2024wsx}. It is shown in more detail how vacuum partition functions and the cosmological constant vanish at all orders of perturbation theory. Further, all-multiplicity higher-loop amplitudes are given and shown to be modular invariant, to have proper factorization, and to be UV-finite at least up to one-loop level, formally even to all levels. Therefore, the model provides a modular invariant and unitary N=8 supergravity theory in twistor space with embedded Super-Yang-Mills and promising UV-finiteness behavior.\\
\end{abstract}

\section{Introduction}
Reference \cite{Kunz:2024wsx} introduced an extension to the massive ambitwistor string supergravity model of \cite{Albonico:2022, Albonico:2023}. The most important difference is that due to an additional worldsheet gauge symmetry the Super-Yang-Mills (SYM) model on the Coulomb branch is embedded in this extension. In \cite{Kunz:2024wsx} the all-multiplicity tree and one-loop amplitudes were calculated and shown to have unitary factorization. It was argued without much detail that the partition functions and the cosmological constant vanish. The purpose of the current article is to present these claims in more detail. In addition all higher-loop amplitudes will be evaluated and demonstrated to be modular invariant, be UV-finite at least up to one-loop level (formally at any level), and factorize correctly for degenerated Riemann surfaces.\\

In order to keep repetitions to a minimum it is assumed in the following that the reader is somewhat familiar with the model of \cite{Kunz:2024wsx} and the same notation will be used. The paper is organized as follows:\\

In section \ref{ExMassATStr} the action of the model is complemented with an auxiliary action and simultaneously has the little group extended from SU(2) to SU(2)$\times$SU(2)$\times$SU(2) making it anomaly-free without affecting the physical spectrum and tree scattering amplitudes. In \cite{Kunz:2024wsx} the existence of such an action was assumed, but now an explicit example will be provided. Actually the new action can be viewed as emanating from reductions performed in a 16-dimensional spinor space. This scenario, being ad-hoc and independent of other results in this article, is relinquished to appendix \ref{Reduction}. In addition to the action the all-multiplicity tree amplitudes from \cite{Kunz:2024wsx} are also revisited, now without omitting the contributions from gauge symmetries deemed trivial or inessential in \cite{Kunz:2024wsx}. This is done in preparation for higher-loop amplitudes where they have to be dealt with in a non-trivial manner.\\

In section \ref{PartitionFunc}, after preliminaries listing basic notations and some general results from the literature, the partition function for Riemann surfaces of genus $g \ge 1$ is calculated. Its measure turns out to be modular invariant, based on chiral splitting of the measure of the classical bosonic string. And it becomes straightforward to see that the partition function and, therefore, the cosmological constant are zero.\\

In section \ref{LoopAmps} loop amplitudes for genus $g \ge 1$ are examined. In particular, the modular invariance of the amplitudes is established. And the integration over the moduli space is well defined, localized to a set of points in a fundamental domain.\\

In subsection \ref{EvenSS} all-multiplicity higher-loop amplitudes on even spin structures are evaluated.\\

In subsection \ref{OddSS} the same is done on odd spin structures. It includes the integration over the single zero mode of the supertwistor.\\

In subsection \ref{NonSeparating} unitary factorization of these amplitudes is demonstrated for non-separating degeneration of Riemann surfaces.\\

In subsection \ref{Separating} the same is done for separating degeneration.\\

In subsection \ref{UVFiniteness} it is shown, with help of a simple scaling argument, that the integration during factorization of a loop amplitude on a non-separating divisor at infinity that has even spin structure is UV-finite. This corroborates the formal claim of UV-finiteness at all loop levels based on modular invariance and, in particular, it proves that a fully reduced tree amplitude is UV-finite at least at the single loop level.\\

Section \ref{Summary} contains summary and discussion.\\

In appendix \ref{Reduction} it is demonstrated in an ad-hoc and naive fashion how all the spinors in the action including the auxiliary spinors could be viewed as originating from a superspinor in 16-dimensional spinor space.\\

Appendix \ref{Original} examines which results of this article are also valid for the original massive ambitwistor string in \cite{Albonico:2022, Albonico:2023}.\\

In appendix \ref{Gauge} the extended little group is gauge-fixed in a non-trivial fashion to simplify not only tree but also loop amplitudes.\\

\section{Extended Massive Ambitwistor String Revisited}
  \label{ExMassATStr}
The twistor action used in \cite{Kunz:2024wsx} is repeated here, but with inclusion of an auxiliary sub-action in twistor space with additional worldsheet supersymmetries to ensure zero central charge of the Virasoro symmetry:
 \begin{align}
  \label{massSugraAction}
  S \;\,&\!=\! \int_\Sigma S_{(\mathcal{Z}, \rho_1, A, B)} + S_{(\lambda, \eta, \rho_2, \tau, D)} + S_{\text{aux}},\nonumber\\
  S_{(\mathcal{Z},\rho,A,B)}\,&\!=\!\int_{\Sigma} \mathcal{Z}^a \!\!\cdot\! \bar\partial \mathcal{Z}_a + A_{ab} \mathcal{Z}^a \!\cdot\! \mathcal{Z}^b + \frac{1}{2} \braket{\rho_A \bar{\partial} \rho_B}\Omega^{AB}\\
  & + a \!\braket{\lambda_A \lambda_B}\!\Omega^{AB} \!+\!  B_{ab} \lambda^a_A \rho^b_B \Omega^{AB} \!+\! b \braket{\lambda_A \rho_B}\Omega^{AB}\,,\nonumber\\
  S_{(\lambda, \eta, \rho, \tau, D)}&\!=\!\int_{\Sigma} \frac{1}{2} (\braket{\rho_A \bar{\partial} \rho_B}\!\Omega^{AB} + \braket{\tau_{\mathcal{I}} \bar{\partial} \tau_{\mathcal{J}}} \!\Omega^{\mathcal{I}\mathcal{J}}) +\ \tilde{a} (\!\braket{\lambda_A \lambda_B}\!\Omega^{AB} +\!\braket{\eta_\mathcal{I} \eta_\mathcal{J}}\!\Omega^{\mathcal{I}\mathcal{J}})\nonumber\\
  & + D_{ab} (\lambda^a_A \rho^b_B \Omega^{AB} \!+\! \eta^a_{\mathcal{I}} \tau^b_{\mathcal{J}}\!\Omega^{\mathcal{I}\mathcal{J}})
  \!+\! d \,(\braket{\lambda_A \rho_B}\Omega^{AB} \!+\! \braket{\eta_{\mathcal{I}} \tau_{\mathcal{J}}} \Omega^{\mathcal{I}\mathcal{J}}) \,,\nonumber\\
  S_{\text{aux}} &\!=\!\! \sum_{i=1}^2 \! \int_{\Sigma}  \!S_{(\mathcal{Z}_i,\rho_{i1},A_i, B_i)} + S_{(\lambda_i, \eta_i, \rho_{i2}, \tau_i, D_i)} + \tilde{\tau}_i^{\prime\mathcal{I}} \bar\partial \tau^\prime_{i\mathcal{I}}\;. \nonumber
  \end{align}
  Before explaining details of the terms in this action, it is instructive to notice that the action can be re-interpreted by extending the little group from SU(2) to SU(2)$\times$SU(2)$\times$SU(2), or rather its complexification G $\!=\!$ SL(2,$\mathbb{C}$)$\times$SL(2,$\mathbb{C}$)$\times$SL(2,$\mathbb{C}$), and presented in a more concise form: 
 \begin{align}
  \label{massSugraActionComb}
  S \;\,\!\!=\!\!\! \int_\Sigma &\!\mathcal{Z}^\mathrm{a} \!\cdot\! \bar\partial \mathcal{Z}_\mathrm{a} + A_{\mathrm{a}\mathrm{b}} \mathcal{Z}^\mathrm{a} \!\cdot\! \mathcal{Z}^\mathrm{b} + S_{\rho_1} + S_{(\!\rho_2, \tau\!)} + S_{\text{aux}}, \nonumber\\
  S_{\rho}\,\!\!=\!\!\!\int_{\Sigma} & \frac{1}{2} \rho_{\mathrm{a}A} \bar{\partial} \rho_B^\mathrm{a}\Omega^{AB} + a \lambda_{\mathrm{a}A} \lambda_B^\mathrm{a}\Omega^{AB} + B_{\mathrm{a}\mathrm{b}} \lambda^{\mathrm{a}}_A \rho^{\mathrm{b}}_B \Omega^{AB}  + b \lambda_{\mathrm{a}A} \rho_B^\mathrm{a}\Omega^{AB}\,,\nonumber\\ 
  S_{(\rho, \tau)}\!\!=\!\!\!\int_{\Sigma} & \frac{1}{2} (\rho_{\mathrm{a}A} \bar{\partial} \rho_B^\mathrm{a}\Omega^{AB} \!+\! \tau_{\mathrm{a}\mathcal{I}} \bar{\partial} \tau^{\mathrm{a}}_{\mathcal{J}}\Omega^{\mathcal{I}\mathcal{J}}) + \tilde{a} (\lambda_{\mathrm{a}A} \lambda_{B}^\mathrm{a}\Omega^{AB} \!+\! \eta_{\mathrm{a}\mathcal{I}} \eta_{\mathcal{J}}^\mathrm{a}\Omega^{\mathcal{I}\mathcal{J}})\\
  & + D_{\mathrm{a}\mathrm{b}} (\lambda^{\mathrm{a}}_A \rho^\mathrm{b}_B \Omega^{AB} \!+\! \eta^\mathrm{a}_{\mathcal{I}} \tau^\mathrm{b}_{\mathcal{J}}\Omega^{\mathcal{I}\mathcal{J}}) + d \,(\lambda_{\mathrm{a}A} \rho_B^\mathrm{a}\Omega^{AB} \!+\! \eta_{\mathrm{a}\mathcal{I}} \tau_{\mathcal{J}}^\mathrm{a} \Omega^{\mathcal{I}\mathcal{J}})\;, \nonumber\\
  S_{\text{aux}} \!=\!\! \sum_{i=1}^2 &\! \int_{\Sigma} \tilde{\tau}_i^{\prime\mathcal{I}} \bar{\partial} \tau^{\prime}_{i\mathcal{I}}\;, \nonumber
  \end{align} 
  where
  \begin{description}
  \item{$\mathrm{a} \!=\! (a,\dot{a},\ddot{a}), (a,\dot{a},\ddot{a} \!=\! 0,1)$ is the G little group index raised and lowered with $\varepsilon^{\mathrm{a}\mathrm{b}} \!=\! \varepsilon^{ab} \oplus \varepsilon^{\dot{a}\dot{b}}\oplus \varepsilon^{\ddot{a}\ddot{b}}$ and $\varepsilon_{\mathrm{a}\mathrm{b}} \!=\! \varepsilon_{ab} \oplus \varepsilon_{\dot{a}\dot{b}}\oplus \varepsilon_{\ddot{a}\ddot{b}}$, respectively. For the individual SL(2,$\mathbb{C}$) index $a$ the contraction notation $\xi_a \zeta_b \varepsilon^{ab} = \xi_a \zeta^a \equiv \braket{\xi \zeta}$ is used.}
  \item{$\mathcal{Z}_{\mathrm{a}}$ are G-indexed supertwistor fields that are worldsheet spinors, repackaged into \textit{Dirac} supertwistors of the form
  \begin{align*}
  \mathcal{Z}\!=\!(\lambda_A,\mu^A,\eta_\iota)\!:&\lambda_A \!=\! (\lambda_\alpha,\tilde \lambda_{\dot\alpha}) , \mu^A \!=\! (\tilde\mu^{\alpha},\mu^{\dot\alpha}),
  \;\eta_{\iota}\!=\!(\eta_\mathcal{I}, \tilde{\eta}^\mathcal{I}), \eta_{\mathcal{I}}\!=\!(\eta_I,\tilde \eta_{\dot{I}}), \tilde{\eta}^\mathcal{I}\!=\!(\tilde \eta^{\prime I}, \eta^{\prime \dot{I}}),\nonumber \\
  &I,\dot{I} = 1 \ldots \frac{\mathcal{N}}{2} , \,\mathcal{I} = 1, \ldots, \mathcal{N}, \,\iota = 1, \ldots, 2\mathcal{N}\,, \,\mathcal{N} = 8\, ,
  \end{align*}
  where $\lambda_A$ and $\mu^A$ are Dirac spinors made up of the homogeneous chiral and anti-chiral components of the twistor $Z = (\lambda_\alpha, \mu^{\dot\alpha})$ and dual twistor $\tilde{Z} = (\tilde \lambda_{\dot\alpha}, \tilde\mu^{\alpha})$, and $\eta_{\iota}\!=\!(\eta_\mathcal{I}, \tilde{\eta}^\mathcal{I})$ are fermionic components with $\mathcal{I}=1,\ldots,\mathcal{N}$ as the R-symmetry index, with $\mathcal{N}\!\!=\!8$ for maximal supergravity.} 
  \item{The inner product is defined as\\
  $\mathcal{Z}_\mathrm{a} \cdot \mathcal{Z}_\mathrm{b} = \frac{1}{2}(\tilde{Z_\mathrm{a}} \cdot Z_\mathrm{b} + \tilde{Z_\mathrm{b}} \cdot Z_\mathrm{a} + \tilde\eta_\mathrm{a}^{\mathcal{I}} \eta_{\mathrm{b}\mathcal{I}} + \tilde\eta_\mathrm{b}^{\mathcal{I}} \eta_{\mathrm{a}\mathcal{I}}),\;\; \tilde{Z_\mathrm{a}} \cdot Z_\mathrm{b} = \tilde\mu_\mathrm{a}^{\alpha}\lambda_{\mathrm{b} \alpha} + \tilde \lambda_{\mathrm{a} \dot\alpha} \mu_\mathrm{b}^{\dot\alpha} = \mu_\mathrm{a}^A \lambda_{\mathrm{b}A}$\,,\\
  with special treatment of $\bar\partial Z$ when taking the dual: $\;\widetilde{\bar\partial Z} = - \bar\partial \tilde{Z}\,$.}
  \item{The indices of the bosonic Dirac spinors are raised and lowered with the skew 4 $\!\times\!$ 4 forms\\
  $\Omega^{AB} = \Omega^{(\alpha,\dot{\alpha})(\beta,\dot\beta)} \xlongequal{e.g.} \varepsilon^{\alpha \beta} \oplus \varepsilon^{\dot{\beta} \dot{\alpha}}$ and $\Omega_{AB} \xlongequal{e.g.}  \varepsilon_{\beta \alpha} \oplus \varepsilon_{\dot{\alpha} \dot{\beta}}$ with $\Omega^{AB} \Omega_{BC} = \delta^A_C$.}
   \item{The symmetric $\mathcal{N} \times \mathcal{N}$ form $\Omega^{\mathcal{I} \mathcal{J}}$ decomposes into two $\frac{\mathcal{N}}{2} \times \frac{\mathcal{N}}{2}$ forms:\footnote{$\Omega^{\mathcal{I} \mathcal{J}}$ is quite different from the corresponding matrices $\Omega^{IJ}$ and $\Omega^{\dot{I}\dot{J}}$ in the original massive ambitwistor string \cite{Albonico:2022, Albonico:2023} where they make up the R-symmetry symplectic metrics. Here the additional worldsheet supersymmetries exclude the fermionic fields $\tilde{\eta}^{\mathcal{I}}$ from the spectrum such that the remaining R-symmetry is not related to a full conventional target space supersymmetry. Note that if one defines $\tilde{\eta}_{\mathcal{I}} \!=\! (\eta^{\prime}_ I, \tilde{\eta}^{\prime}_{\dot{I}}) \!=\! \Omega_{\mathcal{I} \mathcal{J}} \tilde{\eta}^{\mathcal{J}}$ then the fermionic part of the inner product $\mathcal{Z}_\mathrm{a} \!\cdot\! \mathcal{Z}_\mathrm{b}$ can be written as $\!(\tilde{\eta}_{\mathrm{b}\mathcal{I}}, \eta_{\mathrm{b}\mathcal{I}})\Omega^{\iota \varkappa} \begin{pmatrix}
   \tilde{\eta}_{\mathrm{a}\mathcal{K}}\\\eta_{\mathrm{a}\mathcal{K}}
   \end{pmatrix}$ where $\Omega^{\iota \varkappa} \!=\! \begin{pmatrix}
   0 &\Omega^{\mathcal{I}\mathcal{K}}\\
   -\Omega^{\mathcal{I}\mathcal{K}}&0\,
   \end{pmatrix}$ is the symplectic matrix that would correspond to the original ambitwistor string.}
   \begin{equation*}
   \Omega^{\mathcal{I} \mathcal{J}} = \begin{pmatrix}
   0 &\mathbb{1}^{I \dot{J}}\\
   \mathbb{1}^{\dot{I} J} &0\,
   \end{pmatrix}.
   \end{equation*}}
   \item{$\rho^\mathrm{a}_{rA} (r \!=\! 1,2)$ and $\tau^\mathrm{a}_\mathcal{I}$ are auxiliary fermionic and bosonic worldsheet spinors, respectively, whose indices are raised and lowered with the same matrices as their counter parts.}   
  \item{In $S_{\text{aux}} \,\tau^{\prime}_{i\mathcal{I}}$ and $\tilde{\tau}_i^{\prime \mathcal{I}}$ are ungauged spinors ensuring the correct total number of bosonic and fermionic spinors.}
   
  \item{$(B_{\mathrm{a}\mathrm{b}}, b, D_{\mathrm{a}\mathrm{b}}, d)$ are fermionic Lagrange multipliers for the constraints
  \begin{align*}
  &\lambda^{(\mathrm{a}}_A \rho^{\mathrm{b})}_{1B} \Omega^{AB} &&= 0 = && \lambda_{\mathrm{a}A} \rho_{1B}^\mathrm{a}\Omega^{AB},\\
  &\lambda^{(\mathrm{a}}_A \rho^{\mathrm{b})}_{2B} \Omega^{AB} \!+\! \eta^{(\mathrm{a}}_{\mathcal{I}} \tau^{\mathrm{b})}_{\mathcal{J}}\Omega^{\mathcal{I}\mathcal{J}} &&= 0 = &&\lambda_{\mathrm{a}A} \rho_{2B}^\mathrm{a}\Omega^{AB} \!+\! \eta_{\mathrm{a}\mathcal{I}} \tau_{\mathcal{J}}^\mathrm{a} \Omega^{\mathcal{I}\mathcal{J}}
   \end{align*}
   on the supersymmetric gauge currents.}
   
  \item{Little group transformations for the twistors $\mathcal{Z}_\mathrm{a}$ are gauged by the fields $A_{\mathrm{a}\mathrm{b}} = A_{(\mathrm{a}\mathrm{b})}$, and $a, \tilde{a}$ are worldsheet $(0,1)$-forms that also act as Lagrange multipliers, required in order to have a closed current algebra\footnote{Notice that $\lambda_{(\mathrm{a}}^A \lambda_{\mathrm{b})A} \!=\! 0$ because of $\Omega^{AB}$ being skew and also $\eta_{(\mathrm{a}}^{\mathcal{I}} \eta_{\mathrm{b})\mathcal{I}} \!=\!  0$ because of $\Omega^{\mathcal{I}\mathcal{J}}$ being symmetric.}.}
  \end{description}
    
  All spinors obey the usual canonical OPE, e.g. $\mathcal{Z}^A_\mathrm{a}(z) \cdot \mathcal{Z}_B^\mathrm{b}(0) = \frac{\delta^A_B \delta_\mathrm{a}^\mathrm{b}}{z} + \ldots$ and $\tilde{\tau}_\mathrm{a}^\mathcal{I}(z) \tau^\mathrm{b}_\mathcal{J}(0) = \frac{\delta^\mathcal{I}_\mathcal{J} \delta_\mathrm{a}^\mathrm{b}}{z} + \ldots$.\\
    
  Concerning BRST quantization, in addition to the familiar fermionic ($b,c$) ghost system related to worldsheet gravity (the action \eqref{massSugraActionComb} is already written in conformal gauge), the Lagrange multipliers in \eqref{massSugraActionComb} are associated with corresponding (anti-ghost, ghost) pairs:
  \begin{description}
  \item{the bosonic fields $\{A_{\mathrm{a}\mathrm{b}}, a, \tilde{a}\}$ with fermionic ghosts\\
  $\{(M^{\mathrm{a}\mathrm{b}}, N_{\mathrm{a}\mathrm{b}}),(m, n), (\tilde{m}, \tilde{n})\}$ and}
  \item{the fermionic fields $\{B_{\mathrm{a}\mathrm{b}}, b, D_{\mathrm{a}\mathrm{b}}, d\}$ with bosonic ghosts\\
   $\{(\beta^{\mathrm{a}\mathrm{b}}, \!\gamma_{\mathrm{a}\mathrm{b}}),(\beta, \!\gamma), (\tilde{\beta}^{\mathrm{a}\mathrm{b}}, \!\tilde{\gamma}_{\mathrm{a}\mathrm{b}}), (\tilde{\beta}, \tilde{\gamma})\}$.}\\
   \end{description}
  The BRST operator becomes $Q_B= \oint \ud z J_{BRST}(z)$, where the BRST current $J_{BRST}(z)$ is:
 \begin{align*}
 \label{JBRST}
  J_{BRST} \!= &\, c\,T \!+\! N_{\mathrm{a}\mathrm{b}}(J^{\mathrm{a}\mathrm{b}} \!+\! M^{\mathrm{a}}_{\;\mathrm{c}} N^{\mathrm{b}\mathrm{c}}) \!+\! n \lambda^{\mathrm{a}}_A \lambda_{\mathrm{a}}^A \!+\! \tilde{n}(\lambda^{\mathrm{a}}_A \lambda_{\mathrm{a}}^A \!+\!\eta^\mathrm{a}_{\mathcal{I}} \eta_\mathrm{a}^{\mathcal{I}}\!)
   \!+\! \gamma_{\mathrm{a}\mathrm{b}} \lambda^{\mathrm{a}A}\rho^\mathrm{b}_{1A} \!+\! \gamma\lambda_\mathrm{a}^A \rho^\mathrm{a}_{1A}\nonumber\\
  &+\! m \gamma \gamma  \!+\! \tilde{\gamma}_{\mathrm{a}\mathrm{b}} (\lambda^{\mathrm{a}A}\rho^\mathrm{b}_{2A} \!+\! \eta^{\mathrm{a}\mathcal{I}} \tau^\mathrm{b}_{\mathcal{I}}) \!+\! \tilde{\gamma} (\lambda_\mathrm{a}^A \rho^\mathrm{a}_{2A}\!+\!\eta_\mathrm{a}^{\mathcal{I}} \tau^\mathrm{a}_{\mathcal{I}}) \!+\! \tilde{m} \tilde{\gamma}\tilde{\gamma}\,,\nonumber\\ 
  T \!= &\,\mathcal{Z}^\mathrm{a} \!\!\cdot\! \partial \mathcal{Z}_\mathrm{a} \!+\! \frac{1}{2}\!\sum_{r = 1,2} \!\rho^\mathrm{a}_{rA} \partial \rho_{r\mathrm{a}}^A \!+\! \beta^{\mathrm{a}\mathrm{b}} \partial \gamma_{\mathrm{a}\mathrm{b}} \!+\!\beta \partial \gamma \!+\! \frac{1}{2}\tau^\mathrm{a}_{\mathcal{I}} \partial \tau_\mathrm{a}^{\mathcal{I}} \!+\! \tilde{\beta}^{\mathrm{a}\mathrm{b}} \partial \tilde{\gamma}_{\mathrm{a}\mathrm{b}} \!+\! \tilde{\beta} \partial \tilde{\gamma} \nonumber\\
   & +\! M^{\mathrm{a}\mathrm{b}} \partial N_{\mathrm{a}\mathrm{b}} \!+\! m \partial n \!+\! \tilde{m} \partial \tilde{n} \!+\! \frac{1}{2}(b \partial c \!+\! \partial(bc)) \!+\! \sum_{i=1}^2 \tilde{\tau}^{\prime \mathcal{I}}_i \partial \tau^\prime_{i\mathcal{I}} \,, \nonumber\\
  J^{\mathrm{a}\mathrm{b}} \!= & \,\mathcal{Z}^\mathrm{a} \cdot \mathcal{Z}^\mathrm{b}  \!+\! \frac{1}{2}\!\!\sum_{r = 1,2} \!\rho^{(\mathrm{a}}_{rA}  \rho_r^{\mathrm{b})A} \!+\! \beta^{\mathrm{c}(\mathrm{a}}  \gamma^{\mathrm{b})}_\mathrm{c} \!+\! \frac{1}{2} \tau^{(\mathrm{a}}_{\mathcal{I}} \tau^{\mathrm{b})\mathcal{I}}  \!+\! \tilde{\beta}^{\mathrm{c}(\mathrm{a}}  \tilde{\gamma}_\mathrm{c}^{\mathrm{b})}\,,\nonumber
  \end{align*}
  where $T$ is the energy-momentum current and $J^{\mathrm{a}\mathrm{b}}$ the little group G current. The Virasoro central charge vanishes:
    \begin{align*}
    c\!=\!  &-\!3 (8\!-\!2\mathcal{N})_{\mathcal{Z}_{\mathrm{a}}} \!-\! 26_{bc} \!-\! 18_{MN} \!-\! 2_{mn} \!-\! 2_{\tilde{m}\tilde{n}} \!+\! 12_{\rho_1} \!+\!  12_{\rho_2} \!-\! 3\mathcal{N}_\tau \!+\!  20_{\beta\gamma} \!+\!  20_{\tilde{\beta}\tilde{\gamma}} \!-\! \sum_{i=1}^2 \mathcal{N}_{\tau_i \tilde{\tau}_i}\\
    \!=\! &-\! 8 \!+\! \mathcal{N} = 0\,,
    \end{align*}
    and the anomaly coefficient for $J^{\mathrm{a}\mathrm{b}}$ is zero as well ($\text{tr}_{\text{adj}}(t^kt^k) \!=\! 3\!\cdot\! 6 \!=\! 18, \text{tr}_{\text{F}}(t^kt^k) \!=\! 3\!\cdot\!\frac{3}{2}\!=\! \frac{9}{2}$):
    \begin{equation*}
    a^G \!\!=\! \frac{1}{2}\text{tr}_{\text{F}}(t^kt^k) ((8 \!-\! 2\mathcal{N})_{\!\mathcal{Z}} \!\!-\! 8_{\rho_{1,2}} + \mathcal{N}_\tau) \!+\!  \text{tr}_{\text{adj}}(t^kt^k)(\!-1_{\!\mathcal{M}\mathcal{N}} \!+\! 1_{\beta \gamma} \!+\! 1_{\tilde{\beta} \tilde{\gamma}} )  \!\!=\! \frac{9}{4}(8 \!-\! \mathcal{N}) \!=\! 0.
    \end{equation*}
    It follows that the BRST charge $Q_B$ is nilpotent and the model is anomaly-free. Note the similarity of the action in \eqref{massSugraActionComb} without $S_{\text{aux}}$ with the one in \cite{Kunz:2024wsx}. The only difference is the little group. But now there is a new gauge redundancy coming from the little group. In fact, it is known \cite{Okano:2016gya} that whenever massive particles are described by more than two twistors there is a unitary transformation that reduces the description to two twistors and maps all other twistors to zero. In other words, on-shell (but not off-shell!) there is a gauge transformation that eliminates two of the three pairs of twistors, i.e. G (before complexification) contains the 'tiny group' SU(2) $\times$ SU(2)\cite{Geyer:2019ayz} (denoting its complexification SL(2,$\mathbb{C}) \times $SL(2,$\mathbb{C}$) with G$_T$), and, therefore, the change in the little group does not have any impact on the physical spectrum. Of course, SU(2) $\times$ SU(2) $\times$ SU(2) is a small subgroup of the full internal symmetry group ISU(6) of six twistors, i.e. the set of allowed momentum twistors is severely restricted. This will become important when looking at polarized scattering equations for loop amplitudes in section \ref{LoopAmps}.\\
    
    The kinetic term in $S_{\text{aux}}$ containing $\tau^\prime$ and $\tilde{\tau}^\prime$ is decoupled from the rest of the action and, as will be seen shortly when discussing vertex operators, it also does not contribute to the physical spectrum. The main purpose of the little group redundancy and the fields in $S_{\text{aux}}$ is to provide virtual excitations for loop amplitudes and ensure the absence of conformal anomalies. At the end of the section it will be shown that there is no change to tree scattering amplitudes. The non-trivial contributions to loop scattering amplitudes will be taken care of in section \ref{LoopAmps}.\\
      
  Moving on to vertex operators and picture changing operators (PCOs), following \cite{Geyer:2019ayz, Geyer:2020, Albonico:2022, Albonico:2023} fixed vertex operators are of the form
  \begin{equation}
  \label{VO}
  \mathcal{V} \!\!=\!\! \int \!\ud\mu(u) \ud\mu(v) \mathcal{W}(u) \bar{\delta}(\!v^\alpha_{\mathrm{a}} \epsilon^{\mathrm{a}}_\alpha \!-\! 1) \bar{\delta}(\!u^{\alpha}_{\mathrm{a}} \lambda^{\mathrm{a}} _A \!-\! v^{\alpha}_{\mathrm{a}} \kappa^{\mathrm{a}}_A) \bar{\delta}(\!u^{\alpha}_{\mathrm{a}} \eta^{\mathrm{a}}_{\mathcal{I}} \!-\! v^{\alpha}_{\mathrm{a}} \zeta^{\mathrm{a}}_{\mathcal{I}}) \; e^{u_{\alpha}^{\mathrm{a}} \mu_{\mathrm{a}}^A \epsilon^{\alpha}_A + u_{\alpha}^{\mathrm{a}}  \tilde{\eta}_{\mathrm{a}}^{\mathcal{I}} q^{\alpha}_{\mathcal{I}}}\!.\\
  \end{equation}
  Here $\ud\mu(u)$ and $\ud\mu(v)$ are integration measures for the quotient group G $\!/\!$ G$_T \!=\!$ G$_Q$ that is isomorphic to SL(2,$\mathbb{C}$), the spinors $u^\mathrm{a}_\alpha, v^\mathrm{a}_\alpha$ fulfill $u^\mathrm{a}_\alpha u_{\mathrm{a}\beta} \!=\! 0 \!=\! v^\mathrm{a}_\alpha v_{\mathrm{a}\beta} $ with $\alpha$ a 'tiny group' index (not to be confused with $\alpha$ as part of the index $A$ in $\lambda_A^\mathrm{a}$ and $\kappa_A^\mathrm{a}$), $\epsilon^\alpha_\mathrm{a}$ are little group spinors satisfying $\epsilon^\alpha_\mathrm{a} \epsilon^{\mathrm{a}\beta} \!=\! 0$ and forming, together with $v^\alpha_\mathrm{a}$, a basis for G$_Q$, $\kappa^A_\mathrm{a}$ are momentum spinors with associated polarization spinors $\epsilon^A_\alpha \!=\! \epsilon_\alpha^\mathrm{a} \kappa^A_\mathrm{a}$,$\;\zeta^{\mathcal{I}}_\mathrm{a}$ are supermomenta with superpolarization spinors or supercharges $q^{\mathcal{I}}_\alpha \!=\! \epsilon_\alpha^\mathrm{a} \zeta^{\mathcal{I}}_\mathrm{a}$, and the operator $\mathcal{W}(u)$ is a product of fermionic ghost fields and delta functions of bosonic ghost fields, but also depends on $u$ variables:
  \begin{equation}
  \label{fVO}
  \mathcal{W}(u) = c\, n\, \tilde{n}\, \delta(\gamma)\delta(\gamma_{\mathrm{a}\mathrm{b}} u^\mathrm{a}_\alpha u^{\mathrm{b}\alpha})\,\delta(\tilde{\gamma})\delta(\tilde{\gamma}_{\mathrm{a}\mathrm{b}} u^\mathrm{a}_\beta u^{\mathrm{b}\beta}) \,.
  \end{equation}
  As explained in \cite{Geyer:2020} it is sufficient in \eqref{fVO} to force $\gamma_{\mathrm{a}\mathrm{b}}$ and $\tilde{\gamma}_{\mathrm{a}\mathrm{b}}$ to be orthogonal to $u^\mathrm{a}_\alpha u^{\mathrm{b}\alpha}$. The delta functions express the polarized scattering equations. In the following the $\alpha$-index is taken to be from SO(4,$\mathbb{C}$) of which G$_T$ is the double cover. One might ask whether $\epsilon^\alpha_\mathrm{a}$ and $v^\alpha_\mathrm{a}$ also form a basis for SO(4,$\mathbb{C}$) with $v^{\alpha}_\mathrm{c} \epsilon_\beta^\mathrm{c} \!=\!\delta^\alpha_\beta$ but this might not be required or desired in all circumstances, especially as SU(2)$\times$SU(2) is just a small subgroup of SU(4). Appendix \ref{Gauge} shows an example where the basis is limited to a subspace.\\
  
In correlation functions, the exponentials in the vertex operators lead to the following solutions of the equations of motion for the $\lambda$ and $\eta$ fields:
\begin{equation}
\label{fields}
  \lambda_A^\mathrm{a}(\sigma) = \sum_i \frac{u_{i\alpha}^{\mathrm{a}} \epsilon^{\alpha}_{iA}}{\sigma - \sigma_i} , \qquad\eta_{\mathcal{I}}^\mathrm{a}(\sigma) = \sum_i \frac{u_{i\alpha}^{\mathrm{a}} q^{\alpha}_{i\mathcal{I}}}{\sigma - \sigma_i}\,.
\end{equation} 

In the gauge where the momentum spinors and supermomenta are represented only by one pair, with the other two pairs being set to zero, it follows that the dependency on the tiny group can be eliminated and that the $u$ and $v$ variables can be reduced to one pair each. This way the vertex operators reduce to the ones of \cite{Kunz:2024wsx}:
  \begin{equation*}
  \label{VOgauged}
  \mathcal{V} \!=\!\! \int \!\ud^2u \,\ud^2v \mathcal{W}(u) \bar{\delta}(\!\braket{v \epsilon} \!-\!1) \bar{\delta}^{4}(\!\braket{u \lambda_A} \!-\! \braket{v \kappa_A}) \bar{\delta}^{\mathcal{N}}\!(\!\braket{u \eta_{\mathcal{I}}} \!-\! \braket{v \zeta_{\mathcal{I}}}) \; e^{\braket{u \mu^A} \epsilon_A + \braket{u \tilde{\eta}^{\mathcal{I}}} q_{\mathcal{I}}},\\
  \end{equation*}
  where now $\epsilon_a$ are a basis of spinors for the remaining SL(2, $\mathbb{C}$) little group, $\kappa^A_a$ are the non-vanishing momentum spinors with associated polarization spinors $\epsilon^A \!=\! \braket{\epsilon \kappa^A}$,$\;\zeta^{\mathcal{I}}_a$ are the remaining supermomenta with supercharges $q^{\mathcal{I}} \!=\! \braket{\epsilon \zeta^{\mathcal{I}}}$, and the operator $\mathcal{W}(u)$ is just the same as in \cite{Kunz:2024wsx}:
  \begin{equation*}
  \label{fVOgauged}
  \mathcal{W}(u) = c\, n\, \tilde{n}\, \delta(\gamma)\delta(\gamma_{ab} u^a u^b)\,\delta(\tilde{\gamma})\delta(\tilde{\gamma}_{ab} u^a u^b) \,,
  \end{equation*}
  where $\gamma_{ab}$ and $\tilde{\gamma}_{ab}$ got reduced from $\gamma_{\mathrm{a}\mathrm{b}}$ and $\tilde{\gamma}_{\mathrm{a}\mathrm{b}}$, respectively, by setting $u^\mathrm{a}_\alpha \!=\! (u^a,0,0)\delta^0_\alpha$.\\
  
  In principle, there are also vertex operators for the massless $\tau^\prime_{i\mathcal{I}}$ and $\tilde{\tau}^{\prime\mathcal{I}}_i$ fields of the form
  \begin{align*}
  \mathcal{V}_i^\prime \!=\!\ c\, n\, \tilde{n}\, N_{\mathrm{a}\mathrm{b}}\,\delta(\gamma)\delta(\gamma_{\mathrm{a}^\prime\mathrm{b}^\prime})\,\delta(\tilde{\gamma})\delta(\tilde{\gamma}_{\mathrm{a}^{\prime\prime}\mathrm{b}^{\prime\prime}}) \int \frac{\ud t}{t^3} \bar\delta^4(t \tau^\prime_{i I} - \kappa^\prime_I) e^{t \tilde{\tau}^{\prime \dot{I}}_i \kappa^\prime_{\dot{I}}},\\
  \tilde{\mathcal{V}}_i^\prime \!=\!\ c\, n\, \tilde{n}\, N_{\mathrm{a}\mathrm{b}}\,\delta(\gamma)\delta(\gamma_{\mathrm{a}^\prime\mathrm{b}^\prime})\,\delta(\tilde{\gamma})\delta(\tilde{\gamma}_{\mathrm{a}^{\prime\prime}\mathrm{b}^{\prime\prime}}) \int \frac{\ud t}{t^3} \bar\delta^4(t \tau^\prime_{i \dot{I}} - \kappa^\prime_{\dot{I}}) e^{t \tilde{\tau}^{\prime I}_i \kappa^\prime_I},
  \end{align*}
  but corresponding integrated vertex operators will vanish because of some of the PCOs, and an orphan fixed vertex operator in correlation functions would return zero. $\mathcal{V}_i^\prime$ and $\tilde{\mathcal{V}}_i^\prime$ without the ghost fields except $c$ would be valid fixed vertex operators, but they would be completely decoupled from the system. This proves that the $\tau^\prime_{i\mathcal{I}}$ and $\tilde{\tau}^{\prime\mathcal{I}}_i$ fields are absent from the physical spectrum and contribute to scattering amplitudes only through partition functions in loop amplitudes.\\ 
  
  PCOs for fermionic symmetries and insertion operators for bosonic symmetries are
  \begin{equation*}
  \begin{aligned}
  &\Upsilon_{\beta}(z) \!=\! \delta(\beta) [Q_B, \beta], &&\Upsilon_{\hat{u} \beta^{\mathrm{a}\mathrm{b}}}(z) \!=\! \delta(\beta^{\mathrm{a}\mathrm{b}} \hat{u}_\mathrm{a}^\alpha \hat{u}_{\mathrm{b}\alpha}) [Q_B, \beta^{\mathrm{a}\mathrm{b}} \hat{u}_\mathrm{a}^\alpha \hat{u}_{\mathrm{b}\alpha}],\\
  &\Upsilon_{\tilde{\beta}}(z) \!=\! \delta(\tilde{\beta}) [Q_B, \tilde{\beta}], &&\Upsilon_{\hat{u} \tilde{\beta}^{\mathrm{a}\mathrm{b}}}(z) \!=\! \delta(\tilde{\beta}^{\mathrm{a}\mathrm{b}} \hat{u}_\mathrm{a}^\alpha \hat{u}_{\mathrm{b}\alpha}) [Q_B, \tilde{\beta}^{\mathrm{a}\mathrm{b}} \hat{u}_\mathrm{a}^\alpha \hat{u}_{\mathrm{b}\alpha}],\\
  &\Upsilon_{m}(z) \!=\! m \;\delta(\{Q_B, m\}), &&\Upsilon_{\tilde{m}}(z) \!=\! \tilde{m}\; \delta(\{Q_B, \tilde{m}\}),\quad \Upsilon_{M^{\mathrm{a}\mathrm{b}}}(z) \!=\! M^{\mathrm{a}\mathrm{b}} \delta(\{Q_B, M^{\mathrm{a}\mathrm{b}}\}).
  \end{aligned}
  \end{equation*}
  Here $\hat{u}_\mathrm{a}^\alpha$ is a reference little group spinor that forms with $u^\mathrm{a}_\alpha$ a limited basis in SO(4,$\mathbb{C}$) such that $\epsilon_A^\alpha u^{\mathrm{a}}_{\alpha} \hat{u}_{\mathrm{a}\beta} \!=\! U \epsilon_{A\beta}$ and $q_\mathcal{I}^\alpha u^{\mathrm{a}}_{\alpha} \hat{u}_{\mathrm{a}\beta} \!=\! U q_{\mathcal{I}\beta}$ with $U \!\ne\! 0$. In amplitudes $\hat{u}$ will fall out based on solutions of the scattering equations (\cite{Geyer:2020} and see below). For simplicity all of the $\Upsilon$'s will be called PCOs in the remainder of the article, even when of bosonic origin.\\

Integrated vertex operators can now be obtained in straightforward fashion. Applying the PCOs to fixed vertex operators results in:
\begin{align}
\label{iVOw}
  V \!\!=\!\!\! \int \!\!\ud \sigma &\,\ud\mu(u) \,\ud\mu(v) w(u) \bar{\delta}(\!v^\alpha_{\mathrm{a}} \epsilon^{\mathrm{a}}_\alpha \!-\! 1) \bar{\delta}(\!u^{\alpha}_{\mathrm{a}} \lambda^{\mathrm{a}} _A \!\!-\! v^{\alpha}_{\mathrm{a}} \kappa^{\mathrm{a}}_A) \bar{\delta}(\!u^{\alpha}_{\mathrm{a}} \eta^{\mathrm{a}}_{\mathcal{I}} \!-\! v^{\alpha}_{\mathrm{a}} \zeta^{\mathrm{a}}_{\mathcal{I}})  e^{u_{\alpha}^{\mathrm{a}} \mu_{\mathrm{a}}^A \!\epsilon^{\alpha}_A + u_{\alpha}^{\mathrm{a}}  \tilde{\eta}_{\mathrm{a}}^{\mathcal{I}} q^{\alpha}_{\mathcal{I}}}\!,\nonumber\\
  w(u) \!= &\delta(\text{Res}_{\sigma}(\lambda_{A\mathrm{a}} \lambda_B^\mathrm{a}\Omega^{AB})) \delta(\text{Res}_{\sigma}(\eta_{\mathcal{I}\mathrm{a}}\eta^\mathrm{a}_\mathcal{J}\Omega^{\mathcal{I}\mathcal{J}})) \!\left(\frac{\hat{u}^\alpha_\mathrm{a} \lambda^\mathrm{a}_A \epsilon^A_\alpha}{U} \!+\! \epsilon^{\beta A} \frac{u^\mathrm{a}_\beta \rho_{1\mathrm{a}A} \rho^\mathrm{b}_{1B} \hat{u}_\mathrm{b}^\alpha}{U} \epsilon_\alpha^B\!\right)\\
    &\left(\frac{\hat{u}^\alpha_\mathrm{a} \lambda^\mathrm{a}_A \epsilon^A_\alpha}{U} \!+\! \epsilon^{\beta A} \frac{u^\mathrm{a}_\beta \rho_{2\mathrm{a}A} \rho^\mathrm{b}_{2B} \hat{u}_\mathrm{b}^\alpha}{U} \epsilon_\alpha^B \!+\! \frac{\hat{u}^\alpha_\mathrm{a} \eta^\mathrm{a}_\mathcal{I} q_\alpha^\mathcal{I}}{U} \!+\! q^{\beta\mathcal{I}} \frac{u^\mathrm{a}_\beta \tau_{\mathrm{a}\mathcal{I}} \tau^\mathrm{b}_\mathcal{J}\hat{u}_\mathrm{b}^\alpha}{U}\!q_\alpha^\mathcal{J}\!\right).\nonumber  
 \end{align}
 Important to note: $V$ is only BRST-invariant because $u^\mathrm{a}_\alpha, \hat{u}_\mathrm{b}^\beta$ form the above mentioned limited 2-dimensional basis for SO(4,$\mathbb{C}$). Because of the invariant integration measure on G$_Q$ the vertex operators are also invariant under the little group and, therefore, the incorporation of $\Upsilon_{M_{\mathrm{a}\mathrm{b}}}$ into the BRST-invariant integrated vertex operators is already taken care of.\\
 
 For tree amplitudes, based on the solutions of the scattering equations, the residua in the formula for $w(u)$ are just to fix $\kappa^\mathrm{a}_{\alpha} \kappa_\mathrm{a}^{\alpha} \!=\! \tilde{\kappa}^\mathrm{a}_{\dot\alpha} \tilde{\kappa}_\mathrm{a}^{\dot\alpha}$ ($\alpha$ as part of the $A$ index) and $\zeta^\mathrm{a}_I \zeta_\mathrm{a}^I \!=\! -\tilde{\zeta}^\mathrm{a}_{\dot{I}} \tilde{\zeta}_\mathrm{a}^{\dot{I}}$ which can be viewed as little group transformations outside of G \cite{Arkani-Hamed:2017jhn,Penrose:1986}. To take them properly into account the all-multiplicity tree amplitudes need to be multiplied with $\prod_{i=2}^n \delta(\kappa_{i\mathrm{a}}^A \kappa^\mathrm{a}_{iA})\delta(\zeta_{i\mathrm{a}}^\mathcal{I} \zeta^\mathrm{a}_{i\mathcal{I}} )$\footnote{They were deliberately skipped over in \cite{Kunz:2024wsx}.}. The term for $n\!=\!1$ is omitted because of one vertex operator being fixed and is redundant because $\kappa_{1\mathrm{a}}^A \kappa^\mathrm{a}_{1A} \!=\! 0$ follows from the others and momentum conservation $\sum_i \kappa_{i\mathrm{a}}^A \kappa^{\mathrm{a}B}_i \!=\! 0$ baked into the polarized scattering equations \cite{Geyer:2020}. Similarly, $\zeta_{1\mathrm{a}}^\mathcal{I} \zeta^\mathrm{a}_{1\mathcal{I}} \!=\! 0$ from $\sum_i \zeta_{i\mathrm{a}}^\mathcal{I} \zeta^\mathrm{a}_{i\mathcal{I}} \!=\! 0$, again from the scattering equations, this time for the fermionic fields in the supertwistor.\\
 
 When calculating scattering amplitudes, one thing should be noted:\\
 $\lambda^\mathrm{a}_A \epsilon^A_\delta$ is orthogonal to $u^{\mathrm{a}}_\delta$ on the support of the polarized scattering equations: $u_{\mathrm{a}}^\alpha \lambda^\mathrm{a}_A \epsilon^A_\alpha \!=\! v_\mathrm{a}^\alpha \kappa^\mathrm{a}_A \epsilon^\mathrm{b}_\alpha \kappa_{\mathrm{b}}^A \!=\! 0$. Therefore, $\lambda^\mathrm{a}_A \epsilon^A_\beta = -u^{\mathrm{a}}_\alpha H^\alpha_\beta$ for some $H^\alpha_\beta$. Multiplying with $\hat{u}^\alpha_\mathrm{a}$ gives $H ^\alpha_\beta \!=\! \frac{\hat{u}^\alpha_\mathrm{a} \lambda^\mathrm{a}_A \epsilon^A_\beta}{U}$. Defining a spinor $\tilde{\epsilon}^A_\alpha$ through $\epsilon_\alpha^\mathrm{a} \!=\! \kappa^\mathrm{a}_A \tilde{\epsilon}^A_\alpha$, a polarization vector through $e^{AB} \!=\! \epsilon^{\alpha[A} \tilde{\epsilon}^{\,B]}_\alpha$, and a momentum vector through $P_{AB} \!=\! \lambda_{A\mathrm{a}} \lambda^\mathrm{a}_B$ leads to \cite{Geyer:2018}
 \begin{equation*}
e \cdot P \!=\! e^{AB} P_{AB} = \epsilon^{[A}_\alpha \tilde{\epsilon}^{B]\alpha} \lambda_A^\mathrm{a} \lambda_{\mathrm{a}B} \!=\! - u^\mathrm{a}_\beta H^\beta_\alpha \lambda_{\mathrm{a}B} \tilde{\epsilon}^{B\alpha} \!=\! - v^\mathrm{a}_\beta \kappa_{\mathrm{a}B} \tilde{\epsilon}^{B\alpha} H^\beta_\alpha \!=\! - H^\alpha_\alpha \!=\! -\frac{\hat{u}^\alpha_\mathrm{a} \lambda^\mathrm{a}_A \epsilon^A_\alpha}{U}\;.
\end{equation*}
Similarly, by defining $\epsilon_\alpha^\mathrm{a} \!=\! \zeta^\mathrm{a}_\mathcal{I} \tilde{q}^\mathcal{I}_\alpha$ and $Y_{\mathcal{I}\mathcal{J}} \!=\! \eta_{\mathcal{I}\mathrm{a}} \eta^\mathrm{a}_\mathcal{J}$ one arrives at
\begin{equation*}
q_\alpha^\mathcal{I} \tilde{q}^{\mathcal{J}\alpha} Y_{\mathcal{I}\mathcal{J}}  \!=\! -u^\mathrm{a}_\beta G^\beta_\alpha \eta_{\mathrm{a}\mathcal{J}} \tilde{q}^{\mathcal{J}\alpha} \!=\! - v^\mathrm{a}_\beta \zeta_{\mathrm{a}\mathcal{J}} \tilde{q}^{\mathcal{J}\alpha} G^\beta_\alpha \!=\! - G^\alpha_\alpha = -\frac{\hat{u}^\alpha_\mathrm{a} \eta^\mathrm{a}_\mathcal{I} q^\mathcal{I}_\alpha}{U}\;.
\end{equation*}

 It follows that tree amplitudes look like the following generalization of the ones in \cite{Kunz:2024wsx}:
 \begin{align}
 \label{TreeAmps}
 \mathcal{A}_n^{\text{tree}} = \int &\ud \mu_n^{\text{pol}|\mathcal{N}} \;\mathcal{I}_n\,,\nonumber\\
 \ud \mu_n^{\text{pol}|\mathcal{N}} &\!\!:= \;\frac{\prod_l \ud \sigma_l \ud\mu(u_l )\ud\mu(v_l)}{\text{vol}(\text{SL}(2, \mathbb{C})_{\sigma} \!\times\! \text{G}_{Q_u})} \;\prod_{r = 1}^n\bar{\delta}(\!v^\alpha_{r\mathrm{a}} \epsilon^{\mathrm{b}}_{r\alpha} \!-\! 1) &&\bar{\delta}(\!u^{\alpha}_{r\mathrm{a}} \lambda^{\mathrm{a}} _{rA}\!\!-\! v^{\alpha}_{r\mathrm{a}} \kappa^{\mathrm{a}}_{rA}) \;\bar{\delta}(\!u^{\alpha}_{r\mathrm{a}} \eta^{\mathrm{a}}_{r\mathcal{I}} \!-\! v^{\alpha}_{r\mathrm{a}} \zeta^{\mathrm{a}}_{r\mathcal{I}})\nonumber\\
  &\qquad\prod_{r=2}^n \delta(\kappa_{r\mathrm{a}}^A \kappa^\mathrm{a}_{rA})\delta(\zeta_{r\mathrm{a}}^\mathcal{I} \zeta^\mathrm{a}_{r\mathcal{I}} ),\\
    \mathcal{I}_n = &\quad\text{det}^\prime H \; \text{det}^\prime G  \quad = \quad\frac{H^{[kl]}_{[kl]}}{(u_{k\alpha}^\mathrm{a} u_{l\mathrm{a}}^\beta)(u_{k\mathrm{b}}^\alpha u_{l\beta}^\mathrm{b})} && \frac{\text{det} G^{[pr]}_{[pr]}}{(u_{p\alpha}^\mathrm{a} u_{r\mathrm{a}}^\beta)(u_{p\mathrm{b}}^\alpha u_{r\beta}^\mathrm{b})},\nonumber\\
   &H_{ij} = \frac{\epsilon_{iA}^\alpha \epsilon_{j\alpha}^A}{\sigma_{ij}}, && H_{ii} = - e_i \cdot P_i,\nonumber\\
   &G_{ij} = \frac{\epsilon_{iA}^\alpha \epsilon_{j\alpha}^A + q^\alpha_{i\mathcal{I}} q_{j\alpha}^\mathcal{I}}{\sigma_{ij}}, && G_{ii} = - (e_i \cdot P_i \!+\! q^{\alpha\mathcal{I}}_i \tilde{q}^\mathcal{J}_{i\alpha} Y_{i\mathcal{I}\mathcal{J}}),\nonumber
 \end{align}
 where $\sigma_{ij} \!= \sigma_i \!-\! \sigma_j$ and the reference spinor $\hat{u}$ has fallen out, as predicted. The choice of the two rows and columns to remove is arbitrary because the full matrices $H$ and $G$ have co-rank 2:
 \begin{equation*}
 \sum_i U_i H_{ij} \!=\! \hat{u}^\alpha_\mathrm{a} \lambda^\mathrm{a}_A \epsilon^A_{j\alpha} \!+\! \sum_{i\ne j} U_i \frac{\epsilon_{iA}^\alpha \epsilon_{j\alpha}^A}{\sigma_{ij}} \!=\! \hat{u}^\alpha_\mathrm{a} \lambda^\mathrm{a}_A \epsilon^A_{j\alpha} \!+\! (\sum_{i\ne j} \frac{u_{i\alpha}^{\mathrm{a}} \epsilon_{iA}^\alpha}{\sigma_{ij}}) \epsilon_{j\beta}^A \hat{u}^\beta_\mathrm{a} \!=\! 0,
 \end{equation*}
 and this equation can be repeated with a different reference spinor that forms a basis in G$_Q$ together with the first reference spinor. Because of the denominator it seems the two rows and the two columns to remove have to be the same, but this will be relaxed later in section \ref{LoopAmps}. A similar equation applies to the matrix $G_{ij}$.\\
 
 What happens in the gauge where the momenta and supermomenta in the vertex operators are represented by a single pair of twistors \cite{Okano:2016gya}? Then $\epsilon_{i\alpha}^A \!=\! \epsilon_i^A \delta^0_\alpha$ and $q_{i\alpha}^\mathcal{I} \!=\! q_i^\mathcal{I} \delta^0_\alpha$ such that the tree amplitude in \eqref{TreeAmps} reduces to the one in \cite{Kunz:2024wsx}, i.e. the auxiliary action $S_{\text{aux}}$ in \eqref{massSugraAction} has no impact on tree amplitudes, as promised.\\

  \section{Partition Function and Cosmological Constant}
  \label{PartitionFunc}
  This section will use some general results from the original bosonic string and superstring \cite{Verlinde:1986kw, Nelson:1986ab, Eguchi:1986ui, DHoker:1988pdl, Knizhnik:1986kf, DHoker:1989cxq, Polchinski:1998rq} that will be useful in the calculation of the partition function of the ambitwistor string.

  \subsection{Preliminaries}
  \label{Preliminaries}
  For a genus-$g$ Riemann surface $(g\!\ge \!1)$ a homology basis of cycles $a_i$ and $b_i , i = 1, \ldots, g$, can be identified, such that the intersection form is canonical, with $\#(a_i, b_j) \!=\! \delta_{i j} \!=\! - \#(b_j, a_i)$. 
  Dual to this homology class, g linearly independent holomorphic 1-forms $\omega_i$ can be chosen with $a$-periods being normalized and $b$-periods defining a symmetric matrix $\tau_{i j}$:
  \begin{equation}
  \label{abelDiff}
  \oint_{a_i}\!\omega_j = \delta_{ij}\,,\qquad \oint_{b_i}\!\omega_j = \tau_{ij}\,.
  \end{equation}
  $\omega_i$ are known as Abelian differentials of the first kind and $\tau_{ij}$ as the period matrix. Under an element $M$ of the modular group,
  \begin{equation*}
  M =  
  \begin{bmatrix}
  A &B\\
  C &D
  \end{bmatrix} \in Sp(2g, \mathbb{Z}),
  \end{equation*}
  the period matrix transforms as $\tau \rightarrow \tilde{\tau} = (A\tau + B)/(C\tau + D)$ and $M$ acts on the homology basis as $M(\genfrac{}{}{0pt}{}{b_i}{a_i})$ preserving the intersection numbers.\\
  
  An important function defined on a Riemann surface of genus g is the theta function $\theta[\alpha](\mathbf{z})$ \cite{fay1973theta, Geyer:2018xwu}:
  \begin{align}
  \label{theta}
  &\theta[\alpha](\mathbf{z}) = \sum_{m \in \mathbb{Z}^g} \exp \Bigl[ i\pi (m + \alpha^\prime)^T \tau (m + \alpha^{\prime}) + 2\pi i (m + \alpha^\prime)^T (\mathbf{z} + \alpha^{\prime\prime})\Bigr],\\
  & \theta[\alpha](- \mathbf{z}) = (-1)^{4 \alpha^\prime \cdot \alpha^{\prime\prime}} \theta[\alpha](\mathbf{z}), \quad \alpha = \begin{pmatrix}
  \alpha^\prime\\
  \alpha^{\prime\prime}
  \end{pmatrix} , \; \alpha^\prime, \alpha^{\prime\prime} \in (\mathbb{Z}/2\mathbb{Z})^g,\nonumber
  \end{align}
  where the so-called \textit{characteristic} $\alpha$ determines the parity of the theta function and at the same time designates the underlying spin structure as odd or even according to whether $4 \alpha^\prime \cdot \alpha^{\prime\prime}$ is odd or even. The argument $\mathbf{z} \in \mathbb{C}^g$ is defined in linear fashion by the so-called Abel map, given a base point $z_0$:
  \begin{equation*}
  \sum_{j=1}^n c_j z_j \rightarrow \mathbf{z} \in \mathbb{C}^g, \mathbf{z_i} \!=\! \sum_{j=1}^n c_j \int_{z_0}^{z_j} \!\omega_i .
  \end{equation*}
  In the following the Abel map will be assumed implicitly, skipping any bolding of arguments. One quantity occurring multiple times later in chiral determinants is the vector $\Delta$ of Riemann constants (\cite{DHoker:1988pdl}, section 6.E):
  \begin{equation*}
  \Delta \in \mathbb{C}^g\,, \qquad\Delta_i = - \frac{1 + \tau_{i i}}{2} + \sum_{j \ne i} \oint_{a_j} \omega_j(z) \int_{z_0}^z \omega_i,
  \end{equation*}
  where $\Delta_i$ only depends on the base point and not on the path of integration. The vector of Riemann constants is an important quantity in the Riemann vanishing theorem\cite{fay1973theta} for the theta function $\theta(z) \equiv \theta[0](z)$ which can be formulated in the following form:
  \begin{equation}
  \label{RVT}
  \theta(z) = 0 \quad \Leftrightarrow \quad \exists \; z_1,\ldots,z_{g-1} \; \text{with} \; z =  \Delta - \sum_{i=1}^{g-1} z_i\;.
  \end{equation}
  
  The prime form is defined for any odd spin structure $\alpha$ by \cite{DHoker:1989cxq}
  \begin{equation}
  \label{primeF}
  E(z, w) = \frac{\theta[\alpha](z - w)}{h_{\alpha}(z) h_{\alpha}(w)}\,,\qquad h_{\alpha}(z) = \left(\sum_{i=1}^g \partial_i \theta[\alpha](0) \omega_i(z)\right)^{\frac{1}{2}}\,,
  \end{equation}
  which is independent of the choice of odd characteristic $\alpha$ and is a holomorphic $(\!-\frac{1}{2})$ form in both $z$ and $w$ with a unique simple zero for $z \!=\! w$. For more properties of the prime form and the theta function $\theta[\alpha]$ the reader is referred to \cite{fay1973theta}.\\
  
  The Szeg\"{o} kernel for spin structure $\alpha$ is defined as 
  \begin{align}
  \label{Szego}
  &S_{\alpha}(z, w | \tau) = \frac{\theta[\alpha](z-w)}{E(z, w) \theta[\alpha](0)}\,,&\alpha \; \text{even}\,,\nonumber\\
  &S_{\alpha}(z, w | \tau) = \frac{1}{E(z,w)} \frac{\sum_{i=1}^g  \partial_i \theta[\alpha](z-w) \omega_i(y))}{\sum_{i=1}^g \partial_i \theta[\alpha](0) \omega_i(y)}\,, &\alpha \;\; \text{odd}\,.
  \end{align}
  Because of $\theta[\alpha](0) = 0$ for odd $\alpha$, the corresponding Szeg\"{o} kernel needs to be modified from the one for even $\alpha$, and there are some subtleties involved with respect to the choice of modification \cite{DHoker:1988pdl, DHoker:1989cxq}. The one shown in \eqref{Szego} is the one typically used, mainly because it is meromorphic with a simple pole for $z = w$ and the choice of point $y$ is arbitrary.\\
  
  The partition function of a fermionic $b\!-\!c$ system of weight $\lambda$ for $b$ and $1-\lambda$ for $c$ is defined for $g \ge 2, \lambda > 0$ as \cite{Verlinde:1986kw}
  \begin{align}
  \label{ZsDef}
  &Z_{\lambda,\alpha}(z_1, \ldots,z_G) \!=\!\! \int \!\!\mathrm{D} b \mathrm{D} c \; b(z_1) \dots b(z_G) \exp\!\left(\!-\frac{1}{2\pi} \int \!\ud^2 z \sqrt{g} \; b(z) \bar{\nabla}_{1-\lambda}^\alpha c(z) \!\right)\!, &\!\!\!\!\!\lambda > 1,\nonumber\\
  &Z_{1,\alpha}(z_1, \ldots,z_g, w) \!=\! \!\int \!\!\mathrm{D} \beta \mathrm{D} \gamma \, \beta(z_1) \dots \beta(z_g) \gamma(w) \exp\!\left(\!-\frac{1}{2\pi} \!\int \!\ud^2 z \sqrt{g} \,\beta(z) \bar{\nabla}_{1-\lambda}^\alpha \gamma(z) \!\right)\!,\\
  &Z^{\text{even}}_{\frac{1}{2},\alpha}\quad =\! \!\int \!\!\mathrm{D} \chi \mathrm{D} \psi  \exp\!\left(\!-\frac{1}{2\pi} \int \!\ud^2 z \sqrt{g} \; \chi(z) \bar{\nabla}_{\frac{1}{2}}^\alpha \psi(z)\!\right)\!,\nonumber\\
  &Z^{\text{odd}}_{\frac{1}{2},\alpha}(w) =\! \!\int \!\!\mathrm{D} \chi \mathrm{D} \psi \; \chi(w) \,\psi(w) \exp\!\left(\!-\frac{1}{2\pi} \int \!\ud^2 z \sqrt{g} \; \chi(z) \bar{\nabla}_{\frac{1}{2}}^\alpha \psi(z)\!\right)\!,\nonumber
  \end{align}
  where $Q = (2\lambda-1)$, in $Z_{\lambda}$ for $\lambda > 1$ there are $G = Q(g-1)$ zero modes for $b$, in $Z_1\;g$ zero modes for $\beta$ and one for $\gamma$, and for spinors of spin $\frac{1}{2}$ on even spin structures no zero mode and on odd spin structures one zero mode each for $\chi$ and $\psi$. The case $g \!=\!1$ will be handled later separately. These partition functions were calculated in \cite{Verlinde:1986kw, DHoker:1988pdl, Knizhnik:1986kf, DHoker:1989cxq} to be
  \begin{align}
  \label{Zs}
  &Z_{\lambda > 1,\alpha}(z_1, \ldots,z_G) \!=\! \tilde{Z}_1^{-\frac{1}{2}} \theta[\alpha](\sum_{i=1}^{G} \!\!z_i - Q \Delta) \prod_{j > i = 1}^G \!\!E(z_i, z_j) \prod_{i=1}^G \!\sigma(z_i)^Q,\nonumber\\
  &Z_{1,\alpha}(z_1, \ldots,z_g, w) \!=\! \tilde{Z}_1^{-\frac{1}{2}} \theta[\alpha](\sum_{i=1}^g z_i - w - \Delta) \frac{\prod_{j > i = 1}^g \!\!E(z_i, z_j) \prod_{i=1}^g \!\sigma(z_i)}{\prod_{ i = 1}^g \!\!E(z_i, w) \sigma(w)},\nonumber\\
  &Z^{\text{even}}_{\frac{1}{2},\alpha} =\! \tilde{Z}_1^{-\frac{1}{2}} \theta[\alpha](0)\,, \;\qquad\qquad Z^{\text{odd}}_{\frac{1}{2},\alpha}(z) =\! (\tilde{Z}_1^{\frac{1}{2}} \text{det}_{\omega})^{-1} \; h_{\alpha}(z)^2,\\
  &\tilde{Z}_1 = \frac{Z_{1,0}(z_1, \ldots,z_g, w)}{\text{det}_{\omega}}, \qquad\;\; \text{det}_{\omega} \equiv \text{det} ||\omega_i(z_j)||, \qquad \int_{\Sigma}\!\!\ud^2 \!z \;\bar{\omega}_i \omega_j(z) \!= <\!\omega_i|\omega_j \!\!> =\! 2i \text{Im} \,\tau _{ij},\nonumber
  \end{align}
  where $\Delta$ is the vector of Riemann constants, $h_{\alpha}$ has been defined in \eqref{primeF}, $\omega_i(z)$ are the Abelian differentials from \eqref{abelDiff} making up a basis for ker$(\bar{\nabla}_1^\alpha)$, $Z^{\text{odd}}_{\frac{1}{2},\alpha}(z)$ has been calculated with the zero modes factored out, and $\sigma(z)$ is a holomorphic $g/2$ form without zeros or poles and with the property
  \begin{equation*}
  \frac{\sigma(z)}{\sigma(w)} = \frac{\theta(z - \sum_{i=1}^g r_i  -  \Delta)}{\theta(w -  \sum_{i=1}^g r_i  - \Delta)} \prod_{i=1}^g \frac{E(w, r_i)}{E(z, r_i)}\,,
  \end{equation*}
  where $r_i$ are arbitrary points on the worldsheet. According to \cite{Verlinde:1986kw}, the differential $\sigma(z)$ is the carrier of gravitational anomaly, i.e. in correlation functions all powers of $\sigma$ regardless of arguments should sum up to zero. Equation $(5.13)$ in \cite{verlinde_rijksuniversiteit_utrecht_1988} gives a representation of $\sigma(z)$ which will be useful in section \ref{NonSeparating} when dealing with non-separating degeneration:
  \begin{equation}
  \label{sigma}
  \sigma(z) = \exp \Bigl [ \sum_{i=1}^g \oint_{a_i} \ud w \; \omega_i(w)\,\ln E(z, w) \Bigr ]\,.
  \end{equation}
  
   For integer $\lambda$ the spin structure should not matter and typically, particularly for the original superstring, periodic boundary conditions are chosen \cite{Verlinde:1986kw}, i.e. in $Z_{2,\alpha}$ and $Z_{1,\alpha}$ of \eqref{Zs} $\theta[\alpha]$ is replaced by $\theta \!=\! \theta[0]$ like in $\tilde{Z}_1$. The reason for this is that the bosonic part of the string should have periodic boundary conditions, and the related ghosts with integer conformal weight should align to it \cite{Witten:1985cc}. But the ambitwistor string considered here is described by bosonic and fermionic spinors and it will turn out that for the consistency of the model and especially its modular invariance the boundary conditions of all chiral ghosts need to align with the ones of the spinors, similarly to winding chiral bosons and to the bosonic ghosts (of conformal weight $\frac{3}{2}$) in the superstring.\\
  
  Now, assuming that $\{\phi_i\}$ stands for a basis of ker$(\bar{\nabla}_{\lambda}^\alpha)$ and $\{\nu_i\}$ for one of ker$(\bar{\nabla}_{1-\lambda}^\alpha)$, a chiral determinant det$\bar{\nabla}_{1-\lambda}^\alpha$ can be defined as a holomorphic section of a line bundle over the moduli space with fibre generated by $(\nu_1 \wedge \ldots \wedge \nu_n) \otimes (\phi_1 \wedge \ldots \phi_{G+n}), G=Q(g-1)$ and formally represented as a fermionic path integral \cite{Verlinde:1986kw, DHoker:1989cxq}
  \begin{align}
  \label{chiralDets}
  &\text{det} \bar{\nabla}_{1-\lambda}^\alpha = \frac{Z_{\lambda, \alpha}(z_1,\ldots,z_G)}{\text{det}||\phi_i(z_j)||}\cdot \phi_1 \wedge \ldots \wedge \phi_G\,,&\lambda > 1,\nonumber\\
  &\text{det} \bar{\nabla}_0^\alpha = \frac{Z_{1,\alpha}(z_1,\ldots,z_g, w)}{\text{det}_{\omega}\; e}\cdot \omega_1 \wedge \ldots \wedge \omega_g \!\otimes\! e \,,\\
  &\text{det} \bar{\nabla}^{\text{ev}}_{\frac{1}{2},\alpha} = Z^{\text{even}}_{\frac{1}{2},\alpha}\,,\qquad\text{det} \bar{\nabla}^{\text{odd}}_{\frac{1}{2},\alpha}  = \frac{Z^{\text{odd}}_{\frac{1}{2},\alpha}(z)}{h_{\alpha}(z)^2}\,\nonumber \cdot h_{\alpha} \!\otimes\! h_{\alpha}\,.
  \end{align}
  
  As can be expected, for bosonic $b\!-\!c$ systems the general rule is that the chiral determinants are the inverse of the fermionic ones:
  \begin{equation*}
  \text{det}_{\text{[bos]}}\bar{\nabla}_{1-\lambda}^\alpha = (\text{det}_{\text{[ferm]}}\bar{\nabla}_{1-\lambda}^\alpha)^{-1}\,.
  \end{equation*}
  The same applies to the $Z$ functions.\\
  
  Modular transformations $\begin{pmatrix}
  A &B\\
  C &D
  \end{pmatrix} \in$ Sp(2$g, \mathbb{Z}$) of the $Z$ functions have been calculated in \cite{Verlinde:1986kw} to be:
  \begin{align}
  \label{modTrfs}
  &Z_{\lambda,\tilde{\alpha}}(\tilde{z}_1, \ldots, \tilde{z}_G) = \epsilon^{\frac{2}{3}} e^{i\pi\phi(\alpha)} Z_{\lambda,\alpha}(z_1, \ldots, z_G), \qquad\qquad(\lambda \ne \frac{1}{2}),\nonumber\\
  &Z^{\text{odd}}_{\frac{1}{2},\tilde{\alpha}}(\tilde{z}) = \epsilon^{\frac{2}{3}} \;e^{i\pi\phi(\alpha)} \; \text{det}||C\tau + D|| \; Z^{\text{odd}}_{\frac{1}{2},\alpha}(z), &\\
  &Z^{\text{even}}_{\frac{1}{2},\tilde{\alpha}} = \epsilon^{\frac{2}{3}} \; e^{i\pi\phi(\alpha)} \;Z^{\text{even}}_{\frac{1}{2},\alpha} ,\nonumber\\
  &\text{with} \; \tilde{z}_i \!=\! (\tau C^T + D^T)^{-1}_{ij} z_j,\quad \tilde{\omega}_i \!=\! \omega_j (C\tau + D)^{-1}_{ji},\quad \tilde{\Delta} \!-\! \tilde{\tau}\delta_1 \!-\! \delta_2 \!=\! (\tau C^T + D^T)^{-1} \!\Delta, \nonumber\\
  &\qquad \tilde{\alpha} \!=\! \alpha^{\prime} \!+\! \delta, \quad\begin{bmatrix}
  \alpha_1^{\prime}\\
  \alpha_2^{\prime}
  \end{bmatrix} \!=\! \begin{bmatrix}
  D &-C\\
  -B &A
  \end{bmatrix} \begin{bmatrix}
  \alpha_1\\
  \alpha_2
  \end{bmatrix}, \quad\begin{bmatrix}
  \delta_1\\
  \delta_2
  \end{bmatrix} \!=\! \frac{1}{2} \begin{bmatrix}
  \text{diag} C D^T\\
  \text{diag} A B^T
  \end{bmatrix}, \nonumber\\
  &\qquad \epsilon^8 \!=\! 1,\qquad\quad \phi(\alpha) = \alpha_1^{\prime} \cdot \alpha_2^{\prime} - \alpha_1 \cdot \alpha_2 + 2 \alpha_1^{\prime} \cdot \delta_2 \in \frac{\mathbb{Z}_2}{4}\;.\nonumber
  \end{align}
  Even and odd spin structures transform independently, i.e. $\tilde{\alpha}$ is even or odd exactly when $\alpha$ is even or odd, respectively. Further, Szeg\"{o} kernels $S_{\alpha}$ are modular invariant up to a change in the spin structure. Also, $\tilde{\alpha} \!=\! 0$ for $\alpha \!=\! 0$ exactly when $\delta \!=\! 0$, i.e. $Z_{\lambda, 0}$ transforms into $Z_{\lambda, 0}$ up to a phase only for the so-called \textit{theta} subgroup of the modular group, potentially destroying the modular invariance of the system.\\
  
For the special case of $g=1$, the $b\!-\!c$ system with $\lambda \!=\! 2$ has one zero mode for $b$ and one for $c$, like for $\lambda \!=\! 1$ although there is still a difference of picking up an additional factor of $\text{det}_{\omega}$. Also, $\tilde{Z}_1$ has $\theta \!=\! \theta[0] \!=\! \theta_3$ replaced by $-\theta_1$. Therefore,
  \begin{align}
  \label{Zg1_0}
  &\tilde{Z}_1 \!=\!  \tilde{Z}_1^{-\frac{1}{2}} \frac{-\theta_1(z \!-\! w)}{E(z, w)\, \text{det}_{\omega}} \!=\! - \tilde{Z}_1^{-\frac{1}{2}} \theta_1^{\prime}(0) \!=\! Z^{\text{odd}}_{\frac{1}{2},\alpha}(z),\nonumber\\
  &\tilde{Z}_{2,\alpha} \,\text{det}_{\omega} = \frac{Z_{2,\alpha}}{\text{det}_{\omega}} \!=\! Z_{1,\alpha} \!=\!  \tilde{Z}_1^{-\frac{1}{2}} \frac{\theta[\alpha](z \!-\! w \!-\!\Delta_\alpha)}{E(z, w)} \!=\! \tilde{Z}_1^{-\frac{1}{2}}\frac{- \theta_1^{\prime}(0)\,\text{det}_{\omega}}{r_\alpha(z\!-\!w)}
   ,\\
  &\theta_1(z) \!=\! -r_\alpha(z) \theta[\alpha](z - \Delta_\alpha),  \quad r_3(z) \!=\! e^{\frac{1}{4} \pi i\tau + \pi i(z \!+\! \frac{1}{2})} \!=\! r_4(z) , \quad r_2(z) \!=\! 1 \!=\! r_1(z),\nonumber\\
  &\Delta_3 = \Delta =  -\frac{1}{2}(\tau \!+\! 1), \qquad \Delta_4 = -\frac{\tau}{2}, \qquad \Delta_2 = -\frac{1}{2} , \qquad \Delta_1 = 0. \nonumber
  \end{align}
  Without loss of generality, in the following it suffices to consider the case where the location $w$ of the $c$ zero mode is set equal to the location $z$ of the $b$ zero mode such that the partition functions for $g=1$ become independent of the arguments. Because of $\theta_1^{\prime}(0) \!=\! -2 \pi \eta(\tau)^3$, $\eta(\tau)$ being the Dedekind eta function, it follows simply, up to some normalization constants (determined when calculating the partition function):
   \begin{align}
   \label{Zg1}
 &\tilde{Z}_{2,\alpha} \!=\! \frac{Z_{2,\alpha}}{ \text{det}_{\omega}^2 } \!=\! \frac{Z_{1,\alpha}}{ \text{det}_{\omega} } \!=\! q_\alpha^{\!-\frac{1}{4}} \eta(\tau)^2,\qquad Z^{\text{odd}}_{\frac{1}{2}} \!=\! \tilde{Z}_1 \!=\! \eta(\tau)^2, \qquad Z^{\text{even}}_{\frac{1}{2},\alpha} \!=\! \frac{\theta[\alpha](0)}{\eta(\tau)},\\
 &q = e^{\pi i \tau}, \qquad q_\alpha = \begin{cases}
 q, \quad \alpha \!=\! 3,4, \\
 1, \quad \alpha \!=\! 1,2,
 \end{cases}\nonumber
  \end{align}	
  where the $\alpha$ subscript was dropped from $Z^{\text{odd}}_{\frac{1}{2},\alpha}$ because genus $g \!=\!1$ only has one odd spin structure. These formula without the $q_\alpha$ factor look familiar from the usual RNS superstring. The $q_\alpha$ factor will be crucial for the consistency of the ambitwistor string \eqref{massSugraActionComb} when examining loop amplitudes in section \ref{LoopAmps}.\\

  \subsection{Partition Function}
  \label{PartitionFunction}
  A good starting point to find the partition function of this ambitwistor string model for Riemann surfaces of genus $g \!>\! 0$ is the general review of the worldsheet path integral in \cite{Erbin:2021smf}, Chap.2-4, of the bosonization of chiral fermion theories in \cite{Verlinde:1986kw}, and chiral holomorphic splitting of the bosonic and super string in \cite{DHoker:1988pdl}. The first main takeaway is that the modular invariant Weil-Petersson (WP) measure can be split up into the product of a holomorphic and an antiholomorphic part (for $g \ge 2, g=1$ will again be handled later as a special case):
  \begin{equation*}
  \label{WP}
  \ud(\text{WP}) \,(\text{det}\hat{P}_1^{\dagger} \hat{P}_1)^{\frac{1}{2}} = W \wedge \bar{W}, W = \ud m^{3(g-1)} \int\! \!\ud b \,\ud c \!\!\prod_{i=1}^{3(g-1)}\!\!(\mu_i,b) \exp \left[\int \ud^2 \!z\,  b(z) \bar{\nabla}_{\!-1}^\alpha c(z)\,\right],
  \end{equation*}
  where det$\hat{P}_1^{\dagger} \hat{P}_1$ is the determinant of the Laplacian with zero modes excluded and $\mu_i$ are Beltrami differentials corresponding to traceless metric deformations in the Teichm\"{u}ller space covering the moduli space. Of course, the ambitwistor string as chiral theory does not exhibit any chiral splitting, but it can use $W$ to define a modular invariant measure. With help of \eqref{chiralDets} $W$ can be expressed succinctly as
  \begin{equation*}
  W = \ud m^{3(g-1)} \;\frac{\text{det}||(\mu_i , \phi_j)||}{\text{det}||\phi_i(z_j)||} Z_{2,\alpha}(z_1, \ldots, z_{3(g-1)}) = \text{det} \bar{\nabla}_{-1}^\alpha\,.
  \end{equation*}
  Therefore, because according to \eqref{modTrfs} $Z_{2,\alpha}$ and det$\bar{\nabla}_{-1}^\alpha$ are modular invariant up to the same phase and $W \wedge \bar{W}$ is modular invariant, the measure
  \begin{equation*}
  \ud \mu_g \!=\!\frac{\phi_1 \wedge \ldots \wedge \phi_{3(g-1)}}{\text{det}||\phi_i(z_j)||} = \ud m^{3(g-1)} \;\frac{\text{det}||(\mu_i , \phi_j)||}{\text{det}||\phi_i(z_j)||}
  \end{equation*} 
  is modular invariant.\\
 
  This leads to the following formula for the partition function:
  \begin{align*}
  \label{partFct}
  Z_{g\ge2}\! = &\Biggl \langle \prod_{i=1}^{3(g-1)} \!\!\!b(y_i) \!\prod_{j = 1}^g \!\!\Upsilon_{\!m}(\!x_j\!) \Upsilon_{\!\tilde{m}}(\!\tilde{x}_j\!) \hspace{-0.5cm}\prod_{\genfrac{}{}{0pt}{}{\mathrm{a}\mathrm{b}=c_1\oplus c_2 \oplus c_3}{c_i=(\!0,0\!)\!,\!(\!0,1\!)\!,\!(\!1,1\!)}}\hspace{-0.7cm} \Bigl( \Upsilon_{\!\!M^{\mathrm{a}\mathrm{b}}}\!(\!w_{\mathrm{a}\mathrm{b}j}\!)\Upsilon_{\!\!\hat{u}\beta^{\mathrm{a}\mathrm{b}}}(\!z_{\mathrm{a}\mathrm{b}j}\!)\Upsilon_{\!\!\hat{u}\tilde{\beta}^{\mathrm{a}\mathrm{b}}}(\!\tilde{z}_{\mathrm{a}\mathrm{b}j}\!) \Bigr)\;\Upsilon_{\!\beta}(\!z_j\!)\Upsilon_{\tilde{\!\beta}}(\!\tilde{z}_j\!)\!\!\Biggr\rangle\nonumber\\
  = & \sum_{\alpha} \!\int \!\ud \mu_g \frac{Z_{2,\alpha}((y)) Z_{1,\alpha}((x))^2 Z_{1,\alpha}((w))^9}{Z_{1,\alpha}((z))^{10} Z_{1,\alpha}((\tilde{z}))^{10}}\Bigl \langle \prod^g(\bar{\delta}_{\text{tot}} \; j_{\text{tot}})\Bigr\rangle_{\!\!\alpha},\\
   \bar\delta_{\text{tot}}&\!=\! \bar{\delta}(\lambda^A_\mathrm{a} \lambda_A^\mathrm{a} \!+\! \eta^{\mathcal{I}}_\mathrm{a} \eta_{\mathcal{I}}^\mathrm{a} ) \!\ldots\!,\quad j_{\text{tot}} \!=\! (\lambda^A_\mathrm{a} \rho_{2A}^\mathrm{a} \!+\! \eta^{\mathcal{I}}_\mathrm{a} \tau_{\mathcal{I}}^\mathrm{a}) \!\ldots\!\,,\nonumber
  \end{align*}
  where $\bar\delta_{\text{tot}}$ and $j_{\text{tot}}$ are the product of all bosonic and fermionic PCOs without ghost fields, respectively, and where $\langle \ldots \rangle_{\!\alpha}$ still requires to finish the integration over all spinors. $Z_{g\ge2}$ as special case of a loop amplitude will be shown to be modular invariant in section \ref{LoopAmps}.
  \begin{description}
  \item[Even spin structures:]
  When $\alpha$ is even, then any $\lambda^A$ or $\eta^{\mathcal{I}}$ field in $\langle \ldots \rangle_{\!\alpha}$ is orphan (there are no zero modes), i.e. it cannot be contracted. Therefore, the partition function $Z_g$ vanishes on even spin structures for $g \ge 2$.
  \item[Odd spin structures:] On odd spin structures spinors have one zero mode and some of them can be saturated by occurrences in the PCOs in $\langle \ldots \rangle_{\!\alpha}$ and some of them can be canceled between the $Z$ functions for fermionic and bosonic spinors, particularly for the auxiliary spinors. Note that the zero modes of $\rho^\mathrm{a}_{1A}, \rho^\mathrm{a}_{2A},$ and $\tau^\mathrm{a}_{\mathcal{I}}$ should be among the canceled zero modes and not appear in the PCOs, because it would lead to inconsistencies. E.g. the theory should support massless particles, but the zero modes of the auxiliary spinors $\rho^\mathrm{a}_{1A}, \rho^\mathrm{a}_{2A},$ and $\tau^\mathrm{a}_{\mathcal{I}}$ are in the Majorana representation which does not allow the splitting into Weyl spinors for positive and negative helicity in four dimensions. The zero modes of $\tau_{i\mathcal{I}}^\prime$ and $\tilde{\tau}^{\prime\mathcal{I}}_i$ and the ones  of the supertwistor $\mathcal{Z}_\mathrm{a}$ indexed by the tiny group cancel against each other also. Therefore, the remaining uncanceled spinor zero modes are only the G$_Q$-indexed ones of the supertwistor $\mathcal{Z}_\mathrm{a}$. Here the (G$_Q$-indexed) fields $\lambda^\mathrm{a}_A$ and $\eta^\mathrm{a}_{\mathcal{I}}$ that occur in the PCOs should include their zero modes, being dynamic fields, but because the number of the other bosonic components, $\mu^A_\mathrm{a}$, in the supertwistor $\mathcal{Z}_\mathrm{a}$ is smaller than the one of the fermionic ones, $\tilde{\eta}^{\mathcal{I}}_\mathrm{a}\!$, some zero modes do not cancel overall. Therefore, eventually the partition function $Z_g$ vanishes on odd spin structures for $g \ge 2$ as well.
  \end{description}
  
  Genus $g = 1$ is an interesting case for several reasons. $Z_{2,\alpha}$ has one zero mode for each of its fields $b$ and $c$, and the measure for the moduli space of the closed bosonic string needs to be divided by the volume Vol(CKV) of the conformal killing vectors. The first point was already mentioned and taken care of in \eqref{Zg1} and the second point brings up the question of how to divide Vol(CKV) into a product of its chiral, hopefully holomorphic contributions.\\
  
  With help of \eqref{Zg1} and after extracting the $Z$ functions of the spinors from the currents, the partition function can be evaluated to be:
  \begin{align*}
  &Z_{g=1}^{\text{ev}} = \sum_{\alpha} \int \frac{\ud \mu_1 \,Z_{2,\alpha} \, Z_{1,\alpha}^{-9}(Z_{\frac{1}{2},\alpha}^{\text{even}})^8}{\text{Vol(CKV)}_{\text{chir}}}   \Bigl \langle\!\bar{\delta}_{\text{tot}} \; j_{\text{tot}}\!\Bigr\rangle_{\!\!\alpha} = \sum_{\alpha} \!\int\! \frac{\ud \mu_1 \text{det}_{\omega}^{-7}}{\text{Vol(CKV)}_{\text{chir}}} \!\left (\frac{\theta_{\alpha}(0)}{\eta(\tau)}\right )^{\!\!8} \!\frac{\Bigl \langle\!\bar{\delta}_{\text{tot}} \; j_{\text{tot}}\!\Bigr\rangle_{\!\!\alpha}}{q_\alpha^{-2}\eta(\tau)^{16}} ,\\
  &Z_{g=1}^{\text{odd}} =  \int \frac{\;\;\ud \mu_1 Z_{2,1} \, Z_{1,1}^{-9} (Z_{\frac{1}{2}}^{\text{odd}})^8}{\text{Vol(CKV)}_{\text{chir}}} \ud^{\!16} \psi \, \Bigl \langle\!\bar{\delta}_{\text{tot}} \; j_{\text{tot}}\!\Bigr\rangle =  \!\int\! \frac{\ud \mu_1 \text{det}_{\omega}^{-7} \, \ud^{\!16} \psi }{\text{Vol(CKV)}_{\text{chir}}} \Bigl \langle\!\bar{\delta}_{\text{tot}} \; j_{\text{tot}}\!\Bigr\rangle,\,,
  \end{align*}
  where $\ud^{\!16} \psi$ stands for the integration over net 16 fermionic spinor zero modes.
  According to \cite{Polchinski:1998rq} and from \eqref{Zs} it follows that Vol(CKV) $\!=\! 2\text{Im}\, \tau \!=\! \int_{\Sigma}\ud^2 z \!=\! \int_{\Sigma} |\text{det}_{\omega}|^2$, so it makes sense to replace Vol(CKV)$_{\text{chir}}$ with det$_{\omega}$. Because det$_{\omega}$ is a modular form of -1,
  \begin{equation*}
  \ud \mu_1 = \ud \tau \,\text{det}_{\omega}^{-2}
  \end{equation*}
  becomes a modular invariant measure. Both $\ud^{16} \psi$ and $\sum_{\alpha} q_\alpha^2 \theta_{\alpha}(0)^8 \eta(\tau)^{-24}$ are modular forms of weight -8 and later towards the end of the first part of section \ref{LoopAmps}, shortly before subsection \ref{EvenSS}, it will be shown that $\Bigl \langle\!\bar{\delta}_{\text{tot}} \; j_{\text{tot}}\!\Bigr\rangle_{\!\alpha}$ is a modular invariant form (when made also invariant under little group transformations) such that both partition functions $Z_{g=1}^{\text{ev}}$ and $Z_{g=1}^{\text{odd}}$ are modular invariant:
  \begin{align*}
  Z_{g=1}^{\text{ev}} &= \sum_{\tilde{\alpha}} \int \ud \tilde{\tau} \frac{\Bigl \langle\!\bar{\delta}_{\text{tot}} \; j_{\text{tot}}\!\Bigr\rangle_{\!\!\tilde{\alpha}}}{\text{det}_{\tilde{\omega}}^{10}} \tilde{q}_{\tilde{\alpha}}^2 \!\left (\frac{\theta_{\tilde{\alpha}}(0)}{\eta(\tilde{\tau})^3}\right )^{\!\!8} = \sum_{\alpha} \int \frac{\ud \tau}{M_{\tau}^2} \frac{\Bigl \langle\!\bar{\delta}_{\text{tot}} \; j_{\text{tot}}\!\Bigr\rangle_{\!\!\alpha}}{(\text{det}_{\omega} M_{\tau}^{-1})^{10}} q_\alpha^2 \!\left (\frac{\theta_{\alpha}(0)}{\eta(\tau)^3 M_{\tau}}\right )^{\!\!8}\,,\\
  Z_{g=1}^{\text{odd}} &=  \int \ud \tilde{\tau} \frac{\ud^{\!16} \tilde{\psi}}{\text{det}_{\tilde{\omega}}^{10}} \!\Bigl \langle\!\bar{\delta}_{\text{tot}} \; j_{\text{tot}}\!\Bigr\rangle =  \int \frac{\ud \tau}{M_{\tau}^2} \frac{\ud^{\!16} \psi M_{\tau}^{-8}}{(\text{det}_{\omega} M_{\tau}^{-1})^{10}} \!\Bigl \langle\!\bar{\delta}_{\text{tot}} \; j_{\text{tot}}\!\Bigr\rangle\,,\\
  &\tilde{\tau} \!=\! \frac{A\tau + B}{C\tau + D}\,,\qquad M_{\tau} \!=\! (C\tau + D),\qquad \tilde{\omega} \!=\! \frac{\omega}{C\tau + D}, \qquad \tilde{z} \!=\! \frac{z}{C\tau + D},\\
  &\tilde{q}_{\tilde{\alpha}}^2 = \lim_{\tilde{z} \rightarrow 0}(\frac{\theta_1(\tilde{z})}{\theta_{\tilde{\alpha}}(\tilde{z} \!-\! \tilde{\Delta})})^8 \!=\! \lim_{z \rightarrow 0}(\frac{\theta_1(z)}{\theta_\alpha(z \!-\! \Delta)})^8 \!=\! q_\alpha^2\,.
  \end{align*} 
  This corroborates the choice of det$_{\omega}$ for Vol(CKV)$_{\text{chir}}$. Besides that, it is immediately clear, that the partition functions $Z_{g=1}^{\text{ev}}$ and $Z_{g=1}^{\text{odd}}$ vanish for the same reasons as $Z_{g\ge2}$ on their respective type of spin structure.\\
  
  In summary, the partition functions vanish at every order of perturbation theory.\\

  \section{Loop Amplitudes}
  \label{LoopAmps}
  Moving on to correlators that include vertex operators, a multi-loop amplitude for $n$ vertex operators looks like (again first restricted to $g \ge 2$):
  \begin{align}
  \label{genA}
  A_{g,n}& = \left < V_1\ldots V_n \right> = \sum_{\alpha} \int \ud \mu_g \,Z_{g,\alpha}^{\text{chir}}((z),\!(\tilde{z}),\!(y),\!(x),\!(w)) \; A^{\alpha}_g((z),\!(\tilde{z}),\!(x),\!(w))\,,\nonumber\\
  Z_{g,\alpha}^{\text{chir}}&=\! \frac{Z_{2,\alpha}((y)) Z_{1,\alpha}((x))^2 Z_{1,\alpha}((w))^9}{Z_{1,\alpha}((z))^{10} Z_{1,\alpha}((\tilde{z}))^{10} \tilde{Z}_1^4}
  =\!\frac{Y_3(y)Y\!(x)Y\!(\tilde{x}) \prod_{\mathrm{a}\mathrm{b}}\!Y\!(w_{\mathrm{a}\mathrm{b}})}{Y\!(z)Y\!(\tilde{z}) \!\prod_{\mathrm{a}\mathrm{b}} \!Y\!(z_{\mathrm{a}\mathrm{b}}) \!Y\!(\tilde{z}_{\mathrm{a}\mathrm{b}})}\,,\nonumber\\ 
  A^{\alpha}_g(&(z),\!(\tilde{z}),\!(x),\!(w)) = \!\int \!\!\ud m_0\,\tilde{Z}_1^4\Bigl \langle\prod_{j=1}^g ( \bar{\delta}_{x_j} \; \bar{\delta}^{\prime}_{w_j} \; j_{z_j} \; \tilde{j}_{\tilde{z}_j}) V_1\ldots V_n\!\Bigr\rangle_{\!\!\alpha}\,,\\
  &Y_Q(y) \!\equiv\! \theta[\alpha]\!(\Delta_y^Q) \!\!\!\prod_{j > i = 1}^{Q(g-1)} \!\!\!\!\!E(y_i, y_j) \!\!\!\prod_{i=1}^{Q(g-1)} \!\!\!\!\!\sigma(y_i)^Q\,,\; Y(a) \!\equiv\! \theta[\alpha]\!(\Delta_a)  \frac{\prod_{j > i = 1}^g \!\!E(a_i, a_j) \prod_{i=1}^g \!\sigma(a_i)}{\prod_{ i = 1}^g \!\!E(a_i, a_0) \sigma(a_0)}\,, \nonumber\\
  &\Delta_a^Q \!\equiv\! \sum_{i = 1}^{Q(g-1)} \!\!\!\!a_i  - Q \Delta\,,\qquad \Delta_a \!\equiv\! \sum_{i = 1}^g a_i - a_0 - \Delta\,,\nonumber \\
  &\bar{\delta}_{x_j} \!=\! \bar{\delta}(\lambda^A_\mathrm{a}(x_j) \lambda_A^\mathrm{a}(x_j))\;\bar{\delta}(\lambda^A_\mathrm{a}(\tilde{x}_j) \lambda_A^\mathrm{a}(\tilde{x}_j)\!+\!\eta^{\mathcal{I}}_\mathrm{a}(\tilde{x}_j) \eta_{\mathcal{I}}^\mathrm{a}(\tilde{x}_j))\,,\nonumber\\
  &\bar{\delta}^{\prime}_{w_j} = \prod_{\mathrm{a}\mathrm{b}} \bar{\delta}(\{Q_B,M^{\mathrm{a}\mathrm{b}}(w_{\mathrm{a}\mathrm{b}j})\})\,,\nonumber\\
  &j_{z_j} = \lambda^A_\mathrm{a}(z_j) \rho_{1A}^\mathrm{a}(z_j) \prod_{\mathrm{a}\mathrm{b}} (\lambda^{A(\mathrm{a}}(z_{\mathrm{a}\mathrm{b}j})\rho^{\mathrm{b})}_{1A}(z_{\mathrm{a}\mathrm{b}j}))\,,\nonumber\\
  &\tilde{j}_{\tilde{z}_j} = (\lambda^A_\mathrm{a}(\tilde{z}_j) \rho_{2A}^\mathrm{a}(\tilde{z}_j)\!+\!\eta^{\mathcal{I}}_\mathrm{a}(\tilde{z}_j) \tau_{\mathcal{I}}^\mathrm{a}(\tilde{z}_j)) \prod_{\mathrm{a}\mathrm{b}} (\lambda^{A(\mathrm{a}}(\tilde{z}_{\mathrm{a}\mathrm{b}j})\rho^{\mathrm{b})}_{2A}(\tilde{z}_{\mathrm{a}\mathrm{b}j}) \!+\! \eta^{\mathcal{I}(\mathrm{a}}(\tilde{z}_{\mathrm{a}\mathrm{b}j}) \tau^{\mathrm{b})}_{\mathcal{I}}(\tilde{z}_{\mathrm{a}\mathrm{b}j}))\,,\nonumber
   \end{align}
   where $\int \!\ud m_0$ stands for integration over zero modes and only applies to odd spin structures (see subsection \ref{OddSS}), and where the vertex operators are integrated except for one set of fixed ones, one for every gauge symmetry with a ghost zero mode.\\
   
   The amplitude \eqref{genA} is general and applies to all spin structures. As already mentioned in the paragraph after \eqref{Zs}, one important point is that the differential $\sigma(z)$ is the carrier of gravitational anomaly \cite{Verlinde:1986kw} and the number of $\sigma$'s needs to cancel between the numerator and denominator of $Z_{g,\alpha}^{\text{chir}}$ in order to avoid the anomaly. It can easily be verified that this indeed happens. Further, the marked points in the PCOs stem from the chosen Beltrami differentials and because of BRST invariance the amplitude should be independent of the choice of the arguments $(x), (y), (z), (\tilde{z}), (w)$\footnote{In \cite{Verlinde:1987sd} it is shown, using BRST arguments, that this is true up to a total derivative in moduli space. See also \cite{Friedan:1985ge, Polchinski:1998rr}.}. In the actual situation this should be verified. As the worldsheet is a compact Riemann surface this can be done by showing that the amplitude is holomorphic in the marked points without any poles such that it becomes constant in these variables because of the Liouville theorem.\\
   
   $Y_Q(y)$ trivially does not have any poles. Nor does $Y\!(a)$ because of the Riemann vanishing theorem \eqref{RVT}. Therefore, the amplitude does not depend on the choice of $y_i, x_j, \tilde{x_j},$ and $w_{\mathrm{a}\mathrm{b}j}$. For the remaining marked points $z_j, \tilde{z}_j, z_{\mathrm{a}\mathrm{b}j},$ and $\tilde{z}_{\mathrm{a}\mathrm{b}j}$ one needs to look at zeros of the corresponding $Y\!(a)$ and poles of contractions of the currents $j_{z_j}$ and $\tilde{j}_{\tilde{z}_j}$ with each other and with the vertex operators $V_j$.\\
   
   Zeros in $Y\!(z)Y\!(\tilde{z}) \prod_{\mathrm{a}\mathrm{b}} \!Y\!(z_{\mathrm{a}\mathrm{b}}) Y\!(\tilde{z}_{\mathrm{a}\mathrm{b}})$ can result in double or simple poles depending on whether they are paired with contractions between $j_{z_j}$ or $\tilde{j}_{\tilde{z}_j}$ for the same points or not. For the former, the numerator is either trivially zero (e.g. $\lambda^{A(\mathrm{a}}(z_{\mathrm{a}\mathrm{b}j}) \lambda^{\mathrm{b})}_A(z_{\mathrm{a}\mathrm{b}j})$) or is proportional to some $\lambda_\mathrm{a}^A(z_j) \lambda^\mathrm{a}_A(z_j)$ or $\lambda_\mathrm{a}^A(\tilde{z}_j) \lambda^\mathrm{a}_A(\tilde{z}_j) \!+\! \eta_\mathrm{a}^{\mathcal{I}}(\tilde{z}_j) \eta^\mathrm{a}_{\mathcal{I}}(\tilde{z}_j)$ which vanishes by selecting $x_j = z_j$ or $\tilde{x}_j = \tilde{z}_j$ in $\bar{\delta}_{x_j}$. Subleading terms to the double poles and the above mentioned simple poles cancel, too, because of multiplying two Grassmann fields at the same point without contracting between each other. Simple poles for contractions between currents that are not among the zeros of the $Y$'s have zero numerators for the same reasons as the double poles above.\\
   
   Contractions of $j_{z_j}$ or $\tilde{j}_{\tilde{z}_j}$ with integrated vertex operators can also lead to double and simple poles. Double poles occur when both fields $\lambda^{aA}(z_j)$ and $\rho^b_{1A}(z_j)$ from a term in $j_{z_j}$ or both fields $\lambda^{aA}(\tilde{z}_j)$ and $\rho^b_{2A}(\tilde{z}_j)$ or $\eta^{a\mathcal{I}}(\tilde{z}_j)$ and $\tau^b_{\mathcal{I}}(\tilde{z}_j)$ from a term in $\tilde{j}_{\tilde{z}_j}$ contract with the same vertex operator. Assuming that the vertex operator has polarization $\epsilon^A$ and charge $q^{\mathcal{I}}$, then the numerator is multiplied with $\epsilon_\alpha^A \epsilon^\alpha_A$ in the bosonic case and $q_\alpha^{\mathcal{I}} q^\alpha_{\mathcal{I}}$ in the fermionic case, both vanishing because of skew $\Omega^{AB}$ and symmetric $\Omega^{\mathcal{I}\mathcal{J}}$, respectively. Subleading terms and simple poles vanish simply because the PCO terms in $j_{z_j}$ and $\tilde{j}_{\tilde{z}_j}$ arise from commutators between the BRST charge and anti-ghosts and because integrated vertex operators are BRST invariant and do not contain ghost fields.\\
  
  Unfortunately, all gauge symmetries except the worldsheet gravity have one zero mode for the ghost field in addition to the $g$ zero modes for the anti-ghost field and, therefore, among the vertex operators $V_1,\ldots,V_n$ there is one set of fixed vertex operators for all these symmetries like for tree amplitudes. These fixed vertex operators throw a monkey wrench into the argumentation of the previous paragraph. Simple poles from contractions between the PCO currents and fixed vertex operators do not simply cancel. To mitigate the obstruction, one could assume that the locations of the first set of PCOs, say for $j \!=\! 1$, can be selected arbitrarily based on BRST invariance \cite{Friedan:1985ge, Polchinski:1998rr,Verlinde:1987sd}, move them close to the fixed vertex operators to make them integrated, while generating temporary singularities from the OPEs that are not canceled by contractions between ghosts and anti-ghosts which are missing here\footnote{If the correlation function had been given with only fixed vertex operators and additional PCOs one would have encountered the same problem $n$ times and would have had to move as many PCOs close to the fixed vertex operators as necessary to make them all integrated, although only one set of PCOs would have missed out in a ghost--anti-ghost pair to cancel singularities.}. Would it be possible to extract the necessary ghost--anti-ghost pairs from the $Z_1$ functions to avoid these singularities? The answer is no, because these zero modes need to stay within $Z_1$. On the other hand, it will be shown later in the section that at least on even spin structures the amplitude with only integrated vertex operators vanishes in such a way that it cancels the temporary singularities (see paragraph after \eqref{Degenerate}). This demonstrates that on even spin structures general location independence follows from the independence of the locations of one set of PCOs without accompanying anti-ghosts to make all fixed vertex operators integrated. This is not ensured for odd spin structures and one must rely on the BRST arguments for general location independence of the PCOs.\\   
  
  To evaluate the full amplitude $A_{g,n}$  and establish modular invariance, the currents $j_{z_j}$ and $\tilde{j}_{\tilde{z}_j}$ need first to be made invariant under little group transformations, eliminating the $\bar{\delta}^{\prime}_{w_j}$'s. A similar approach to the one used for the vertex operators can be adopted:
  \begin{align}
  \label{su2Currs}
  &\bar{\delta}^{\prime}_{w_j} j_{z_j} \tilde{j}_{\tilde{z}_j} \rightarrow  j^u_{z_j} \tilde{j}^u_{\tilde{z}_j},\\
  &j^u_{z_j} = \lambda_\mathrm{a}^A(z_j) \rho^\mathrm{a}_{1A}(z_j)\! \int \!\frac{ \prod_{\mathrm{a}\mathrm{b} = c_1 \oplus c_2 \oplus c_3} \ud^2 u_{(\mathrm{a}\mathrm{b})j} \;e^{ ||u_{(\mathrm{a}\mathrm{b})j}||^2}\;u_{(\mathrm{a}\mathrm{b})j}^\mathrm{a} \lambda_{\mathrm{a}A}(z_{\mathrm{a}\mathrm{b}j}) \; u_{(\mathrm{a}\mathrm{b})j}^\mathrm{b}\rho^A_{1\mathrm{b}}(z_{\mathrm{a}\mathrm{b}j})} {\pi^{18}\prod_{i=1}^3 \prod_{c_{i1} \in c_i}^{c_{i3}} (u^\mathrm{a}_{(c_{i1})j} u_{(c_{i2})j\mathrm{a}}) (u^\mathrm{a}_{(c_{i2})j}u_{(c_{i3})j\mathrm{a}}) (u^\mathrm{a}_{(c_{i3})j} u_{(c_{i1})j\mathrm{a}})}\,,\nonumber\\
  &\tilde{j}^u_{\tilde{z}_j} =\! (\!\lambda_\mathrm{a}^A\!(\tilde{z}_j\!) \rho^\mathrm{a}_{2A}\!(\tilde{z}_j\!)\!+\! \eta_\mathrm{a}^{\mathcal{I}}\!(\tilde{z}_j\!) \tau^\mathrm{a}_\mathcal{I}\!(\tilde{z}_j\!)\!)\!\! \int \!\!\!\frac{\pi^{-18}\prod_{\mathrm{a}\mathrm{b}= c_1 \oplus c_2 \oplus c_3} \ud^2 \tilde{u}_{(\mathrm{a}\mathrm{b})j} \;e^{ ||\tilde{u}_{(\mathrm{a}\mathrm{b})j}||^2}}{\prod_{i=1}^3 \!\prod_{c_{i1} \!\in c_i}^{c_{i3}} (\tilde{u}^\mathrm{a}_{(\!c_{i1}\!)j} \tilde{u}_{(\!c_{i2}\!)j\mathrm{a}}) (\tilde{u}^\mathrm{a}_{(\!c_{i2}\!)j}\tilde{u}_{(\!c_{i3}\!)j\mathrm{a}}) (\tilde{u}^\mathrm{a}_{(\!c_{i3}\!)j} \tilde{u}_{(\!c_{i1}\!)j\mathrm{a}})}\nonumber\\
  &\qquad\qquad\qquad \prod_{\mathrm{a}\mathrm{b}} (\tilde{u}_{(\mathrm{a}\mathrm{b})j}^\mathrm{a} \lambda_{\mathrm{a}A}(z_{\mathrm{a}\mathrm{b}j}) \; \tilde{u}_{(\mathrm{a}\mathrm{b})j}^\mathrm{b}\rho^A_{2\mathrm{b}}(z_{\mathrm{a}\mathrm{b}j})\!+\!\tilde{u}_{(\mathrm{a}\mathrm{b})j}^\mathrm{a} \eta_\mathrm{a}^{\mathcal{I}}(\tilde{z}_{\mathrm{a}\mathrm{b}j}) \tilde{u}_{(\mathrm{a}\mathrm{b})j}^\mathrm{b} \tau_{\mathrm{b}\mathcal{I}}(\tilde{z}_{\mathrm{a}\mathrm{b}j}))\,,\nonumber
  \end{align}
  where the exponentials stem from the fact that the Haar measure on SU(2), in terms of spinors, is given by a product of Gaussian measures \cite{Livine:2011gp} and they make the integrals finite at infinity. $||u||^2$ is defined with help of the conjugate spinor $u^c_\mathrm{a} \!=\! \varepsilon^{\mathrm{a}\mathrm{b}} \bar{u}_\mathrm{b}$ as $||u||^2 \!=\! u^c_\mathrm{a} u^\mathrm{a} \!=\! \bar{u}_1 u_1 + \bar{u}_2 u_2 + \bar{u}_{\dot{1}} u_{\dot{1}} + \bar{u}_{\dot{2}} u_{\dot{2}} +\bar{u}_{\ddot{1}} u_{\ddot{1}} + \bar{u}_{\ddot{2}} u_{\ddot{2}}$. The nine $u_{(\mathrm{a}\mathrm{b})j}$ are not completely general but make up a basis for the little group: $u_{(\!c_{1k}\!)j}^\mathrm{a} \!=\! (u_{(\!c_{1k}\!)j}^a,\!0,\!0), u_{(\!c_{2k}\!)j}^\mathrm{a} \!=\! (0,\!u_{(\!c_{2k}\!)j}^{\dot{a}},\!0), u_{(\!c_{3k}\!)j}^\mathrm{a} \!=\! (0,\!0,\!u_{(\!c_{3k}\!)j}^{\ddot{a}}), k \!=\! 1,2,3$. The same applies for $\tilde{u}_{(\mathrm{a}\mathrm{b})j}$. The integrands have spurious singularities arising from the denominator when two spinors become orthogonal to each other, but then the numerator contains the product of two fermionic $\rho$ fields multiplied with proportional $u$ spinors making it vanish when the arguments of the $\rho$ fields are taken to be the same. On the other hand, this means that the order of integration is not arbitrary: the integration over the $u$ spinors in \eqref{su2Currs} needs to be done after the integrand becomes independent of the arguments of the PCOs. This way, in the integrand the $u_{(\mathrm{a}\mathrm{b})j}$ always form a basis except when spurious singularities occur. Again, the same is true for $\tilde{u}_{(\mathrm{a}\mathrm{b})j}$.\\
  
  In order to calculate $A^{\alpha}_g((z),\!(\tilde{z}),\!(x),\!(w))$ in \eqref{genA}, first, it can be noticed that integrating out the $\mu^A, \lambda_A, \tilde{\eta}^{\mathcal{I}},$ and $\eta_{\mathcal{I}}$ fields leads, on the support of the delta functions in the $n$ vertex operators, to solutions of the equations of motion for $\lambda_A$ and $\eta_{\mathcal{I}}$ and to polarized scattering equations at loop level:
  \begin{equation}
  \label{ScattEqs}
  \begin{aligned}
  &v_{r\mathrm{a}}^\alpha \kappa_{rA}^\mathrm{a} \! = \! u_{r\mathrm{a}}^\alpha \lambda^\mathrm{a}_A(\sigma_r), \; r=1 ,\ldots, n\;,&\lambda^\mathrm{a}_A(\sigma) = \lambda^\mathrm{a}_{0A} \frac{h_{\alpha}(\sigma)}{\sqrt{\ud \sigma}} + \sum_{l =1}^n u_l^{\mathrm{a}\beta} \epsilon_{l A\beta} \frac{S_{\alpha}(\sigma, \sigma_l | \tau)}{\sqrt{\ud \sigma}\sqrt{\ud \sigma_l}}, \\
  &v_{r\mathrm{a}}^\alpha \zeta_{r\mathcal{I}}^\mathrm{a}  = \! u_{r\mathrm{a}}^\alpha \,\eta_{\mathcal{I}}^\mathrm{a}(\sigma_r), \; r=1, \ldots ,n\;, &\eta^\mathrm{a}_{\mathcal{I}}(\sigma) = \eta^\mathrm{a}_{0\mathcal{I}}\, \frac{h_{\alpha}(\sigma)}{\sqrt{\ud \sigma}} + \sum_{l =1}^n u_l^{\mathrm{a}\beta} q_{l\mathcal{I}\beta} \frac{S_{\alpha}(\sigma, \sigma_l | \tau)}{\sqrt{\ud \sigma}\sqrt{\ud \sigma_l}},
  \end{aligned}
  \end{equation}
  where for even spin structures the zero mode terms with $\lambda^\mathrm{a}_{0A} $ and $\eta^\mathrm{a}_{0A}$ are missing\footnote{As explained in section \ref{PartitionFunction} the only uncanceled zero modes considered on odd spin structures are the ones from the supertwistor $\mathcal{Z}_\mathrm{a}$.}. Additionally one gets contributions from 8 remaining $\tilde{Z}_1^{\frac{1}{2}}Z_{\frac{1}{2}\alpha}$ which also are different for even and odd spin structures.\\
  
  Before continuing, the space of allowed momentum twistors and the nature of solutions to the polarized scattering equations need to be examined more closely. Starting with a general massive momentum represented by 3 pairs of twistors $\kappa_\mathrm{a}^A \!=\! (\kappa_a^A, \kappa_{\dot{a}}^A, \kappa_{\ddot{a}}^A)$ with $\kappa_{\dot{0}} \!=\! 0 \!=\! \kappa_{\ddot{1}}$ (such that $k^{AB} \!=\! \kappa_\mathrm{a}^A \kappa^{\mathrm{a}B} \!=\! \braket{\kappa^A \kappa^B}$) and applying G$_T$ transformations to mix $\kappa_{\dot{a}}^A$ and $\kappa_{\ddot{a}}^A$ leads to a representation where $\kappa_{\dot{a}}^A \kappa^{\dot{a}B} \!=\! \dot{k}^{AB}  \!=\! -\ddot{k}^{AB} \!=\! \kappa_{\ddot{a}}^A \kappa^{\ddot{a}B}$. This is the most general situation when selecting the undotted $a$-indices to represent the two twistors making up the on-shell momentum. Concerning solutions of the polarized scattering equations one can work from a solution of the ordinary scattering equations $k_i \!\cdot\! k_j \,S_{\alpha}(\sigma_i, \sigma_j | \tau) \!=\! 0$ and then solve the polarized form for the undotted $a$-indices first. It becomes apparent that for the dotted $a$-indices there are no solutions unless all particles have the same momenta $\dot{k}_i^{AB}  \!=\! - \ddot{k}_i^{AB} \!=\! \pm k_i^{AB} \!=\! \pm \kappa_{ia}^A \kappa_i^{aB}$ up to a sign or the dotted momenta are all zero except for two or three particles for which the $u_{\dot{a}}$ and $u_{\ddot{a}}$ can be selected independently from $u_a$. The same holds for the supermomenta $(\zeta_a^\mathcal{I}, \zeta_{\dot{a}}^\mathcal{I}, \zeta_{\ddot{a}}^\mathcal{I})$. Appendix \ref{Gauge} gives an example for two particles. In all cases the three $(a, \dot{a}, \ddot{a})$-indexed solutions can be kept disjointed by mapping each of them to a separate SO(4,$\mathbb{C}$)-index, e.g. $u^\mathrm{a}_\alpha \!=\! (u^a, 0, u^{\dot{a}}, u^{\ddot{a}})_\alpha$. They can receive parallel treatment by introducing the 'metric' $D^{\alpha\beta} \!=\! \text{diag}(1,1,1,-1)$. Then the reference spinors $\hat{u}^\mathrm{a}_\alpha$ can be chosen such that
  \begin{equation}
  \label{basis}
  \frac{1}{U} u^\mathrm{[a}_\alpha \hat{u}^\mathrm{b]}_\beta \!=\! \varepsilon^{ab} \delta^0_\alpha \delta^0_\beta \!+\! \varepsilon^{\dot{a}\dot{b}} \delta^2_\alpha \delta^2_\beta \!+\! \varepsilon^{\ddot{a}\ddot{b}} \delta^3_\alpha \delta^3_\beta \quad\text{with}\quad U \!=\! u_{\mathrm{a} \alpha} \hat{u}^\mathrm{a}_\beta D^{\alpha \beta} \!=\!  u_a \hat{u}^a \!\equiv\! \braket{u \hat{u}}\;.
  \end{equation}
  The same applies to $v_\alpha^{[\mathrm{a}} \epsilon_\beta^{\mathrm{b}]}$. It follows that in \eqref{TreeAmps} $\epsilon_{iA}^\alpha \epsilon_{j\alpha}^A \!=\! \epsilon_{iA} \epsilon_j^A, q^\alpha_{i\mathcal{I}} q_{j\alpha}^\mathcal{I} \!=\! q_{i\mathcal{I}} q_j^\mathcal{I},$ and  $(u_{k\alpha}^\mathrm{a} u_{l\mathrm{a}}^\beta)(u_{k\mathrm{b}}^\alpha u_{l\beta}^\mathrm{b}) \!=\! \braket{u_k u_l}^2$, i.e. tree amplitudes are manifestly the same as for the simple little group SL(2,$\mathbb{C}$). The integrated vertex operators \eqref{iVOw} can be simplified to
  \begin{align}
  \label{iVOw_simpl}
  w(u) \!= &\delta(\text{Res}_{\sigma}(\lambda_{A\mathrm{a}} \lambda_B^\mathrm{a}\Omega^{AB})) \delta(\text{Res}_{\sigma}(\eta_{\mathcal{I}\mathrm{a}}\eta^\mathrm{a}_\mathcal{J}\Omega^{\mathcal{I}\mathcal{J}}))\\
   &\Bigl(\frac{\hat{u}^\alpha_\mathrm{a} \lambda^\mathrm{a}_A \epsilon^A_\alpha}{U} \!+\! \epsilon^A_0 \rho_{1aA} \rho^a_{1B} \epsilon_0^B \!+\! \epsilon^A_2 \rho_{1\dot{a}A} \rho^{\dot{a}}_{1B} \epsilon_2^B  \!+\! \epsilon^A_3 \rho_{1\ddot{a}A} \rho^{\ddot{a}}_{1B} \epsilon_3^B\!\Bigr)\nonumber\\
   &\Bigl(\frac{\hat{u}^\alpha_\mathrm{a} \lambda^\mathrm{a}_A \epsilon^A_\alpha}{U} \!+\! \epsilon^A_0 \rho_{2aA} \rho^a_{2B} \epsilon_0^B \!+\! \epsilon^A_2 \rho_{2\dot{a}A} \rho^{\dot{a}}_{2B} \epsilon_2^B \!+\! \epsilon^A_3 \rho_{2\ddot{a}A} \rho^{\ddot{a}}_{2B} \epsilon_3^B \nonumber\\ 
   &\;+\! \frac{\hat{u}^\alpha_\mathrm{a} \eta^\mathrm{a}_\mathcal{I} q_\alpha^\mathcal{I}}{U} \!+\! q^{\mathcal{I}}_0 \tau_{a\mathcal{I}} \tau^a_\mathcal{J} q_2^\mathcal{J} \!+\! q^{\mathcal{I}}_2 \tau_{\dot{a}\mathcal{I}} \tau^{\dot{a}}_\mathcal{J} q_2^\mathcal{J} \!+\! q^{\mathcal{I}}_3 \tau_{\ddot{a}\mathcal{I}} \tau^{\ddot{a}}_\mathcal{J} q_3^\mathcal{J}\!\Bigr).\nonumber  
 \end{align}
 BRST-invariance using the PCOs in $j^u_{z_j}$ and $\tilde{j}^u_{\tilde{z}_j}$ can be reestablished by reinserting \eqref{basis}, with e.g. $\hat{u}^{\dot{b}}_2 \!=\! u_{(\!c_{2k}\!)j}^{\dot{b}}$ from \eqref{su2Currs} etc.\\
    
  The form \eqref{iVOw_simpl} for the integrated vertex operators will make integrating out the auxiliary spinors from $S_{\rho_1}$ much simpler. Consider the following matrix:
  \begin{align*}
  &\mathcal{B} = \begin{pmatrix}
  A &C\\
  -C^T &H
  \end{pmatrix},\\
  &A_{ij} = \Lambda^\mathrm{a}_{iA}(z^{\prime}_i) \Lambda_{j\mathrm{a}}^A(z^{\prime}_j) \frac{S_{\alpha}(z^{\prime}_i, z^{\prime}_j |\tau)}{\sqrt{\ud z^{\prime}_i}\sqrt{\ud z^{\prime}_j}}, \quad A\!=\! 10g\!\times\!10g \;\text{antisymmetric matrix},\nonumber\\
  &C_{ij}^{\mathrm{a}\beta} = \;\;\Lambda_{iA}^\mathrm{a}(z^{\prime}_i) \epsilon_j^{A\beta} \;\; \frac{S_{\alpha}(z^{\prime}_i, \sigma_j |\tau)}{\sqrt{\ud z^{\prime}_i}\sqrt{\ud \sigma_j}},\nonumber\\
  &H_{ij} = \;\;\epsilon^\beta_{iA} \epsilon_{j\beta}^A \;\;\, \frac{S_{\alpha}(\sigma_i, \sigma_j | \tau)}{\sqrt{\ud \sigma_i}\sqrt{\ud \sigma_j}},\;(i \!\ne\!j \!=\! 1,\ldots,n),\qquad H_{ii} = -  e \cdot P(\sigma_i)\,,\nonumber\\
  &\Lambda_{iA}^\mathrm{a}(z^\prime_i) = \begin{cases}
  \lambda_A^\mathrm{a}(z_k), i \!=\! 10k,\\
  u_{(k)l}^\mathrm{a} (u^\mathrm{b}_{(k)l} \lambda_{A\mathrm{b}}(z_{(k)l})), i \!=\! 10k+l, l\!=\!1,\ldots, 9\;,
  \end{cases}k=1,\ldots,g\;.\nonumber
  \end{align*} 
  Contractions of the auxiliary spinors do not give the determinant of this matrix $\mathcal{B}$ because $A$ and $C$ have to be treated like Pfaffians whereas this is not the case for the symmetric $H$ matrix. Rather one has to produce a type of reduced determinant by using the following rules:
  \begin{enumerate}
  \item{When selecting an element $A_{kl}$, remove rows and columns $k$ and $l$ from $\mathcal{B}$. Collect all indices $i,j$ of unselected elements $A_{ij}$ of $A$ (the unremoved rows and columns of $A$) and put all indices $i$ into set $I$ and all indices $j$ into $J$. Then $I \cap J \!=\! \emptyset$. Let $A^I_J$ denote the subset of remaining selected elements in $A$ (corresponding to the complement of $I \cup J$). Restrict the number of items in $I$ or $J$ to be less than $n-1$.}
  \item{For unused rows (has to be an even number) in $C$ select all possible mutually exclusive combinations $C^{\mathrm{a}\alpha}_{kd}, C^{\mathrm{b}\beta}_{rd}$ with $k,r \in I \cup J$ and $d \in D \subseteq \{3,\ldots,n\}$, assuming that fixed vertex operators have indices $1$ and $2$.}
  \item{Compute the (reduced) determinant of $H$ where every selected combination $C^{\mathrm{a}\alpha}_{kd}, C^{\mathrm{b}\beta}_{rd}$ replaces all of the $H_{id}$ elements with $\pm(C_{ri\alpha}^\mathrm{a} C^\alpha_{kd\mathrm{a}} - C_{ki\alpha}^\mathrm{a} C^\alpha_{rd\mathrm{a}}), i \!=\! 3,\ldots,n$. The correct sign of the replacement depends on the location of the fields that get contracted. Denote the resulting $H$ with $(H^{[12]}_{[12]})^I_{DJ}$.}
  \item{Multiply the Pfaffian of the selected sub-matrix $A^I_J$ of $A$ with the reduced determinant \emph{det}$^\prime$(see \eqref{TreeAmps}) of $H$ in point 3 and sum over all possible selections.}
  \end{enumerate}
  When calculating the determinant of $H$ in point 3 the presence of fixed vertex operators without ghost fields has to be taken into account by removing two rows and two columns. On even spin structures the choice of removal is arbitrary because $H$ is degenerate with co-rank 2:
  \begin{align}
  \label{Degen}
  \sum_i U_i C_{ri\alpha}^\mathrm{b} C_{kd\mathrm{b}}^\alpha &= \Lambda_{kA}^\mathrm{b}(z^{\prime}_k) \epsilon_{d\gamma}^A \hat{u}^\gamma_\mathrm{a} \frac{S_{\alpha}(z^{\prime}_k, \sigma_d |\tau)}{\sqrt{\ud z^{\prime}_k}\sqrt{\ud \sigma_d}} \Lambda_{rB\mathrm{b}}(z^{\prime}_r) \sum_i u_{i\beta}^\mathrm{a} \epsilon_i^{B\beta} \frac{S_{\alpha}(z^{\prime}_r, \sigma_i |\tau)}{\sqrt{\ud z^{\prime}_r}\sqrt{\ud \sigma_i}}\nonumber\\
  &= \frac{S_{\alpha}(z^{\prime}_k, \sigma_d |\tau)}{\sqrt{\ud z^{\prime}_k}\sqrt{\ud \sigma_d}} \epsilon_{d\gamma}^A \hat{u}^\gamma_\mathrm{a} \Lambda_{kA}^\mathrm{b}(z^{\prime}_k) \Lambda_{rB\mathrm{b}}(z^{\prime}_r)\lambda^{B\mathrm{a}}(z^{\prime}_r).
  \end{align}
  Subtracting $\sum_i U_i C_{ki\beta}^\mathrm{b} C_{rd\mathrm{b}}^\beta$ and setting $z^{\prime}_k \!=\! z^{\prime}_r$ leads to:
  \begin{equation}
   \label{Degenerate}
   \sum_i U_i (C_{ri\beta}^\mathrm{b} C_{kd\mathrm{b}}^\beta \!-\! C_{ki\beta}^\mathrm{b} C_{rd\mathrm{b}}^\beta) \sim \varepsilon_{ABCD}\epsilon_{d\gamma}^D \lambda^C_\mathrm{c}(\!z^{\prime}_r\!) \lambda^B_\mathrm{b}(\!z^{\prime}_r\!) \lambda^A_\mathrm{a}(\!z^{\prime}_r\!) \!=\! 0 \;\;\text{regardless of} \,(\mathrm{a}, \!\mathrm{b}, \!\mathrm{c}, \!\mathrm{d}).
  \end{equation}
  To show this, assume first that in both $C_{ri\beta}^\mathrm{b}$ and $C_{kd\mathrm{a}}^\alpha$ $\Lambda_{* A}^\mathrm{a}(z^\prime_i) \!=\! \lambda_A^\mathrm{a}(z_k)$. Then the last equation vanishes because for G$_Q \cong$ SL(2,$\mathbb{C}$) there are really only two $\mathrm{a}$ indices of $\lambda^A_\mathrm{a}(\sigma)$ to distinguish. When one of the $\Lambda_{* A}^\mathrm{a}(z^\prime_i)$ is different from $\lambda_A^\mathrm{a}(z_k)$ then the $\mathrm{b}$ index in \eqref{Degen} is restricted to one of the SL(2,$\mathbb{C}$)s in the little group G. Two reference spinors $\hat{u}$ can be chosen such that for each of them $\epsilon_{iA}^\alpha u^a_{i\alpha} \hat{u}_a^\beta \!=\! U_i \epsilon_{iA}^\beta$ with the index $a$ restricted to the single SL(2,$\mathbb{C}$) where they form a basis. Then the last equation above is zero because all little group indices come from the same SL(2,$\mathbb{C}$). Explicitly this can be verified by choosing $\kappa_{iA}^{\dot{a}} \!=\! \kappa_{iA}^{\ddot{a}} \!=\! 0$ for all vertex operators except for two\footnote{Because some PCOs contain $\lambda_A^\mathrm{a}$ with $\lambda_A^\mathrm{a}\!=\! (0,\lambda_A^{\dot{a}},0)$ or $\lambda_A^\mathrm{a}\!=\! (0,0,\lambda_A^{\ddot{a}})$, not all vertex operators can have $\kappa_{iA}^{\dot{a}} \!=\! \kappa_{iA}^{\ddot{a}} \!=\! 0$ (see appendix \ref{Gauge} for more details).}. This proves that on even spin structures there are enough degeneracies to cancel spurious singularities when all vertex operators are made integrated, as mentioned in the two paragraphs before \eqref{su2Currs}. Whether the choice of the two rows and columns to remove is arbitrary for odd spin structures is less clear.\\
  
  It also follows that the contributions from the dotted indices to the substitutions $\pm(C_{ri\alpha}^\mathrm{a} C^\alpha_{kd\mathrm{a}} - C_{ki\alpha}^\mathrm{a} C^\alpha_{rd\mathrm{a}}), i \!=\! 3,\ldots,n$ cancel out. This is obvious when all momenta are the same across the $\mathrm{a}$-indices because of symmetry and the form of the matrix $D^{\alpha\beta}$. When only two particles have dotted indices, they can be selected to be the ones with fixed vertex operators, but the direct cancellation is also evident from appendix \ref{Gauge}. For the case of three particles two of them can be chosen for the fixed vertex operators and the dotted indices of the integrated vertex operator of the third particle cancel out immediately. This means that the dotted indices only appear in the Pfaffian above. Section \ref{NonSeparating} will show that ultimately they will completely fall out of the amplitudes.\\ 
  
  Contractions of the auxiliary spinors of the other worldsheet supersymmetries in $S_{(\rho_2, \tau)}$ can be handled in the same way, using the matrix
  \begin{align*}
  &\mathcal{\tilde{B}} = \begin{pmatrix}
  \tilde{A} &\tilde{C}\\
  -\tilde{C}^T &G
  \end{pmatrix},\\
  &\tilde{A}_{ij} = (\tilde{\Lambda}^\mathrm{a}_{iA}(z^{\prime}_i) \tilde{\Lambda}_{j\mathrm{a}}^A(z^{\prime}_j) + \tilde{H}^\mathrm{a}_{i\mathcal{I}}(\tilde{z}^{\prime}_i) \tilde{H}_{j\mathrm{a}}^\mathcal{I}(\tilde{z}^{\prime}_j)) \frac{S_{\alpha}(\tilde{z}^{\prime}_i, \tilde{z}^{\prime}_j |\tau)}{\sqrt{\ud \tilde{z}^{\prime}_i}\sqrt{\ud \tilde{z}^{\prime}_j}}, \tilde{A}\!=\! 10g\!\times\!10g \;\text{antisymmetric matrix},\nonumber\\
  &\tilde{C}^{\mathrm{a}\beta}_{ij} = \;\;(\tilde{\Lambda}_{iA}^\mathrm{a}(\tilde{z}^{\prime}_i) \epsilon_j^{A\beta} + \tilde{H}_{i\mathcal{I}}^\mathrm{a}(\tilde{z}^{\prime}_i)q_j^{\mathcal{I}\beta})\;\; \frac{S_{\alpha}(\tilde{z}^{\prime}_i, \sigma_j |\tau)}{\sqrt{\ud \tilde{z}^{\prime}_i}\sqrt{\ud \sigma_j}},\nonumber\\
  &\tilde{C}^\mathrm{a}_{ij\beta} \tilde{C}^\beta_{kl\mathrm{a}} \equiv \Bigl(\epsilon_{j\beta}^A \tilde{\Lambda}^\mathrm{a}_{iA}(\tilde{z}^{\prime}_i) \tilde{\Lambda}_{kB\mathrm{a}}(\tilde{z}^{\prime}_k) \epsilon_l^{B\beta} + q_{j\beta}^\mathcal{I} \tilde{H}^\mathrm{a}_{i\mathcal{I}}(\tilde{z}^{\prime}_i)\tilde{H}_{k\mathcal{J}\mathrm{a}}(\tilde{z}^{\prime}_k) q_l^{\mathcal{J}\beta}\Bigr)\frac{S_{\alpha}(\tilde{z}^{\prime}_i, \sigma_j |\tau)}{\sqrt{\ud \tilde{z}^{\prime}_i}\sqrt{\ud \sigma_j}}\frac{S_{\alpha}(\tilde{z}^{\prime}_k, \sigma_l |\tau)}{\sqrt{\ud \tilde{z}^{\prime}_k}\sqrt{\ud \sigma_l}},\nonumber\\
  &G_{ij} =(\epsilon^\beta_{iA} \epsilon_{j\beta}^A + q^\beta_{i\mathcal{I}} q_{j\beta}^\mathcal{I}) \frac{S_{\alpha}(\sigma_i, \sigma_j | \tau)}{\sqrt{\ud \sigma_i}\sqrt{\ud \sigma_j}}, \;\; G_{ii} = - (e_i \cdot P_i \!+\! q^{\beta\mathcal{I}}_i \tilde{q}^\mathcal{J}_{i\beta} Y_{i\mathcal{I}\mathcal{J}})\,,\nonumber\\
  &\tilde{\Lambda}_{iA}^\mathrm{a}(\!\tilde{z}^\prime_i) \!=\! \begin{cases}
  \lambda_A^\mathrm{a}(\!\tilde{z}_k),\\
  \tilde{u}_{(k)l}^\mathrm{a} \tilde{u}^\mathrm{b}_{(k)l}\lambda_{A\mathrm{b}}(\!\tilde{z}_{(k)l}\!),
  \end{cases},\;\; \tilde{H}_{i{\mathcal{I}}}^\mathrm{a}(\!\tilde{z}^\prime_i) \!=\! \begin{cases}
  \eta_{\mathcal{I}}^\mathrm{a}(\!\tilde{z}_k), \,\quad\qquad\qquad i \!=\! 10k,\\
  \tilde{u}_{(k)l}^\mathrm{a} \tilde{u}^\mathrm{b}_{(k)l}\eta_{\mathcal{I}\mathrm{b}}(\!\tilde{z}_{(k)l}\!),\;i \!=\! 10k\!+\!l, l\!=\!1,\ldots,9.
  \end{cases} \nonumber\\
  \end{align*}
  Like before $H$, the resulting matrix $G$ on even spin structures is degenerate with co-rank 2 as well. And dotted indices also only appear in the Pfaffian.\\
  
  The outcome from integrating out the spinor fields can be summarized as:
  \begin{align}
  \label{AV}
  A^\alpha_g \!=\!& \!\int \!\ud m_0\,\tilde{Z}_1^4 (Z_{\frac{1}{2},\alpha})^8 \prod_{j=1}^g \bar\delta_{x_j} A^{\alpha}_{g,V,n}\,,\nonumber\\
  A^{\alpha}_{g,V,n}&\!=\!\;\Bigl \langle \prod_{j=1}^g ( j^u_{z_j} \tilde{j}^u_{\tilde{z}_j} ) V_1\ldots V_n\!\Bigr\rangle_{\!\!\alpha}\nonumber\\
  \!=\!&\prod_{l=1}^n \frac{\ud \sigma_l \ud\mu(u_l )\ud\mu(v_l)}{\text{vol}(\text{G}_Q)_u}\!\!\prod_{r = 1}^n \!\bar{\delta}(\!v^\alpha_{r\mathrm{a}} \epsilon^{\mathrm{a}}_{r\alpha} \!-\! 1) \bar{\delta}(\!u^{\alpha}_{r\mathrm{a}} \lambda^{\mathrm{a}} _{rA}\!\!-\! v^{\alpha}_{r\mathrm{a}} \kappa^{\mathrm{a}}_{rA}) \bar{\delta}(\!u^{\alpha}_{r\mathrm{a}} \eta^{\mathrm{a}}_{r\mathcal{I}} \!-\! v^{\alpha}_{r\mathrm{a}} \zeta^{\mathrm{a}}_{r\mathcal{I}})\bar\delta_r\mathcal{H} \mathcal{G},\nonumber\\
  &\bar\delta_r = \prod_{r = 2}^n \bar\delta(\text{Res}_{\sigma_r}(\lambda^\mathrm{a}_A \lambda_{B\mathrm{a}}\Omega^{AB})) \bar\delta(\text{Res}_{\sigma_r}(\eta^\mathrm{a}_\mathcal{I} \eta_{\mathcal{J}\mathrm{a}} \Omega^{\mathcal{I}\mathcal{J}})),\nonumber\\
  &\mathcal{H} = \prod_{j=1}^g \int \!\!\frac{\prod_{\mathrm{a}\mathrm{b}= c_1 \oplus c_2 \oplus c_3} \ud^2 u_{(\mathrm{a}\mathrm{b})j} \;e^{ ||u_{(\mathrm{a}\mathrm{b})j}||^2}\; \sum_{I,J} \text{Pf}(A^I_J) \sum_D \text{qDet}(H^{[ij]}_{[kl]})^I_{DJ}}{\pi^{18} \prod_{i=1}^3 \!\prod_{c_{i1} \!\in c_i}^{c_{i3}} (u^\mathrm{a}_{(\!c_{i1}\!)j} u_{(\!c_{i2}\!)j\mathrm{a}}) (u^\mathrm{a}_{(\!c_{i2}\!)j}u_{(\!c_{i3}\!)j\mathrm{a}}) (u^\mathrm{a}_{(\!c_{i3}\!)j} u_{(\!c_{i1}\!)j\mathrm{a}})},\\
  &\mathcal{G} = \prod_{j=1}^g \int \!\!\frac{\prod_{\mathrm{a}\mathrm{b}= c_1 \oplus c_2 \oplus c_3} \ud^2 \tilde{u}_{(\mathrm{a}\mathrm{b})j} \;e^{ ||\tilde{u}_{(\mathrm{a}\mathrm{b})j}||^2}\; \sum_{I,J} \text{Pf}(\tilde{A}^I_J) \sum_D \text{qDet} (G^{[pr]}_{[st]})^I_{DJ}}{\pi^{18}  \prod_{i=1}^3 \!\prod_{c_{i1} \!\in c_i}^{c_{i3}} (\tilde{u}^\mathrm{a}_{(\!c_{i1}\!)j} \tilde{u}_{(\!c_{i2}\!)j\mathrm{a}}) (\tilde{u}^\mathrm{a}_{(\!c_{i2}\!)j}\tilde{u}_{(\!c_{i3}\!)j\mathrm{a}}) (\tilde{u}^\mathrm{a}_{(\!c_{i3}\!)j} \tilde{u}_{(\!c_{i1}\!)j\mathrm{a}})},\nonumber\\
  &H_{ij} = \epsilon^\beta_{iA} \epsilon_{j\beta}^A \frac{S_{\alpha}(\sigma_i, \sigma_j | \tau)}{\sqrt{\ud \sigma_i}\sqrt{\ud \sigma_j}}, \qquad\qquad\qquad\qquad\;\, H_{ii} = - e_i \cdot P_i,\nonumber\\
  &G_{ij} =(\epsilon^\beta_{iA} \epsilon_{\beta}^A + q^\beta_{i\mathcal{I}} q_{j\beta}^\mathcal{I}) \frac{S_{\alpha}(\sigma_i, \sigma_j | \tau)}{\sqrt{\ud \sigma_i}\sqrt{\ud \sigma_j}}, \qquad\qquad\; G_{ii} = - (e_i \cdot P_i \!+\! q^{\beta\mathcal{I}}_i \tilde{q}^\mathcal{J}_{i\beta} Y_{i\mathcal{I}\mathcal{J}}).\nonumber
  \end{align}
  
  There remains two major points to get resolved, namely the integration over moduli space and modular invariance of the amplitude. Starting with the latter, it follows from \eqref{modTrfs} that $Z_{g,\alpha}^{\text{chir}} \tilde{Z}_1^4 (Z^{\text{even}}_{\frac{1}{2},\alpha})^8$ is modular invariant up to a change in the (even) spin structure. For odd spin structures and for the remainder of the amplitude one has to observe that the fields have been treated as functions, not as differential forms. Therefore, the fields and the integration variables transform as modular forms with a weight, namely all spinor fields with weight $\frac{1}{2}$, $\sigma$ with $-1$, and $u$ with $-\frac{1}{2}$\footnote{The conjugate spinor $u^c$ has modular weight $\frac{1}{2}$ such that $||u||^2 \!=\! u^c_\mathrm{a} u^\mathrm{a}$ in the exponentials in \eqref{su2Currs} and \eqref{AV} has weight 0.}. By observing that $h_{\alpha}^2$ in $Z^{\text{odd}}_{\frac{1}{2},\alpha}$ transforms like $\theta(0)$, det$_{\omega}$ as modular form of weight $-1$, and $\ud m_0$ as one of weight $-8$ it follows that for odd spin structures $\ud m_0 \,Z_{g,\alpha}^{\text{chir}} \tilde{Z}_1^4 (Z^{\text{odd}}_{\frac{1}{2},\alpha})^8$ is modular invariant up to a change in the (odd) spin structure. $j^u_{z_j}$ and $\tilde{j}^u_{\tilde{z}_j}$ have weight $1$, $\bar{\delta}_{x_j}$ weight $-2$, and $\bar{\delta}^{\prime}_{w_j}$ disappeared during $\bar{\delta}^{\prime}_{w_j} j_{z_j} \tilde{j}_{\tilde{z}_j} \rightarrow  j^u_{z_j} \tilde{j}^u_{\tilde{z}_j}$. Therefore, the weights of the currents and the $\bar{\delta}_{x_j}$'s cancel each other. Integrated vertex operators have zero modular weight. The product of two fixed vertex operators $s$ and $t$ without the ghosts is divided by $(u^\mathrm{a}_{s\alpha} u^\alpha_{t\mathrm{a}})^4$ of their little group integration variables, giving it also vanishing modular weight. When doing contractions note that Szeg\"{o} kernels are modular invariant up to a change in spin structure. Therefore,  the modular invariance of the overall amplitude $A_{g,n}$ has been established, allowing the integration to be restricted to a fundamental domain. Note that invariance under the little group was essential.\\
  
  It is important to point out that $Z_{g,\alpha}^{\text{chir}} \tilde{Z}_1^4 (Z^{\text{even}}_{\frac{1}{2},\alpha})^8$ and $\ud m_0 \,Z_{g,\alpha}^{\text{chir}} \tilde{Z}_1^4 (Z^{\text{odd}}_{\frac{1}{2},\alpha})^8$ are manifestly modular invariant only because of selecting the same boundary conditions for all integer-spin ghosts and spinors. This is in contrast to the original RNS superstring where bosonic matter and $Z_2$ use periodic boundary conditions for $g \!\ge\! 1$. Here, such periodic boundary conditions for integer-spin $Z$ functions would only allow modular invariance for the \textit{theta} subgroup, making the model much less useful. For the RNS superstring this $\theta$ modular invariance can be lifted to full modular invariance by choosing special moduli-dependent locations in the partition functions (for $g \!=\! 1$ by choosing $z \!=\! -\frac{\tau}{4} \!-\! \frac{1}{2}$ for $r_3(z)$ in \eqref{Zg1_0}, for $g \!=\! 2$ proven in an elaborate manner in \cite{DHoker:2001jaf}). This might be possible even here for the ambitwistor model, but these moduli-dependent locations would introduce infrared divergences on the compactified moduli space due to some $q^{-2}$ factor similarly to the appearance of the tachyon in the bosonic string and make the model inconsistent.\\
  
  For the integration over a fundamental domain in moduli space it is important that the moduli get localized on a set of points for solutions of the polarized scattering equations \eqref{ScattEqs} such that the integration path is well defined when being taken around the boundary \cite{Geyer:2018xwu}. The $g$ pairs of delta functions 
  \begin{equation*}
  \bar{\delta}(\lambda^A_\mathrm{a}(x_j) \lambda_A^\mathrm{a}(x_j))\;\bar{\delta}(\lambda^A_\mathrm{a}(\tilde{x}_j) \lambda_A^\mathrm{a}(\tilde{x}_j)\!+\!\eta^{\mathcal{I}}_\mathrm{a}(\tilde{x}_j) \eta_{\mathcal{I}}^\mathrm{a}(\tilde{x}_j))
  \end{equation*}
 in the $\bar{\delta}_{x_j}$'s serve this purpose. Considering that the two terms in the second delta function are different algebraic entities and, therefore, present two independent conditions the delta functions provide $3g$ constraints for the $3(g\!-\!1)\!+\! \delta_{1g}$ moduli in $\ud \mu_g$, three more than necessary for $g \!>\! 1$, but one of those three is necessary for $g \!=\! 1$. Actually, these additional conditions are needed. For tree amplitudes, because of global symmetries, three $\sigma$ and three $u$ integration variables are redundant to satisfy the polarized scattering equations, reflecting the fact that momentum conservation is built into the scattering equations. For loop amplitudes on even spin structures, solutions of the polarized scattering equations still include momentum conservation (shown in the next section) but there are only 3 $u$ variables free and no $\sigma$ variable except for one when $g \!=\! 1$ so that 3 additional conditions are needed for $g \!>\! 1$ and 2 for $g \!=\! 1$. On odd spin structures momentum conservation is no longer part of the solutions to the scattering equations but the zero modes need to satisfy additional conditions. Therefore, the number of constraints is exactly the one required.\\
  
  In summary, the loop amplitudes for $g \ge 2$ are modular invariant and well defined on moduli space, with the moduli of the integrand localized on a set of points of a fundamental domain.\\

  \subsection{Loop Amplitudes on Even Spin Structures}
  \label{EvenSS}
  The main difference between even and odd spin structures is that spin $\frac{1}{2}$ fields do not have zero modes for the former. 
  
  One consequence is that momentum and supercharge conservation are built into the polarized scattering equations, like on the tree level, in contrast to odd spin structures (see next section \ref{OddSS}):
  \begin{align*}
  0 = \sum_i \epsilon_{i\gamma}^{[A}&(v^{\mathrm{a}\gamma}_i \kappa_{i\mathrm{a}}^{B]} - u^{\mathrm{a}\gamma}_i \lambda_\mathrm{a}^{B]}\!(\sigma_i)) = \sum_i \!\kappa_{i\mathrm{a}}^A \kappa_i^{\mathrm{a}B} - \sum_{ij} u^{\mathrm{a}\gamma}_i u^\beta_{j\mathrm{a}} \epsilon_{i\gamma}^{[A} \epsilon_{j\beta}^{B]} S_{\alpha}(\sigma_i, \sigma_j | \tau) = \sum_i k_i^{AB},\\
  \sum_i \kappa_i^{A\mathrm{a}} \zeta_{i\mathrm{a}}^\mathcal{I} &= \sum_i((\kappa_i^{A\mathrm{a}}  v_{i\mathrm{a}}^\gamma) (\epsilon_{i\gamma}^\mathrm{b} \zeta_{i\mathrm{b}}^\mathcal{I}) - (\kappa_i^{A\mathrm{a}} \epsilon_{i\mathrm{a}}^\gamma) (v^\mathrm{b}_{i\gamma} \zeta_{i\mathrm{b}}^\mathcal{I})) = \sum_i (u_{i\mathrm{a}}^\gamma \lambda_\mathrm{a}^A(\sigma_i)) q_{i\gamma}^\mathcal{I} - \epsilon_i^{A\gamma} (u^\mathrm{b}_{i\gamma} \eta^\mathcal{I}_\mathrm{b}(\sigma_i))\\
  & = \sum_{ij} (\epsilon_j^{A\beta} q_i^{\mathcal{I}\gamma} - \epsilon_i^{A\gamma} q_j^{\mathcal{I}\beta}) u_{i\gamma}^\mathrm{a} u_{j\mathrm{a}\beta} S_{\alpha}(\sigma_i, \sigma_j | \tau) = 0.
  \end{align*}
  Here antisymmetry in the arguments of the Szeg\"{o} kernel has been taken advantage of and that $v^\mathrm{a}_\gamma$ and $\epsilon_\mathrm{b}^\beta$ form a basis in G$_Q$. Therefore, the two conservation laws do not need to be explicitly displayed in the amplitude.\\
  
  The all-multiplicity loop amplitude for genus $g \!\ge\! 2$ becomes
  \begin{align}
  \label{LoopAmplEven}
  A^{\text{even}}_{g,n}&\!= \!\sum_{\alpha} \! \int \!\!\ud \mu_g \,Z_g^{\alpha}((z),\!(\tilde{z}),\!(y),\!(x),\!(w)) (\prod_{j=1}^g \!\bar{\delta}_{x_j}) A^{\alpha}_{g,V,n}\;,\\
  Z_g^{\alpha} &\!= \frac{Z_{2,\alpha}((y)) Z_{1,\alpha}((x))^2 Z_{1,\alpha}((w))^9 (Z^{\text{even}}_{\frac{1}{2},\alpha})^8 }{Z_{1,\alpha}((z))^{10} Z_{1,\alpha}((\tilde{z}))^{10}} = \frac{Y_3(y)Y\!(x)Y\!(\tilde{x}) \prod_{\mathrm{a}\mathrm{b}}\!Y\!(w_{\mathrm{a}\mathrm{b}})\theta[\alpha](0)^8}{Y\!(z)Y\!(\tilde{z}) \!\prod_{\mathrm{a}\mathrm{b}} \!Y\!(z_{\mathrm{a}\mathrm{b}}) \!Y\!(\tilde{z}_{\mathrm{a}\mathrm{b}})}\,,\nonumber
  \end{align}
  with the same notation as in \eqref{genA} and \eqref{AV}, but in \eqref{LoopAmplEven} all occurrences of $\lambda^a_A(\sigma)$ and $\eta^a_\mathcal{I}(\sigma)$ being replaced with the expressions from \eqref{ScattEqs} with the zero modes excluded. Note that $\bar\delta_r = \prod_{r=2}^n \bar\delta(\kappa_{r\mathrm{a}}^A \kappa^\mathrm{a}_{rA})\bar\delta(\zeta_{r\mathrm{a}}^\mathcal{I} \zeta^\mathrm{a}_{r\mathcal{I}})$ in \eqref{AV} like for tree amplitudes. This has the consequence, together with momentum conservation and after solving the polarized scattering equations, that $\lambda^\mathrm{a}_A(\sigma)\lambda_\mathrm{a}^A(\sigma) \!=\! \sum_{r=1}^n k_{rA}^A S_{\alpha}(\sigma, \sigma_r|\tau) \!=\! 0$ for all $\sigma$'s such that $\bar\delta(\lambda^\mathrm{a}_A(\sigma)\lambda_\mathrm{a}^A(\sigma))$ can no longer be used to fix moduli and is actually singular. In other words, it is crucial to hold off with satisfying the conditions in $\bar\delta_r$ till at least the moduli integrations have been done and maybe even till the very end to avoid spurious singularities.\\
  
  One-loop amplitudes present a special case because the $(b,c)$ ghost system now also has one zero mode for both ghost and anti-ghost. Taking into account from subsection \ref{PartitionFunction} that $\ud \mu_1$ is to be replaced by $\ud \tau \, \text{det}_{\omega}^{-2}$ and $A^{\alpha}_{1,V,n}$ loses one $\ud \sigma$ integration and is divided by $\ud z \!=\! \text{det}_{\omega} \!=\! \text{Vol(CKV)}_{\text{chir}}$ and by vol(GL(1,$\mathbb{C}$))$_{\sigma}$, the one-loop amplitude becomes
  \begin{align}
  \label{1LoopAmplEven}
  &A^{\text{even}}_{1,n} \!= \!\sum_{\alpha} \! \int \!\!\frac{\ud \tau}{\text{det}_{\omega}^{10}} q_\alpha^2 \Bigl(\frac{\theta[\alpha](0)}{\eta(\tau)^3}\Bigr)^8 \,\bar{\delta}_{x_1}  \frac{A^{\alpha}_{1,V,n}}{\text{vol}(\text{GL}(1,\mathbb{C}))_{\sigma}}\,, &&q_\alpha^2 \!=\! \lim_{z \rightarrow 0}(\frac{\theta_1(z)}{\theta_\alpha(z \!-\! \Delta)})^8\;.
  \end{align}
  Here the modulus integration is localized by one of the three conditions in $\bar\delta_{x_1}$.\\
  
  When evaluating the integral one can set det$_\omega \!=\! \omega(z) \!=\! 1(\ud z)$ in \eqref{1LoopAmplEven} which has been used in \cite{Kunz:2024wsx}.\\

  \subsection{Loop Amplitudes on Odd Spin Structures}
  \label{OddSS} 
  For odd spin structures the zero modes of spinors $\mu^{\mathrm{a}A}, \lambda^\mathrm{a}_A, \tilde{\eta}^{\mathrm{a}\mathcal{I}},$ and $\eta^\mathrm{a}_{\mathcal{I}}$ need to be taken into account. $\lambda^\mathrm{a}_A$ and $\eta^\mathrm{a}_{\mathcal{I}}$ are taken care of by simply including the zero modes when replacing the fields with the expressions from \eqref{ScattEqs}. On the other hand, the zero modes of $\mu^{\mathrm{a}A}$ and $\tilde{\eta}^{\mathrm{a}\mathcal{I}}$ appear in the exponentials of the vertex operators \eqref{VO} by replacing $\mu^{\mathrm{a}A}$ and $\tilde{\eta}^{\mathrm{a}\mathcal{I}}$ with $\mu^{\mathrm{a}A} \!-\! \mu_0^{\mathrm{a}A} h_{\alpha}/\sqrt{\ud \sigma}$ and $\tilde{\eta}^{\mathrm{a}\mathcal{I}} \!-\! \tilde{\eta}_0^{\mathrm{a}\mathcal{I}}h_{\alpha}/\sqrt{\ud \sigma}$, respectively. As discussed in \ref{PartitionFunction} all other zero modes and the G$_T$-indexed zero modes of the supertwistors cancel against each other such that the remaining zero modes are the G$_Q$-indexed ones considered in this section that have only 2 degrees of freedom in the $\mathrm{a}$ components, like the remaining part of the fields.\\
  
  Using the incidence relations \cite{Geyer:2018} in four dimension \cite{Albonico:2020}, with slight modifications that take into account that the $\tilde{\eta}_\mathrm{a}^\mathcal{I}$'s, like the $\mu_\mathrm{a}^A$'s, are missing from the spectrum  but not the $\eta_{\mathrm{a}\mathcal{I}}$'s,\footnote{In this section $\beta$ and $\dot\beta$ refer to components of $\lambda_A$, $\mu^A$, and $x_0^{\beta\dot\beta}$, while $\xi$ refers to indices from the tiny group.} 
  \begin{align*}
  &\mu_0^{\mathrm{a}\dot\beta} \!\frac{h_{\alpha}}{\sqrt{\ud \sigma}} \!=\! x_0^{\beta \dot\beta} \lambda^\mathrm{a}_{\beta} \!-\! \Theta_0^{\mathcal{I}\dot\beta} \eta_{\mathcal{I}}^\mathrm{a}, &&\tilde{\mu}_0^{\mathrm{a} \beta} \!\frac{h_{\alpha}}{\sqrt{\ud \sigma}} \!=\! -x_0^{\beta \dot\beta} \tilde{\lambda}^\mathrm{a}_{\dot\beta} \!-\! \Theta_0^{\mathcal{I}\beta} \eta_{\mathcal{I}}^\mathrm{a}, &&\tilde{\eta}_0^{\mathrm{a}\mathcal{I}}\frac{h_{\alpha}}{\sqrt{\ud \sigma}} \!=\! \Theta_0^{\mathcal{I}A} \lambda^\mathrm{a}_A\,,
  \end{align*}
  together with the scattering equations \eqref{ScattEqs}, the integral over the zero modes gives
  \begin{align}
  \label{momCons}
  &\int \frac{\ud \mu(\mu_0^{\mathrm{a}A}) \; \ud \mu(\tilde{\eta}_0^{\mathrm{a}\mathcal{I}})}{\text{vol}(\text{G}_Q \times \mathbb{C})_{\mu_0}}\exp\left[-\sum_{i=1}^n U_i \right ]\;,\\
  &U_i = ((u_{i\xi}^\mathrm{a} \mu_{0\mathrm{a}}^A) \epsilon^\xi_{iA} + (u_{i\xi}^\mathrm{a} \tilde{\eta}_{0\mathrm{a}} ^{\mathcal{I}}) q^\xi_{i\mathcal{I}})\frac{h_{\alpha}}{\sqrt{\ud \sigma}} \nonumber\\
  &\quad =  x_0^{\beta \dot\beta} (\epsilon^\xi_{i \dot\beta} (u_{i\xi}^\mathrm{a} \lambda^\mathrm{a}_{i \beta})  \!-\! \epsilon^\xi_{i \beta} (u_{i\xi}^\mathrm{a} \tilde{\lambda}^\mathrm{a}_{i \dot\beta})) \!-\! \Theta_0^{\mathcal{I}A} (u_{i\xi}^\mathrm{a} \eta_{\mathcal{I}\mathrm{a}}) \epsilon^\xi_{iA}  \!+\! \Theta_0^{\mathcal{I}A} (u_{i\xi}^\mathrm{a} \lambda _{iA\mathrm{a}}) q^\xi_{i\mathcal{I}}\nonumber\\
  &\quad =  x_0^{\beta \dot\beta} ((v_{i\mathrm{a}\xi} \kappa^\mathrm{a}_{i \beta})(\epsilon^\xi_{i\mathrm{b}} \tilde{\kappa}^\mathrm{b}_{i \dot\beta})  \!-\! (\epsilon^\xi_{i\mathrm{a}} \kappa^\mathrm{a}_{i \beta}) (v_{i\mathrm{b}\xi} \tilde{\kappa}^\mathrm{b}_{i \dot\beta})) \!+\! \Theta_0^{\mathcal{I}A} (-(\epsilon^\xi_{i\mathrm{a}} \kappa^\mathrm{a}_{i \beta})(v_{i\mathrm{b}\xi} \zeta^\mathrm{b}_{\mathcal{I}}) \!+\! (v_{i\mathrm{a}\xi} \kappa^\mathrm{a}_{i \beta})(\epsilon^\xi_{i\mathrm{b}} \zeta^\mathrm{b}_{\mathcal{I}})) \nonumber\\
  &\quad = x_0^{\beta \dot\beta} \kappa^\mathrm{a}_{i\beta} \tilde{\kappa}_{i \dot\beta \mathrm{a}} \!+\! \Theta_0^{\mathcal{I}A} \kappa_{iA\mathrm{a}} \zeta^\mathrm{a}_{i\mathcal{I}} = x_0 \!\cdot\! k_i \!+\! \Theta_0^{\mathcal{I}A} \kappa_{iA\mathrm{a}} \zeta^\mathrm{a}_{i\mathcal{I}}\,\nonumber,
   \end{align}
  where $\ud \mu(\mu_0^{A\mathrm{a}})$ and $\ud \mu(\tilde{\eta}_0^{\mathrm{a}\mathcal{I}})$ are G$_Q$-invariant measures for each $A$ component of $\mu_0^{A\mathrm{a}}$ and each $\mathcal{I}$ component of $\tilde{\eta}_0^{\mathrm{a}\mathcal{I}}$, $k_i^{\beta \dot\beta} \!=\! \kappa^\mathrm{a}_{i\beta} \tilde{\kappa}_{i \dot\beta \mathrm{a}} = k_i^{\nu} \sigma_{\nu}^{\beta \dot\beta}$ is the $i$th energy momentum, $\sigma_{\nu}$ being the Pauli matrices. In the next step the integration over $\mu_0^{\mathrm{a}A}$ with four degrees of freedom is replaced with integration over $x_0^{\mu}$ disregarding the Jacobian\footnote{This can be accomplished by inserting $1\!=\! \int \ud^4x^{\nu} \delta(x^{\nu} \!\!-\! \!\sum_i \!\!u^\mathrm{a}_{i\xi}\mu_{0\mathrm{a}}^A \epsilon^\xi_{iA} \frac{h_{\alpha}}{\sqrt{\ud \sigma}} \frac{k^{\nu}}{k^2})$ with $k^{\nu} \!=\! \sum k_i^{\nu}$ and integrating over $\mu_0^{\mathrm{a}A}$ after rescaling.}, providing a delta function $\delta(\sum_{i=1}^n k_i^{\mu})$ stating energy-momentum conservation of the amplitude. In similar manner the integration over $\tilde{\eta}_0^{\mathcal{I}\mathrm{a}}$ is replaced by the integration over $\Theta_0^{\mathcal{I}A}$, resulting in a delta function $\delta(\sum_{i=1}^n \kappa_{iA\mathrm{a}} \zeta^\mathrm{a}_{i\mathcal{I}})$. \\
  
  This way the conservation laws that no longer follow from the scattering equations alone arise from the integration over half of the zero modes and the all-multiplicity loop amplitude for genus $g \!\ge\! 2$ can be represented as
  \begin{align}
  \label{LoopAmplOdd}
  A^{\text{odd}}_{g,n}&\!= \delta(\sum_{i=1}^n \!k_i) \delta(\sum_{i=1}^n \kappa_{iA\mathrm{a}} \zeta^\mathrm{a}_{i\mathcal{I}}) \!\sum_{\alpha} \! \int \!\!\ud \mu_g \,Z_g^{\alpha}((z),\!(\tilde{z}),\!(y),\!(x),\!(w)) \!\!\int\!\!\ud m_0 (\prod_{j=1}^g \!\bar{\delta}_{x_j}) A^{\alpha}_{g,V,n}\,,\nonumber\\
  Z_g^{\alpha} &\!= \frac{Z_{2,\alpha}(\!(y)\!) Z_{1,\alpha}(\!(x)\!)^2 Z_{1,\alpha}(\!(w)\!)^9 (Z^{\text{odd}}_{\frac{1}{2},\alpha})^8 }{Z_{1,\alpha}(\!(z)\!)^{10} Z_{1,\alpha}(\!(\tilde{z})\!)^{10}} \!=\! \frac{Y_3(y)Y\!(x)Y\!(\tilde{x}) \prod_{\mathrm{a}\mathrm{b}}\!Y\!(w_{\mathrm{a}\mathrm{b}})}{Y\!(z)Y\!(\tilde{z}) \!\prod_{\mathrm{a}\mathrm{b}} \!Y\!(z_{\mathrm{a}\mathrm{b}}) \!Y\!(\tilde{z}_{\mathrm{a}\mathrm{b}})}\!\!\left(\!\frac{h_{\alpha}(z)^2}{\text{det}_\omega}\!\right)^{\!\!8}\!,\\
  \ud m_0 &=  \frac{\ud \mu(\!\lambda_{0A}^\mathrm{a}) \ud \mu(\!\eta_{0\mathcal{I}}^\mathrm{a})}{\text{vol}(\text{G}_Q \times \mathbb{C})_{\lambda_0}}\,,\nonumber
  \end{align}
 where again the notation has been borrowed from \eqref{genA} and \eqref{AV}, but in \eqref{LoopAmplOdd} all occurrences of $\lambda^\mathrm{a}_A(\sigma)$ and $\eta^\mathrm{a}_\mathcal{I}(\sigma)$ being replaced with the expressions from \eqref{ScattEqs} with the zero modes included. In the measure for the zero modes its invariance under the little group G$_Q$ has been taking into account by dividing by the volume of the quotient group. Besides momentum and supercharge conservation no longer being guaranteed implicitly, $\bar\delta_r$ in \eqref{AV} does not simplify as much as for even spin structures but keeps a linear dependence on the zero modes. But again $\lambda^\mathrm{a}_A(\sigma)\lambda_\mathrm{a}^A(\sigma) \!=\! 0$ follows from $\bar\delta_r$ and the polarized scattering equations i.e. it is still a good idea to wait till the end of the amplitude evaluation to enforce $\bar\delta_r$.\\
 
 Modular invariance of the amplitude is easily re-established by observing that $Z_g^{\alpha}$ is a modular form with weight 8 which cancels against $\ud m_0$ and the integration measure of \eqref{momCons} each of which have weight -4.\\
 
 One-loop amplitudes are obtained using the same steps as in the even spin structures case:
 \begin{align}
 \label{1LoopAmplOdd}
 &A^{\text{odd}}_{1,n}\!= \delta(\sum_{i=1}^n \!k_i) \delta(\sum_{i=1}^n \!\kappa^\mathrm{a}_{iA} \zeta_{i\mathrm{a}\mathcal{I}}) \! \int \!\frac{\ud \tau}{\text{det}_{\omega}^{10}} \frac{\ud \mu(\!\lambda_{0A}^\mathrm{a}) \ud \mu(\!\eta_{0\mathcal{I}}^\mathrm{a})}{\text{vol}(\text{G}_Q \times \mathbb{C})_{\lambda_0}}  \,\bar{\delta}_{x_1}  \frac{A_{1,V,n}}{\text{vol}(\text{GL}(1,\mathbb{C}))_{\sigma}}\;.
 \end{align}
 Considering that on the torus $\omega \!=\! \ud z$ and the spinors are $\frac{1}{2}-$forms, this can be rewritten in a manifestly modular invariant form:
 \begin{align*}
 &A^{\text{odd}}_{1,n}\!=\! \delta(\sum_{i=1}^n \!k_i) \delta(\sum_{i=1}^n \kappa^\mathrm{a}_{iA} \zeta_{i\mathrm{a}\mathcal{I}}) \!\! \int \!\!\frac{\ud \tau}{\ud^2 z}\frac{\ud \mu(\!\lambda_{0A}^\mathrm{a}\!\sqrt{\!\ud z}) \ud \mu(\!\eta_{0\mathcal{I}}^\mathrm{a}\!\sqrt{\!\ud z})}{(\text{vol}(\text{G}_Q \times \mathbb{C})_{\lambda_0}\ud^4 z)}\,\bar{\delta}_{x_1}  \frac{A^{\alpha}_{1,V,n}}{\text{vol}(\text{GL}(1,\mathbb{C}))_{\sigma}}\;.
 \end{align*}
 
 In sections \ref{NonSeparating} and \ref{Separating} the loop amplitudes will be investigated for the non-separating and separating cases, respectively.\\

\subsection{Non-separating Degeneration}
\label{NonSeparating}
Non-separating degeneration happens when the amplitude is evaluated on a non-separating \textit{divisor at infinity} $\mathcal{D}^{\text{nonsep}}_{g-1,n+2}$ at the boundary of the compactified moduli space $\widehat{\mathcal{M}}_{g,n}$  corresponding to a reduced moduli space $\widehat{\mathcal{M}}_{g-1,n+2}$ \cite{Witten:2012bh}. To establish unitarity of the theory the amplitude needs to factorize properly on such a divisor.\\

In order to show this, consider the non-separating degeneration of the last homology $a_g$-cycle, and denote the corresponding modular parameter by $q_{gg} \!=\! e^{\pi i t}$ \footnote{$t$ is one of the coordinates in $\ud \mu_g$ and does not necessarily coincide with the $\tau_{gg}$ element in the period matrix but the difference is at most a constant for $q \!\rightarrow\! 0$ \cite{fay1973theta} which will be chosen to be zero.}. At the boundary divisor, the cycle $a_g$ shrinks to a point pinching an infinitely thin tube connecting a pair of locations $\sigma_{+}$ and $\sigma_{-}$ on the worldsheet and where $q_{gg}$ needs to go to zero. This can be accomplished by switching the integration path of $t$ to one that encircles the singularity at infinity, described by $\oint \!\frac{\ud q_{gg}} {q_{gg}}$. The period matrix behaves as \cite{fay1973theta}
\begin{equation}
\label{periodMatrix}
\tau = \begin{pmatrix}
\tau^{(g-1)}_{ij} &\int_{\sigma_{-}}^{\sigma_{+}}\omega_i^{(g-1)}\\
\int_{\sigma_{-}}^{\sigma_{+}}\omega_j^{(g-1)} &\frac{1}{\pi i} \text{ln} q_{gg} + \text{const}
\end{pmatrix}
+ O(q_{gg}^2),
\end{equation}
where $\tau^{(g-1)}$ is the period matrix and $\omega_i^{(g-1)}$ the holomorphic differential for the Riemann surface of genus $g \!-\! 1$.\\

There are four different scenarios: the degeneration happens on an odd or even cycle on an odd or even spin structure. They will be examined in the following order:

\begin{description}

\item [Odd cycle of an odd spin structure:]
According to \cite{fay1973theta, Tuite:2010}, the differentials $\omega_i$ and Szeg\"{o} kernels $S_{\alpha}$ will degenerate on the divisor $\mathcal{D}^{\text{nonsep}}_{g-1,n+2}$ the following way:
\begin{align}
\label{Degeneration}
&\omega_i = \omega_i^{(g-1)} + O(q_{gg}^2)\; \text{for} \; i<g\,,\nonumber\\
&\omega_g = \omega _{+ -}^{(g-1)} + O(q_{gg}^2),\\
&S_{\alpha}(\sigma, \sigma_{*}|\tau) = S_{\alpha^{(g-1)}}(\sigma, \sigma_{*}|\tau^{(g-1)}) + O(q_{gg}) \equiv S_{\alpha}^{(g-1)}(\sigma, \sigma_{*}|\tau)+ O(q_{gg}),\nonumber
\end{align}
where $\omega_i$ is an Abelian homomorphic differential of the first kind, and $\omega_{+-}$ is an Abelian differential of the third kind, with residue 1 at $\sigma \!=\!  \sigma_{+}$ and residue $-1$ at $\sigma \!=\! \sigma_{-}$.\\
 
To find out what happens to the zero modes of $\lambda_A^\mathrm{a}(\sigma) $ and $\eta_\mathcal{I}^\mathrm{a}(\sigma)$, note that for the RNS ambitwistor string the total energy-momentum operator fulfills as solution of the equation of motion \cite{Geyer:2018xwu}:
\begin{align*}
&P^\mu = \sum_{I=1}^g c_i^\mu \omega_i + \sum_{j=1}^n k_j^\mu \omega_{j*}, &&\omega_{j*} = S_{\alpha}(\sigma, \sigma_j | \tau) - S_{\alpha}(\sigma, \sigma_{*}|\tau),
\end{align*}
where $c_i^\mu$ are $g$ constants of integration of the homogeneous equation and the term with the arbitrary point $\sigma_{*}$ vanishes because of momentum conservation. Then, according to \eqref{Degeneration}, $P^\mu$ changes on the divisor to
\begin{align*}
P^\mu &\rightarrow \sum_{I=1}^{g-1} c_i^\mu \omega_i^{(g-1)}  + k_{+}^\mu S_{\alpha}^{(g-1)}(\sigma, \sigma_{+} | \tau) + k_{-}^\mu S_{\alpha}^{(g-1)}(\sigma, \sigma_{-} | \tau) + \sum_{j=1}^n k_j^\mu S_{\alpha}^{(g-1)}(\sigma, \sigma_j | \tau)\\
&= \sum_{I=1}^{g-1} c_i^\mu \omega_i^{(g-1)} + \sum_{j=1}^{n+2} k_j^\mu S_{\alpha}^{(g-1)}(\sigma, \sigma_j | \tau),
\end{align*}
where $k_{n+2}^\mu \!=\! k_{-}^\mu \!=\! - k_{+}^\mu \!=\! - k_{n+1}^\mu, \sigma_{n+1} \!=\! \sigma_{+},$ and $\sigma_{n+2} \!=\! \sigma_{-}$, i.e. $P^\mu$ becomes a solution of the equation of motion on $\mathcal{D}^{\text{nonsep}}_{g-1,n+2}$.\\

Similarly, $\lambda(\sigma)$ and $\eta(\sigma)$ from \eqref{ScattEqs} with single zero modes will degenerate accordingly (taking into account that the little group is the quotient group G$_Q$):
\begin{align}
\label{LambdaOnDiv}
\lambda_A^\mathrm{a}(\sigma) \!\rightarrow \sum_{i=1}^{n+2} u_{i\beta}^\mathrm{a} \epsilon^\beta_{iA} \frac{S_{\alpha}^{(g-1)}\!(\!\sigma, \!\sigma_i | \tau)}{\sqrt{\ud \sigma}\sqrt{\ud \sigma_i}}, &\kappa_{n+1 A}^\mathrm{a} \!=\! \kappa_{+A}^\mathrm{a}, \kappa_{n+2 A}^\mathrm{a} \!=\! \kappa_{-A}^\mathrm{a}, \kappa^\mathrm{a}_{+A} \kappa_{+B\mathrm{a}} \!=\! -\kappa^\mathrm{a}_{-A} \kappa_{-B\mathrm{a}},\\
\eta_\mathcal{I}^\mathrm{a}(\sigma) \!\rightarrow \sum_{i=1}^{n+2} u_{i\beta}^\mathrm{a} q^\beta_{i\mathcal{I}} \frac{S_{\alpha}^{(g-1)}(\sigma, \sigma_i | \tau)}{\sqrt{\ud \sigma}\sqrt{\ud \sigma_i}},& \; \zeta_{n+1 \mathcal{I}}^\mathrm{a} \!=\! \zeta_{+\mathcal{I}}^\mathrm{a},  \zeta_{n+2 \mathcal{I}}^\mathrm{a} \!=\! \zeta_{-\mathcal{I}}^\mathrm{a} , \zeta_{+\mathcal{I}}^\mathrm{a} \zeta_{+\mathcal{J}\mathrm{a}} \!=\! - \zeta_{-\mathcal{I}}^\mathrm{a} \zeta_{-\mathcal{J}\mathrm{a}},\nonumber\\
&\;\kappa^\mathrm{a}_{+A} \zeta_{+\mathcal{I}\mathrm{a}} \!=\! -\kappa^\mathrm{a}_{-A} \zeta_{-\mathcal{I}\mathrm{a}},\nonumber
\end{align}
where the polarization spinors $\epsilon_{\pm} \!=\! \epsilon_{n+1,2}$ are defined as usually by $\epsilon_{n+1,2\beta}^A \!=\! \epsilon_{\pm\beta}^A \!=\! \epsilon^\mathrm{a}_{\pm\beta} \kappa_{\pm\mathrm{a}}^A$, here with $\epsilon^\mathrm{a}_{\pm\beta}$ making up a basis for the little group G$_Q$ with normalization $\epsilon^\mathrm{a}_{+\beta} \epsilon^\beta_{-\mathrm{a}} \!=\! 1$, and where $u_{+\beta}^\mathrm{a} \!=\! u_{n+1\,\beta}^\mathrm{a}$ and $u_{-\beta}^\mathrm{a} \!=\! u_{n+2\,\beta}^\mathrm{a}$ fulfill $u^\mathrm{a}_{+\xi} u_{-\mathrm{a}}^\xi \!\ne\! 0$. This is required to avoid singularities as will be seen shortly. Defining a momentum operator similarly to the RNS ambitwistor as
\begin{equation*}
P^{AB}\!(\sigma) = \lambda_\mathrm{a}^A(\sigma) \lambda^{B\mathrm{a}}(\sigma),
\end{equation*}
it follows by inserting from \eqref{LambdaOnDiv} and using the polarized scattering equations for the first $n$ particles with $\lambda_A^a$ given on the divisor by \eqref{LambdaOnDiv} that $P^{AB}(\sigma)$ does not have double poles and can be represented as:
\begin{align*}
P^{AB}\!(\sigma) \!=\, &\epsilon_{+}^{\beta[A} \!(\!u^\mathrm{a}_{+\beta} \lambda_\mathrm{a}^{\!B]}(\!\sigma_{+}\!)\!) \!\frac{S_{\alpha}(\!\sigma, \!\sigma_{+}|\tau)}{\sqrt{\!\ud \sigma}\sqrt{\ud \sigma_{+}}} \!+\! \epsilon_{-}^{\beta[A} \!(\!u^\mathrm{a}_{-\beta} \lambda_\mathrm{a}^{\!B]}(\!\sigma_{-}\!)\!) \!\frac{S_{\alpha}(\!\sigma, \!\sigma_{-}|\tau)}{\sqrt{\!\ud \sigma}\sqrt{\ud \sigma_{-}}}  \!+\! \!\sum_{i=1}^n \!k_i^{AB} \!\frac{S_{\alpha}(\!\sigma, \!\sigma_i|\tau)}{\sqrt{\!\ud \sigma}\sqrt{\ud \sigma_i}},\\
P^A_A(\sigma) \!=\! 0&,\quad P^2(\sigma) \!=\! \varepsilon_{ABCD}\;\lambda_\mathrm{a}^A(\sigma) \lambda^{B\mathrm{a}}(\sigma)\lambda_\mathrm{b}^C(\sigma) \lambda^{D\mathrm{b}}(\sigma) \!=\! 0,\nonumber
\end{align*}
where in the first equation of the second line the presence of $\bar{\delta}(\lambda_\mathrm{a}^A(\sigma) \lambda^\mathrm{a}_A(\sigma))$ has been used and in the second equation the fact that there are only two distinguishable $\mathrm{a}$-indices in G$_Q$. $P^A_A(\sigma) \!=\! 0$ leads to $\epsilon_{\pm\alpha}^A (\!u^\alpha_{\pm\mathrm{a}} \lambda^\mathrm{a}_A(\sigma_{\pm})) \!=\! 0$ and $P^2(\sigma)$, like $P(\sigma)$, does not have double poles. Thus, the scattering equations $k_i \cdot P(\sigma_i) \!=\! 0$ hold for $i \!=\! 1,\ldots,n$ and also $\epsilon_{\pm}^{\beta[A} \!(\!u^\mathrm{a}_{\pm\beta} \lambda_\mathrm{a}^{B]}(\sigma_{\pm}\!)) P_{AB}(\sigma_{\pm})\!=\! 0$. The freedom in selecting $u_{\pm}$ is now used to extend the polarized scattering equations to all $n\!+\!2$ particles on the divisor:
\begin{align}
\label{ScattEqsOnDiv}
&v_{\pm\alpha}^\mathrm{a} \kappa_{\pm\mathrm{a}}^A \!=\! u_{\pm\alpha}^\mathrm{a} \lambda^A_\mathrm{a}(\sigma_{\pm}), &&k_{\pm} \cdot P(\sigma_{\pm}) = 0.
\end{align}
Moreover, the corresponding scattering equations for $\eta_i, i=1,\ldots,n\!+\!2$ are consistent with the condition $\bar\delta(\eta^\mathrm{a}_\mathcal{I} \eta^{\mathcal{I}\mathrm{a}})$.\\

How does this affect the amplitude? To reflect the fact that there are now $n\!+\!2$ punctures and scattering equations for $n\!+\!2$ particles the following equation is added, initially for $g \!>\! 1$:
\begin{align}
\label{EquivCompl}
  \delta^4& \Bigl (\sum_{i=1}^n \!k_i^\mu \Bigr ) \,\delta^{4\mathcal{N}}\!\Bigl(\sum_{i=1}^n \!\kappa^\mathrm{a}_{iA} \zeta_{i\mathrm{a}\mathcal{I}}\Bigr )\frac{\ud \mu(\!\lambda_{0A}^\mathrm{a}) \ud \mu(\!\eta_{0\mathcal{I}}^\mathrm{a})}{\text{vol}(\text{G}_Q \times \mathbb{C})_{\lambda_0}}\frac{1}{(2\pi i)^3}\frac{\ud q_{gg}}{q_{gg}}\frac{\ud q_{1}}{q_{1}}\frac{\ud q_{2}}{q_{2}} \bar{\delta}(\lambda_\mathrm{a}^A(x_j) \lambda^\mathrm{a}_A(x_j)\!)\nonumber\\
  &\bar{\delta}(\lambda_\mathrm{a}^A(\tilde{x}_j) \lambda^\mathrm{a}_A(\tilde{x}_j)) \bar{\delta}(\eta_\mathrm{a}^{\mathcal{I}}(\tilde{x}_j) \eta^\mathrm{a}_{\mathcal{I}}(\tilde{x}_j))\nonumber\\ 
  &\!=\! \int \!\!\frac{\ud \mu(\kappa_{n+1\mathrm{a}}^A) \,\ud \mu(\zeta_{n+1\mathrm{a}}^{\mathcal{I}})}{\text{vol}(\text{G}_Q \times \mathbb{C})_{\kappa}} \;\frac{\prod_{l=n+1}^{n+2} \!\ud \sigma_l \ud\mu(u_l )\ud\mu(v_l)}{\text{vol}(\text{SL}(2, \mathbb{C})_{\sigma} \!\times\! \text{G}_{Q_u})}
   \frac{1}{(u^\mathrm{a}_{n\!+\!1\alpha} u^\alpha_{n\!+\!2\,\mathrm{a}})^4}\nonumber\\
   &\prod_{r = n\!+\!1}^{n\!+\!2} \!\bar{\delta}(\!v^\alpha_{r\mathrm{a}} \epsilon^{\mathrm{a}}_{r\alpha} \!-\! 1) \bar{\delta}(\!u^{\alpha}_{r\mathrm{a}} \lambda^{\mathrm{a}} _{rA}\!\!-\! v^{\alpha}_{r\mathrm{a}} \kappa^{\mathrm{a}}_{rA}) \bar{\delta}(\!u^{\alpha}_{r\mathrm{a}} \eta^{\mathrm{a}}_{r\mathcal{I}} \!-\! v^{\alpha}_{r\mathrm{a}} \zeta^{\mathrm{a}}_{r\mathcal{I}})\\
   &\bar{\delta}(\lambda_\mathrm{a}^A(x_j) \lambda^\mathrm{a}_A(x_j)\!) \bar{\delta}(\lambda_\mathrm{a}^A(\tilde{x}_j) \lambda^\mathrm{a}_A(\tilde{x}_j) \!+\! \eta_\mathrm{a}^{\mathcal{I}}(\tilde{x}_j) \eta^\mathrm{a}_{\mathcal{I}}(\tilde{x}_j)).\nonumber
  \end{align}
On the left hand side the three $\bar\delta$'s at the end originally fixed the 3 moduli, but because of switching the moduli integration paths to the singularity at infinity these delta functions get released. They are replaced on the right hand side by just two, because these are the proper PCOs. The delta functions for the conservation laws on the left hand side are built into the right hand side when taking together with the delta functions in $A^{\alpha}_{g,V,n}$ and equations in \eqref{LambdaOnDiv} and \eqref{ScattEqsOnDiv}. Recognizing that the $\ud \sigma_l$'s are fixed due to the gauge symmetry, the number of delta functions balance out on both sides, and finally $(u^\mathrm{a}_{n\!+\!1\alpha} u^\alpha_{n\!+\!2\,\mathrm{a}})^{4-\mathcal{N}} \!\!=\! (u^\mathrm{a}_{n\!+\!1\alpha} u^\alpha_{n\!+\!2\,\mathrm{a}})^{-4}$ is the Jacobian going from $\ud \mu(\!\lambda_{0A}^\mathrm{a}) \ud \mu(\!\eta_{0\mathcal{I}}^\mathrm{a})$ to $\prod_l \ud \mu(\kappa_{l\mathrm{a}}^A) \,\ud \mu(\zeta_{l\mathrm{a}}^{\mathcal{I}})$, making use of the normalization $\epsilon^\mathrm{a}_{n+1\beta} \epsilon^\beta_{n+2\mathrm{a}} \!=\! 1$. $\mathcal{N} = 8$ is crucial here.\\

For $g \!=\! 1$ there is only one modulus $\ud q_{gg}/q_{gg}$ on the left hand side with the same two $\bar\delta$ functions as on the right hand side (enough to fix the single modulus), but also a division by vol(GL(1,$\mathbb{C})_{\sigma}$) because of one $c$ zero mode of the $b\!-\!c$ system.\\

The important point here to see is that the right hand side is equivalent to a completeness relation for a set of physical states represented by fixed vertex operators:
\begin{align}
\label{Completeness}
1 \!=\!\! \int \;&\frac{\ud \mu(\kappa_{n+1\mathrm{a}}^A) \,\ud \mu(\zeta_{n+1\mathrm{a}}^{\mathcal{I}})}{\text{vol}(\text{G}_Q \times \mathbb{C})_{\kappa}} \;\frac{\mathcal{W}_{\text{vac}}}{\ud \sigma_{*} \ud^3 u} \prod_{l= n\!+\!1}^{n\!+\!2} \!\!\!\ud \mu(u_l)\ud \mu(v_l)\;\bar{\delta}(\!v^\alpha_{l\mathrm{a}} \epsilon^{\mathrm{a}}_{l\alpha} \!-\! 1)\nonumber\\
&\bar{\delta}(\!u^{\alpha}_{l\mathrm{a}} \lambda^{\mathrm{a}} _{lA}\!\!-\! v^{\alpha}_{l\mathrm{a}} \kappa^{\mathrm{a}}_{lA}) \bar{\delta}(\!u^{\alpha}_{l\mathrm{a}} \eta^{\mathrm{a}}_{l\mathcal{I}} \!-\! v^{\alpha}_{l\mathrm{a}} \zeta^{\mathrm{a}}_{l\mathcal{I}})e^{u_{l\alpha}^{\mathrm{a}} \mu_{\mathrm{a}}^A(\sigma_l) \epsilon^{l\alpha}_A + u_{l\alpha}^{\mathrm{a}}  \tilde{\eta}_{\mathrm{a}}^{\mathcal{I}}(\sigma_l) q^{l\alpha}_{\mathcal{I}}},\\
\kappa^{A\mathrm{a}}_{n+1} & \kappa^B_{n+1\mathrm{a}} \!=\! -\kappa^{A\mathrm{a}}_{n+2} \kappa^B_{n+2\mathrm{a}},\;\;\zeta_{n+1}^{\mathcal{I}\mathrm{a}} \zeta_{n+1\mathrm{a}}^\mathcal{J} \!=\! - \zeta_{n+2}^{\mathcal{I}\mathrm{a}} \zeta_{n+2\mathrm{a}}^\mathcal{J},\;\;\kappa^\mathrm{a}_{n+1A} \zeta_{n+1\mathcal{I}\mathrm{a}} \!=\! -\kappa^\mathrm{a}_{n+2A} \zeta_{n+2\mathcal{I}\mathrm{a}},\nonumber\\
\mathcal{W}_{\text{vac}}(&\sigma_{n+1}, \sigma_{n+2}, \sigma_{*})\!\!=\! c(\sigma_{n+1})c(\sigma_{n+2})c(\sigma_{*}) n \tilde{n} \delta(\gamma)\delta(\tilde{\gamma})\!\!\prod_{\mathrm{a}\mathrm{b}}\!\Bigl(\delta(\gamma_{\mathrm{a}\mathrm{b}}) \delta(\tilde{\gamma}_{\mathrm{a}\mathrm{b}})N_{\mathrm{a}\mathrm{b}})\Bigr),\nonumber\\
\bra{0}\mathcal{W}&\-_{\text{vac}}(0,0,0)\ket{0} = 1, \qquad\qquad \epsilon_{n+1\mathrm{a}}^\alpha \epsilon^\mathrm{a}_{{n+2}\alpha} \!=\! 1,\nonumber
\end{align}
where the ghosts in $\mathcal{W}_{\text{vac}}$ can be considered to originate from the vacuum normalization or, alternatively, from the vertex operators with an additional $c(\sigma_{*})$ ghost from zero mode normalization. Division by $\ud^3 u$ means that 3 values of $u$ can be chosen arbitrarily. All fields in $\mathcal{W}_{\text{vac}}$ except $c$ appear just once and their location can be either $\sigma_{n+1}$ or $\sigma_{n+2}$. The $c$ field needs to be taken at three different locations. Utilizing state-operator isomorphism \cite{Polchinski:1998rq} vacuum normalization can be expressed by moving all fields in $\mathcal{W}_{\text{vac}}$ to the origin as the same location, using $c(\sigma_{n+1})c(\sigma_{n+2})c(\sigma_{*}) \!\rightarrow\! c(0) \partial c(0) \frac{1}{2} \partial^2 c(0) = c_1 c_0 c_{-1}$. In \eqref{EquivCompl} the quantity $\ud^2 \sigma / \Bigl(\text{vol}(\text{SL}(2, \mathbb{C})_{\sigma} \!\times\! \text{G}_{Q_u}) (u^\mathrm{a}_{n\!+\!1\alpha} u^\alpha_{n\!+\!2\,\mathrm{a}})^4\Bigr)$ is the outcome when integrating out locally the ghosts in \eqref{Completeness} as zero modes \footnote{Strictly speaking, this is only correct on the tree level, because correlators of three $c$ ghosts for $g \!\ge\! 1$ would vanish, i.e. it would have been more proper to skip \eqref{EquivCompl} as intermediate equation.}. Actually the presence of these ghosts is required to end up with the correct number of ghosts and PCOs on the divisor as the next paragraph will show. The exponential in \eqref{Completeness} ensures that \eqref{LambdaOnDiv} and \eqref{ScattEqsOnDiv} are satisfied as solutions to the equations of motion and do not need to be used as external conditions. The comparison of \eqref{EquivCompl} with \eqref{Completeness} makes evident that the twistor zero modes, when moved to the divisor, are equivalent to the insertion of a complete set of physical states.\\

Continuing with the process of moving the components of the amplitude to the divisor $\mathcal{D}^{\text{nonsep}}_{g-1,n+2}$ one can observe that, after temporarily moving back $j^u_{z_g} \tilde{j}^u_{\tilde{z}_g}\rightarrow \bar{\delta}^{\prime}_{w_g} j_{z_g} \tilde{j}_{\tilde{z}_g}$, for each PCO in $\bar{\delta}^{\prime}_{w_g} j_{z_g} \tilde{j}_{\tilde{z}_g}$ the corresponding former anti-ghost zero mode $b_g$ in $Z_g^\alpha$ becomes available for contractions with ghosts. When all of them are moved to the locations $\sigma_{n+1}$ and $\sigma_{n+2}$, including the three $b$'s from $Z_2$, the fixed vertex operators can be changed to being integrated, taking advantage of the anti-ghosts no longer being locked up in partition functions. With regard to the worldsheet supersymmetries, one typically works with one fully fixed and two partially fixed vertex operators \cite{Geyer:2020, Kunz:2024wsx}, but it is also possible to use just two fully fixed vertex operators by applying two of the three PCOs $\delta(\beta^{\mathrm{a}\mathrm{b}})\lambda_A^\mathrm{a} \rho_{1}^{\mathrm{b}A}$ to the same vertex operator\footnote{This has been used also for the tree amplitudes in \eqref{TreeAmps}.}:
\begin{align*}
\Bigl(\!\delta(\gamma_{\mathrm{a}\mathrm{b}}u_{1\alpha}^\mathrm{a} u_1^{\mathrm{b}\alpha}) \delta(\beta_{\mathrm{a}\mathrm{b}}&\hat{u}_\alpha^\mathrm{a} \hat{u}^{\mathrm{b}\alpha}) \lambda^\mathrm{a}_A \rho_{1\mathrm{a}}^A \hat{u}_\mathrm{a}^\alpha  \hat{u}^\mathrm{a}_\alpha\!\Bigr) \Bigl(\!\delta(\gamma)\delta(\beta)\lambda^\mathrm{a}_A(\sigma_1\!) \rho_{1\mathrm{a}}^A\!\Bigr) e^{u_{1\alpha}^\mathrm{a} \mu_\mathrm{a}^A(\!\sigma_1\!) \epsilon^\alpha_{1A}}\;\quad\quad \Rightarrow\\
\Bigl(\frac{\hat{u}_\mathrm{a}^\alpha \lambda^\mathrm{a}_A(\sigma_1) \epsilon_{1\alpha}^A}{u_{1\alpha}^\mathrm{a} \hat{u}^\alpha_\mathrm{a}}& + \frac{\rho_{1\mathrm{a}}^A(\sigma_1\!) \rho_1^{B\mathrm{a}}(\sigma_1\!)}{2}\epsilon^\alpha_{1A}\epsilon_{1B\alpha}\Bigr)e^{u_{1\alpha}^\mathrm{a} \mu_\mathrm{a}^A(\!\sigma_1\!) \epsilon^\alpha_{1A}},\\
\Bigl(\!\delta(\gamma_{\mathrm{a}\mathrm{b}}u_{2\alpha}^\mathrm{a} u_2^{\mathrm{b}\alpha}) \delta(\beta_{\mathrm{a}\mathrm{b}}& \hat{u}_\alpha^\mathrm{a} \hat{u}^{\mathrm{b}\alpha}) \lambda^\mathrm{a}_A \rho_{1\mathrm{a}}^A  \hat{u}_\mathrm{a}^\alpha \hat{u}^\mathrm{a}_\alpha\!\Bigr)\\
& \Bigl(\!(u_{2\mathrm{a}}^\alpha \hat{u}^\mathrm{a}_\alpha) \delta(\gamma_{\mathrm{a}\mathrm{b}}u_2^\mathrm{a} \hat{u}^\mathrm{b}) \delta(\beta_{\mathrm{a}\mathrm{b}}\hat{u}^\mathrm{a} u_2^\mathrm{b}) \lambda^\mathrm{a}_A \hat{u}_\mathrm{a}^\alpha \rho_{1\mathrm{a}}^A u^\mathrm{a}_{2\alpha} \!\Bigr) e^{u_{2\alpha}^\mathrm{a} \mu_\mathrm{a}^A(\!\sigma_2\!) \epsilon^\alpha_{2A}} \; \Rightarrow\\
 \Bigl(\frac{\hat{u}_\mathrm{a}^\alpha \lambda^\mathrm{a}_A(\sigma_2) \epsilon_{2\alpha}^A}{u_{2\alpha}^\mathrm{a} \hat{u}^\alpha_\mathrm{a}}& + \frac{\rho_{1\mathrm{a}}^A(\sigma_2\!) \rho_1^{B\mathrm{a}}(\sigma_2\!)}{2}\epsilon^\alpha_{2A}\epsilon_{2B\alpha}\Bigr) e^{u_{2\alpha}^\mathrm{a} \mu_\mathrm{a}^A(\!\sigma_2\!) \epsilon^\alpha_{2A}}\;.
\end{align*}
This can be done in a very similar way for the other supersymmetry involving $\rho_{2A}^a$ and $\tau_{\mathcal{I}}^a$.
Further, $\bar{\delta}_{x_g}$ contributes to, say, $r \!=\! n+2$ in $\bar\delta_r$ of  $A^{\alpha}_{g-1,V,n+2}$ in \eqref{AV}. Then $r \!=\! n+1$ in $\bar\delta_r$ arises for $\bar\delta(\text{Res}_{\sigma_{n+1}}(\lambda^\mathrm{a}_A \lambda^A_\mathrm{a}))$ from vol($\mathbb{C}$) in the integration measure which, therefore, will be left out from now on, and for $\bar\delta(\text{Res}_{\sigma_r}(\eta^\mathrm{a}_\mathcal{I} \eta^\mathcal{I}_\mathrm{a}))$ by assigning $\zeta_{n+1\mathrm{a}}^\mathcal{I} \!=\! \pm \zeta_{n+2\mathrm{a}}^\mathcal{I}$ directly in such a way that $\zeta_{n+1}^{\mathcal{I}\mathrm{a}} \zeta_{n+1\mathrm{a}}^\mathcal{J} \!=\! - \zeta_{n+2}^{\mathcal{I}\mathrm{a}} \zeta_{n+2\mathrm{a}}^\mathcal{J}$ in the forward direction. So indeed the two fixed vertex operators end up fully integrated.\\

Because the Szeg\"{o} kernels asymptotic behavior, according to \eqref{Degeneration}, is $S_{\alpha}^{(g)} \rightarrow S_{\alpha}^{(g-1)} + O(q_{gg})$ on the divisor, the same is true for $A^{\alpha}_{g,V,n}$. Then, so far it has been demonstrated the following transition on the divisor:
\begin{equation*}
 \delta(\sum_{i=1}^n \!k_i) \delta(\sum_{i=1}^n \kappa_{iA\mathrm{a}} \zeta^\mathrm{a}_{i\mathcal{I}}) \int \ud m_0 \bar\delta(x_g) A^{\alpha}_{g,V,n} \rightarrow \int \frac{\ud \mu(\kappa_{n+1\mathrm{a}}^A) \,\ud \mu(\zeta_{n+1\mathrm{a}}^{\mathcal{I}})}{\text{vol}(\text{G}_Q)_{\kappa} \delta(k_{n+2 A}^A) } A^{\alpha}_{g-1,V,n + 2}\,.
\end{equation*}
By removing the $b$ zero modes in $Z_2$ and all of the $Z_1$s, $Z_g^\alpha$ in \eqref{LoopAmplOdd} has a smooth transition for $q_{gg} \!\rightarrow\! 0$ from $g$ to $g \!-\! 1$ with finite results\footnote{For $g \!=\!2$ there are only two $b$ anti-ghosts released from $Z_2$ instead of three, which leaves one vertex operator fixed with regard to the $b\!-\!c$ ghosts and makes one $c$ ghost available for the smooth transition of $Z_2$ to the $g \!=\!1$ level with one ghost and anti-ghost each. Then during degeneration for $g \!=\!1$ this single $b$ anti-ghost and single $c$-ghost get released from $Z_2$ with the consequence that the $\ud \sigma_{*}$ in \eqref{Completeness} does not get canceled and changes one of the integrated vertex operators to fixed with regard to the $b\!-\!c$ ghosts such that the resulting tree amplitude has the correct number (3) of $c$-ghosts and corresponding fixed vertex operators.}. This follows directly from the defining integrals for the $Z$ functions in \eqref{ZsDef}, but can be also argued alternatively:
\begin{itemize}
\item
By temporarily setting $y_{3(g\!-\! 1)} \!=\! y_{3g \!-\! 4} \!=\! y_{3g \!-\! 5} \!=\! x_g \!=\! \tilde{x}_g \!=\! w_{* g} \!=\! z_{* g} \!=\! \tilde{z}_{* g}\!\equiv\! *_g$, all the $\sigma(*_g)$'s cancel against each other.
\item
The remaining $\sigma(z)$'s change to $g \!-\! 1$ level smoothly by taking into account that according to \eqref{Degeneration} $\oint_{a_g} \!\ud w \,\omega_g \ln E(z_g,w) \!\rightarrow\! \oint_{a_g} \!\ud w \,\omega_{+-}^{(g-1)} \ln E(z_g,w) \!=\! 0$ in equation \eqref{sigma} for $\sigma(z)$, due to the vanishing of an a-cycle integration of an Abelian differential of the third kind.
\item
By temporarily setting the arguments $a$ of the prime forms $E(a, *_g)$'s in the denominator of $Z_g^\alpha$ to be the same as the ones in the numerator, the $E(a, *_g)$'s cancel against each other except for the three $E(y_{*}, y_{*})$'s between the 3 points $y_{3(g\!-\! 1)}, y_{3g \!-\! 4}, y_{3g \!-\! 5}$ arising from correlations between the corresponding $b$ zero modes during bosonization, but which are broken up when these $b$ zero modes are released from $Z_2$. Setting a $E(z_i, *_g)$ equal to a $E(w_j, *_g)$ in $Z_1$ functions needs to be done carefully in order not to trigger the Riemann vanishing theorem, by choosing $z_i$ and $w_j$ from different $Z_1$ functions.
\item
It remains to show that the arguments of the $\theta$ functions adjust to the $g \!-\! 1$ level. For spin structures with $\alpha^\prime_g \!=\! 0$ in \eqref{theta} this can be easily achieved by setting $*_g \!=\! 0$. For the other spin structures one has to additionally ensure that for $q_{gg} \!\rightarrow\! 0$ the $g$-component of the theta function becomes a multiplicative factor $q_{gg}^{1/4}$ without any additional arbitrary phase which can be achieved by choosing appropriate locations. Note that it does not matter whether an individual $\theta[\alpha] \!\sim\! q_{gg}^{1/4}$ ($h_\alpha^2 \!\sim\! q_{gg}^{1/4}$) or $\theta[\alpha] \!\sim\! 1$ ($h_\alpha^2 \!\sim\! 1$), because there are the same number of $\theta[\alpha]$'s or $h_\alpha^2$'s in the numerator and denominator.
\end{itemize}

The three moduli that were used to fix $\bar{\delta}_{x_g}$, with one of them assigned to $q_{gg}$, become decoupled and lead to simple integrals $\oint \frac{\ud q}{2\pi i q} = 1$.
As can be gleaned from section \ref{PartitionFunction} the transition of $Z_2$ to the $g \!-\! 1$ level by removing $b$ zero modes also ensures the transition of the modular invariant measure $\ud \mu_g$ to $\ud \mu_{g-1}$. The final result is, up to a constant:
 \begin{align}
 \label{NonSepAmplOddOnOdd}
  (A^{\text{odd}}_{g,n})^{\text{odd}}_{\text{cyc}} \!= &\frac{1}{4}\!\int\!\!\ud m_{\text{fw}} A^{\text{even}}_{g-1,n+2} \\
  \!= &\frac{1}{4}\!\sum_{\alpha} \!\! \int \!\!\ud \mu_{g-1} Z_{g-1}^{\alpha}((z),\!(\tilde{z}),\!(y),\!(x),\!(w)) \!\!\int\!\!\ud m_{\text{fw}} \prod_{j=1}^{g-1} \! \bar{\delta}_{x_j} \,A^{\alpha}_{g-1,V,n+2}\;,\nonumber\\
  &\ud m_{\text{fw}}\, \delta(k_{n+2 A}^A) = \frac{\ud \mu(\kappa_{n+1\mathrm{a}}^A) \,\ud \mu(\zeta_{n+1\mathrm{a}}^{\mathcal{I}})}{\text{vol}(\text{G}_Q)_{\kappa}} \,.\nonumber
  \end{align}
Note that for $g\!=\!1$ this reduces to simply $A^{\text{odd}}_{1,n} \!=\! \frac{1}{4}\int\!\!\ud m_{\text{fw}} A^{\alpha}_{0,V,n+2}$ where the right hand side is a tree amplitude.\\

Much of the work done in this subsection can be reused and the other cases can be handled much faster now.

\item [Even cycle of an odd spin structure:]
In the amplitude the two points $\sigma_{+}$ and $\sigma_{-}$ are declared as two new punctures $\sigma_{n+1}$ and $\sigma_{n+2}$, though this time not based on a degeneration of the zero modes, but by inserting a complete set of physical states at the divisor $\mathcal{D}^{\text{nonsep}}_{g-1,n+2}$ using equation \eqref{Completeness}.
Continuing like before, making the inserted vertex operators integrated and performing a smooth transition on the divisor, one obtains: 
 \begin{align}
  \label{NonSepAmplEvenOnOdd}
  (A^{\text{odd}}_{g,n})^{\text{even}}_{\text{cyc}} \!= &\frac{3}{4}\!\int\!\!\ud m_{\text{fw}}A^{\text{odd}}_{g-1,n+2} \quad=\quad \delta(\sum_{i=1}^{n+2} \!k_i) \delta(\sum_{i=1}^{n+2} \kappa^\mathrm{a}_{iA} \zeta_{i\mathcal{I}\mathrm{a}}) \\
  &\frac{3}{4}\!\sum_{\alpha} \! \int \!\ud \mu_{g-1} \,Z_{g-1}^{\alpha}((z),\!(\tilde{z}),\!(y),\!(x),\!(w))
 \!\int\!\!\ud m_0 \;\ud m_{\text{fw}} \prod_{j=1}^{g-1} \; \bar{\delta}_{x_j} \,A^{\alpha}_{g-1,V,n+2}\;,\nonumber\\
  &\ud m_0 = \frac{\ud \mu(\!\lambda_{0A}^\mathrm{a}) \ud \mu(\!\eta_{0\mathcal{I}}^\mathrm{a})}{\text{vol}(\text{G}_Q \times \mathbb{C})_{\lambda_0}},\quad \ud m_{\text{fw}}\,\delta(k_{n+2 A}^A) = \frac{\ud \mu(\kappa_{n+1\mathrm{a}}^A) \,\ud \mu(\zeta_{n+1\mathrm{a}}^{\mathcal{I}})}{\text{vol}(\text{G}_Q)_{\kappa}}\,,\nonumber
  \end{align}
  where the delta functions for momentum and supercharge conservation has been extended from $n$ to $n\!+\!2$ because of $k_{n+2} \!=\! - k_{n+1}$ and $\kappa^\mathrm{a}_{n+2A} \zeta_{n+2\mathcal{I}\mathrm{a}} \!=\! -\kappa^\mathrm{a}_{n+1A} \zeta_{n+1\mathcal{I}\mathrm{a}}$ in the forward direction.
    
  \item [Even cycle of an even spin structure:]
  This case is basically the same as the previous one, except there are no spinor zero modes:
  \begin{align}
  \label{NonSepAmplEvenOnEven}
  (A^{\text{even}}_{g,n})^{\text{even}}_{\text{cyc}} &=\!\frac{3}{4}\!\int\!\!\ud m_{\text{fw}}A^{\text{even}}_{g-1,n+2}\\
  &=\!\frac{3}{4}\!\sum_{\alpha} \!\! \int \!\!\ud \mu_{g-1} \!Z_{g-1}^{\alpha}((z),\!(\tilde{z}),\!(y),\!(x),\!(w)) \!\!\int\!\!\ud m_{\text{fw}} \prod_{j=1}^{g-1} \!\bar{\delta}_{x_j}\,A^{\alpha}_{g-1,V,n+2}\,\nonumber.
  \end{align}
  Note that again for $g\!=\!1$ this reduces to $A^{\text{even}}_{1,n} \!=\! \frac{3}{4}\int\!\!\ud m_{\text{fw}} A^{\alpha}_{0,V,n+2}$ with the right hand side being a tree amplitude.
  
\item [Odd cycle of an even spin structure:]
There is not much difference to the previous case, except for the complication that the spinor fields would develop zero modes on the divisor $\mathcal{D}^{\text{nonsep}}_{g-1,n+2}$. 
\begin{align}
  \label{NonSepAmplOddOnEven}
  (A^{\text{even}}_{g,n})^{\text{odd}}_{\text{cyc}} \!= &\frac{1}{4}\!\int\!\!\ud m_{\text{fw}}A^{\text{odd}}_{g-1,n+2} \quad=\quad\delta(\sum_{i=1}^{n+2} \!k_i) \delta(\sum_{i=1}^{n+2} \kappa^\mathrm{a}_{iA} \zeta_{i\mathcal{I}\mathrm{a}}) \\
  &\frac{1}{4}\!\sum_{\alpha} \! \int \!\ud \mu_{g-1} \,Z_{g-1}^{\alpha}((z),\!(\tilde{z}),\!(y),\!(x),\!(w))
 \!\int\!\!\ud m_0 \;\ud m_{\text{fw}} \prod_{j=1}^{g-1} \; \bar{\delta}_{x_j} \,A^{\alpha}_{g-1,V,n+2}\;.\nonumber
  \end{align}
\end{description}

  A couple of remarks:
  \begin{description}
  \item [Equal Amplitudes:] The analysis for the non-separating degeneration has revealed that all cycles, 3 even and 1 odd, contribute each with the same outcome. Also note that according to \cite{Alvarez-Gaume:1986rcs}, exploiting modular invariance, any even spin structure can be transformed into one that contains no odd cycle and any odd spin structure into one with exactly one odd cycle at the very end such that the spin structure on the non-separating divisor can always be chosen to be even with no odd cycles left.\\
  
  As a quick sanity check, consider one-loop amplitudes \eqref{1LoopAmplEven} and \eqref{1LoopAmplOdd} without extracting the ghosts and anti-ghosts from the partition functions during evaluation on the divisor. For even spin structures $q_\alpha^2 (\theta[\alpha](0)/\eta(\tau)^3)^8 \!\rightarrow\! 1$ for $q \!\rightarrow\! 0$ like (trivially) for the odd spin structure, and one ends up with one additional set of fixed vertex operators too much. Gauge invariance then requires to insert ghost $\!\!-\!\!$ anti-ghost pairs that are used, together with the remaining set of PCOs, to change them to integrated vertex operators. This leads to the tree amplitudes with integration over a complete set of particle $\!\!-\!\!$ anti-particle pairs in the forward direction, as expected. Note, that without the $q_\alpha^2$ factor two of the even spin structures would have led to a $q^{-2}$ singularity, making the model inconsistent.\\ 
   
  \item [Reversability:] Formally, \eqref{NonSepAmplOddOnOdd}, \eqref{NonSepAmplEvenOnOdd}, \eqref{NonSepAmplEvenOnEven}, and \eqref{NonSepAmplOddOnEven} are exact equations. This means that the process of factorization is reversible, taking the sum over all cycles. This fact will be used in the next section for factorization on separating degeneration.\\
  \end{description}

 \subsection{Separating Degeneration}                                    
 \label{Separating}
 Pinching a separating cycle involves amplitude evaluation on a separating divisor $\mathcal{D}^{\text{sep}}$ at the boundary of compactified $\widehat{\mathcal{M}}_{g,n}$  corresponding to a reduced moduli space $\widehat{\mathcal{M}}_{g_1,n_1+1} \times \widehat{\mathcal{M}}_{g_2,n_2+1}$ with $g \!=\! g_1 \!+\! g_2, n \!=\! n_1 \!+\! n_2$\cite{Witten:2012bh}.\\
  
  Because the previous section on non-separating degeneration has shown that the process of reducing loop amplitudes to tree amplitudes is reversible, in order to demonstrate that general amplitudes of arbitrary genus factorize properly for separating degeneration it is sufficient to do this just for tree amplitudes, already done in \cite{Kunz:2024wsx, Albonico:2020}. Surely it is possible to examine the separating degeneration of an amplitude of general genus directly but going through a reduction to tree amplitudes poses a substantial simplification.\\

   \subsection{UV Finiteness}
   \label{UVFiniteness}
   Because of modular invariance a scattering amplitude can be evaluated on a fundamental domain that does not include the origin of any coordinate $t$ in $\ud \mu_g$ avoiding any UV divergence usually associated with small $t$'s. In other words, any UV divergences encountered in one domain can be mapped to IR divergences in a different domain. On odd spin structures one also has to contend with one integration over zero modes, although this can be reduced to an integration over momenta on a non-separating divisor with an even spin structure which will be shown to be UV-finite. Does this mean that the theory is manifestly UV-complete and only IR-divergent for large $t$ values? On a formal level yes, but the argument is not strong enough to be considered a proof.\\
   
   Before continuing, an effort is made to simplify the situation by fixing the gauge of the tiny group G$_T$ in order to avoid rather complicated formalism. This was already discussed generally earlier in the main part of this section but is also explicitly done in appendix \ref{Gauge} in such a way that for the remainder of the section it is allowed to assume that only the first twistor pair of momenta and supermomenta is contributing, in other words, to replace the quotient group G$_Q$ with simple SL(2,$\mathbb{C}$).\\
   
   As a first step to find more substantiated evidence, a simple scaling argument can show that the theory is at least UV finite for a single integration on a non-separating divisor. As mentioned at the end of section \ref{NonSeparating} in the remark on \textit{Equal Amplitudes} it is sufficient to consider the cases \eqref{NonSepAmplOddOnOdd} and \eqref{NonSepAmplEvenOnEven}, which includes the reduction to tree amplitudes. $\int\!\ud m_{\text{fw}} A^{\text{even}}_{g,n}$ is considered to be the imaginary part of an off-shell integral over the loop momentum $p^{AB} \!=\! p^{\mu} \gamma_{\mu}^{AB} \!=\! \kappa_{n+1\mathrm{a}}^A \kappa_{n+1}^{B\mathrm{a}} \!=\! k^{AB}_{n+1} \!=\! -k^{AB}_{n+2}$ where $\ud m_{\text{fw}}$ is replaced by
  \begin{equation*}
  \ud m_{\text{fw}} \,\delta(k_{n+1 A}^A)\delta(k_{n+2 A}^A) \rightarrow \frac{\ud^4 p^\mu}{p^2} \ud^\mathcal{N} \!q_{n+1}^{\mathcal{I}}\,\ud^\mathcal{N} \!q_{n+2}^{\mathcal{I}}\,.
  \end{equation*}
  For the sake of simplicity, in the remainder of the section $n\!+\!1$ and $n\!+\!2$ are replaced by $+$ and $-$, respectively. What happens when the momentum $p$ goes to infinity with scaling parameter $\Lambda$? Then there are solutions to the massive scattering equations \cite{Naculich:2014naa} that, in the forward direction $k_{-} \!=\! - k_{+}$, require  that the difference $\sigma_{+} \!\!-\! \sigma_{-}$ scales like $\Lambda^{-1}$. Assume for the moment that all the other differences of $\sigma_i$'s do not scale. The polarized scattering equations then demand that the $u_{\pm}$ must scale as $\Lambda^{-1/2}$ to make the individual terms finite. The two delta functions $\bar\delta^4(\braket{v_{\pm} \kappa_{\pm}^A} - \ldots)$ in \eqref{AV} scale together as $\Lambda^{-4}$ which will go into the Jacobian when solving the scattering equations. This together with $\Lambda^{-2}$ from $p^{-2}$ and $\Lambda^{-2}$ from $\ud^2 u_{+} \ud^2 u_{-}$ gives an overall factor of $\Lambda^{-8}$ so far.\\
  
  Integration over the odd parameters $\ud q^{\mathcal{I}}_{\pm}$ multiplies the amplitude with factors of the form $\braket{u_{\pm}u_i} S_{\alpha}(\sigma_{\pm}, \sigma_i | \tau) (\text{including}\; i \!=\! \mp)$ which does not make the scaling behavior worse.\\
  
  But it remains to investigate the scaling behavior of the two determinants in \eqref{AV}. Exploiting the fact that the matrices contain fixed vertex operators a convenient way to check the UV behavior of one determinant is to eliminate the two rows and columns containing the two points $\sigma_{\pm}$ being connected. The reduced determinants are then
  \begin{equation*}
  \text{det}^\prime H^I_{DJ} \text{det}^\prime G^I_{DJ} = \frac{\text{det} (H_{[+-]}^{[+-]})^I_{DJ} \text{det} (G_{[+-]}^{[+-]})^I_{DJ}}{\braket{u_{+} u_{-}}^4}.
  \end{equation*}
 This scales as $\Lambda^4$ from the denominator.\\
  
 But on even spin structures including the tree level the reduced determinants degenerate even further asymptotically. For an unsubstituted column in $(H_{[+-]}^{[+-]})^I_{DJ}$
 \begin{align*}
 \label{Degeneracy}
  (c u_{+a} &+ u_{-a})\sum_i u_i^a ((H_{[+-]}^{[+-]})^I_{DJ})_{ij} = (c u_{+a} + u_{-a})\Bigl(- \frac{u_{+}^a \epsilon_{+}^B\epsilon_{jB}}{\sigma_{+j}} - \frac{u_{-}^a \epsilon_{-}^B \epsilon_{jB}}{\sigma_{-j}} \Bigr)\nonumber\\
  &= \braket{u_{+} u_{-}}\Bigl(\frac{\epsilon_{+}^B\epsilon_{jB}}{\sigma_{+j}} - c \frac{\epsilon_{-}^B \epsilon_{jB}}{\sigma_{-j}} \Bigr),
  \qquad\frac{1}{\sigma_{\pm j}} = \frac{S_{\alpha}(\sigma_{\pm}, \sigma_j | \tau)}{\sqrt{\sigma_{\pm}}\sqrt{\sigma_j}},\nonumber
  \end{align*}
  where $c$ is a constant such that $\epsilon_+^A \!=\! c \epsilon_-^A$. This can be assumed without loss of generality, e.g. by choosing ($c \!=\! e^{i\phi}$)
  \begin{align*}
  &\epsilon_+^a \!=\! (1/\sqrt{2})(1,i),\quad\epsilon_-^a \!=\! (1/\sqrt{2})(i,1), &&\braket{\epsilon_+ \epsilon_-}\!=\! 1,\\
  &\kappa_{-0}^A \!=\! -\sin(\phi) \kappa_{+0}^A -\cos(\phi)\kappa_{+1}^A, &&\kappa_{-1}^A \!=\! -\cos(\phi) \kappa_{+0}^A + \sin(\phi) \kappa_{+1}^A,
  \end{align*}
  or just simply $\epsilon_+^a \!=\! (1,0), \epsilon_-^a \!=\! (0,1),$ and $\kappa_{-a}^A \!=\! -\kappa_{+(1-a)}^A$ such that $\epsilon_+^A \!=\! \epsilon_-^A$. Then it follows:
  \begin{align*}
 (cu_{+a} &+ u_{-a})\sum_i u_i^a ((H_{[+-]}^{[+-]})^I_{DJ})_{ij} \!=\! \braket{u_{+} u_{-}} \epsilon_{+}^B\epsilon_{jB} \Bigl(\frac{1}{\sigma_{+j}} - \frac{1}{\sigma_{-j}} \Bigr) = \braket{u_{+} u_{-}} \epsilon_{+}^B\epsilon_{jB} \;O(\Lambda^{-1}).
  \end{align*}
The last equation holds because of $\sigma_{+} \!-\! \sigma_{-}$ scaling as $\Lambda^{-1}$. For substituted columns it follows from \eqref{Degen}, simplified because of the limitation to just the undotted indices:
\begin{equation*}
\sum_i u_i^a (C_{ri0}^b C^0_{kjb} \!-\! C_{ki0}^b C^0_{rjb}) \!=\! \epsilon_j^A\Xi_{AB}(z)\!\Bigl(\!- \frac{u_{+}^a \epsilon_{+}^B}{\sigma_{z+}} - \frac{u_{-}^a \epsilon_{-}^B}{\sigma_{z-}} \!\Bigr),\;\; \Xi_{AB}(z) \!=\! \frac{1}{\sigma_{zj}} \Lambda_{A[k}^b(z) \Lambda_{r]Bb}(z).
\end{equation*}
Therefore, the same degeneracy still prevails:
\begin{align*}
(cu_{+a} &+ u_{-a}) \sum_i u_i^a (C_{ri0}^b C^0_{kjb} \!-\! C_{ki0}^b C^0_{rjb}) = (cu_{+a} + u_{-a})\epsilon_j^A\Xi_{AB}(z)\Bigl(- \frac{u_{+}^a \epsilon_{+}^B}{\sigma_{z+}} - \frac{u_{-}^a \epsilon_{-}^B}{\sigma_{z-}} \Bigr)\\
  &= (\epsilon_j^A \Xi_{AB}(z) \epsilon_{+}^B) \braket{u_{+} u_{-}} O(\Lambda^{-1}).
 \end{align*}
The same applies to $\text{det}G_{[+-]}^{[+-]}$ where the $q_{\mathcal{I}}$ contributions can be ignored asymptotically unless they are the only ones (for SYM amplitudes \cite{Kunz:2024wsx}) but then the Parke-Taylor factor does not have a worse behavior than $O(\Lambda)$, and according to the photon decoupling identity \cite{Geyer:2016} even only $O(1)$. Therefore, $\text{det}^\prime H^I_{DJ} \text{det}^\prime G^I_{DJ}$ scales at most as $\Lambda^2$, and the amplitude $\!\int\!\!\ud m_{\text{fw}} A^{\text{even}}_{g,n}$ behaves as $\ud^4 (\Lambda p) \Lambda^{-6}$ or better for the loop integral on the divisor.\\

Can scaling behavior of other integration variables $\sigma_i$ or $u_i$ improve the overall scaling or make it worse? The polarized scattering equations impose strong conditions, because the $v_i$'s have to fulfill $\braket{v_i \epsilon_i} \!=\! 1$ and, therefore, cannot scale. Consequences are
\begin{enumerate}
\item{If one of the $u_i$'s goes like $O(\Lambda^{-x})$ with $x \!>\! 0$, then there must be some other $j$ (could be $\pm$ when $x\!=\!1/2$ but not necessarily) with $\sigma_i \!-\! \sigma_j \!\sim\! O(\Lambda^{-2x})$ and $u_j \!\sim\! O(\Lambda^{-x})$. Then the same analysis as before for $u_{\pm}$ can be applied to $u_i$ and $u_j$ resulting in $\ud^2 u_i \ud^2 u_j  \!\sim\! O(\Lambda^{-4x})$ and the reduced determinants $\sim\! O(\Lambda^{4x})$, i.e. no change overall for any pair $i,j$ with $u_{i,j} \!\sim\! O(\Lambda^{-x})$ and $\sigma_i \!-\! \sigma_j \!\sim\! O(\Lambda^{-2x})$. The same is true when $u_i \!\sim\! O(\Lambda^{-1/2})$ and $\sigma_i \!-\! \sigma_j \!\sim\! O(\Lambda^{-1})$ for all $i,j$.}
\item{$\sigma_i \!\sim\! O(\Lambda^{-x})$ with $x \!>\! 0$ just by itself with $u_i \!\sim\! O(1)$ results in $\Lambda^{-x}$, an overall improvement, but any difference $\sigma_i \!-\! \sigma_j$ would have to be $O(1)$ which is unlikely unless $\sigma_i \!\sim\! O(\Lambda^{-x})$ is a single occurrence.}
\item{On the tree level $\sigma_i \!\sim\! O(\Lambda^x)$ with $x \!>\! 0$ would require $u_i \sim O(\Lambda^x)$ (though unlikely in view of $u_{\pm} \sim O(\Lambda^{-\frac{1}{2}})$) and vice versa, leading to reduced determinants $\!\sim\! O(\Lambda^{-4x})$, i.e. also giving an overall improvement of $\Lambda^{-x}$, but it would have to be a single occurrence (any pair of such $\sigma_i$'s would blow up the scattering equations). On the loop level the case would be even trickier and depend on the asymptotic behavior of the Szeg\"{o} kernel for large arguments which is not clear. Therefore, it is better to assume that this case cannot occur.}
\end{enumerate}
In summary, there is no change to the overall scaling of $\Lambda^{-6}$. Unfortunately, this only proves UV finiteness for a single integration in isolation. When the amplitude is fully reduced to the tree level and all loop integrations are performed together, the situation is more intricate when propagators share multiple loop momenta. In this regard it is quite possible that supersymmetry improves the UV-behavior when integrating over $\ud q^{\mathcal{I}}_{\pm}$ and treating external particles as singlets with respect to the R-symmetry but to look into this is beyond the scope of this work.\\

On a related note, a $\Lambda^{-8}$ behavior for single loops would have been even better because this would have made sure that only loop integrals over 4 or more propagators contribute, in line with the \textit{no-one-loop-triangle} conjecture of conventional $N\!=\!8$ supergravity \cite{Bern:2018jmv}. If the integral over loop momenta could actually be mapped to a Feynman diagram with 3 or more propagators, one could argue that, because the scattering equations are solved on-shell, two propagators adjacent to the $p^{-2}$ propagator, being embedded in 6D kinematics, must be of the form $((p \!-\! k)^2 )^{-1} \xlongequal{\text{on-shell}} (\!-\! 2p\!\cdot\! k)^{-1}$ because of $p^{AB}p_{AB} \!=\! 0 \!=\! k^{AB} k_{AB}$ on-shell such that by using the off-shell form on the left side for $p$ the single loop behavior would increase to $\Lambda^{-8}$. It would be interesting to find out whether this conjecture could be proven rigorously.\\

\section{Summary and Outlook}
  \label{Summary} 
  
  This continuation of the previous work \cite{Kunz:2024wsx} filled in more details which were skipped there and established some new results. In particular, in section \ref{ExMassATStr} an additional auxiliary action was added to make the full action anomaly-free, basically by adding a couple of free spinors that only contribute to loops without appearing as external particles and by expanding the little group to one that on-shell includes a redundant 'tiny group' which then was used to generate on-shell a quotient group that was isomorphic to the original little group SU(2). Although the choice of such an action might not be at all unique, the selected action seems to be minimal and natural, and has some desired features, namely, essentially not changing vertex operators and tree amplitudes (section \ref{ExMassATStr}) and allowing to evaluate loop amplitudes while keeping modular invariance and unitary factorization, with expected results when they are reduced on separating and non-separating divisors (section \ref{LoopAmps}, \ref{NonSeparating}, and \ref{Separating}). And as shown in appendix \ref{Gauge}, the model can be gauge-fixed in such a way that it becomes virtually indistinguishable from the one with just SU(2) as little group.\\
  
  Vacuum partition functions and the cosmological constant all vanish at every loop level (section \ref{PartitionFunction}). Loop amplitudes can be explicitly displayed on even and odd spin structures (section \ref{EvenSS} and \ref{OddSS}, respectively). When they are evaluated on non-separating divisors of the compactified moduli space it turns out that all cycles, three even and one odd, contribute equal amounts to the same reduced amplitude. Important to note is that this calculation is exact and reversible\footnote{When location independence of the PCOs is guaranteed.}, a fact which is used to demonstrate successful factorization on separating divisors (section \ref{Separating}). Fully reduced to the tree-level the amplitudes look in line with expectations from generalized unitarity. They are formally UV-complete before reduction and it is shown with a scaling argument that UV-finiteness holds at least on the single-loop level for reduced amplitudes (section \ref{UVFiniteness}).\\ 
   
   All these properties indicate that this theory is a compelling modular invariant and unitary $N \!=\! 8$ supergravity model that has an embedded version of Super-Yang-Mills. It would be interesting to see whether UV-finiteness could be sharpened to a $\Lambda^{-8}$ behavior and/or could get extended to multiple loop integrations beyond formal arguments. In order to handle IR-divergences due to massless particles one area of investigation would be to go off-shell and define a string field theory in twistor space allowing proper IR-renormalization. Another possibility for future work is to do research on symmetry breakdown scenarios for the SU(4) $\!\times\!$ SU(4) R-symmetry of this model. Further, the fact that SYM is embedded in supergravity leads to the question, whether and how \textit{double copy} is built into this theory.\\

\appendix

\section{Reduction from 16-dimensional Spinor Space}
\label{Reduction}
This appendix contains an ad-hoc reduction from 16-dimensional spinor space without providing details on an actual mechanism of performing it.\\

In 16 spinor dimensions, the little group is typically taken to be SO(8) \cite{Geyer:2019ayz}, but here the existence of a $8\!+\!2\!=\!10$-dimensional spacetime, incidence relation, or scattering equations is left open. Starting with a 16-dimensional superspinor $\mathcal{Z}_a \!=\!(\lambda_a^\mathcal{A}| \eta_a^{\mathcal{J}})$ with $\mathcal{A}\!=\!1,\ldots,16, a \!=\! 1,\ldots,8$ an SO(8) index, and $\mathcal{J}$ unspecified, a reduction can be achieved by branching SO(8) as \cite{Feger:2019tvk}
\begin{align*}
\text{SO}(8) &\rightarrow \text{SU}(2) \times \text{SU}(2) \times \text{SU}(2) \times \text{SU}(2),\\
8_s &= (2,1,2,1) + (1,2,1,2),
\end{align*}
breaking this 8-dimensional representation into 4 2-dimensional SU(2) ones. In terms of irreducible representations of Lie algebras one possible outcome of the branching is the direct sum of four su(2)s:
\begin{equation*}
so(8) \rightarrow su(2) \oplus su(2) \oplus su(2) \oplus su(2)\,.
\end{equation*}
One of them is assumed to be broken off into two singlet ones and the other three are kept together as algebra of a new little group G = SU(2) $ \!\times\!$ SU(2) $ \!\times\!$ SU(2). Now, by decomposing the superspinors into 2 8-dimensional ones and by selecting the number of odd components independently, one obtains for the little group G the superspinors
$\mathcal{Z}_\mathrm{a} \!=\!(\lambda^\mathrm{a}_A, \mu_\mathrm{a}^A | \eta^\mathrm{a}_\iota)$ and $(\tau^\mathrm{a}_\mathcal{I} | \rho^\mathrm{a}_{1A}, \rho^\mathrm{a}_{2A})$ with $\eta_{\iota}^\mathrm{a}\!=\!(\eta_\mathcal{I}^\mathrm{a}, \tilde{\eta}^{\mathcal{I}\mathrm{a}}), A \!=\! 1,\ldots,4,\, \mathcal{I} \!=\! 1,\ldots,8$, and little group index $\mathrm{a} \!=\! (a,\dot{a}, \ddot{a}), a,\dot{a}, \ddot{a} \!=\!1,2$,
and for the two singlet representations simple spinors $(\tau^\prime_{i\mathcal{I}} |)$ and $(\tilde{\tau}^{\prime\mathcal{I}}_i |)$ with no odd components.\\
After introducing supersymmetric gauge symmetries between the superspinors on the worldsheet, the action is chosen to be the one in \eqref{massSugraActionComb}.\\
 
This scenario indicates how all spinors in the action \eqref{massSugraActionComb} including the auxiliary ones could originate from a single spinor in a higher-dimensional spinor space. Without an actual theory in that space it is not particularly convincing but still tantalizing.\\

\section{Original Massive Ambitwistor String}
\label{Original}
It is interesting to note that the same $S_{\text{aux}}$ in \eqref{massSugraAction} (except for missing $\tau^\mathrm{a}_\mathcal{I}$ and $\tilde{\eta}_\mathrm{a}^\mathcal{I}$) could be added to the action of the original massive ambitwistor string of \cite{Albonico:2022, Albonico:2023} to give it zero central charge. Then the question arises which results of this article are valid also for this model. The partition function and the cosmological constant are zero on all spin structures for the same reasons. Loop amplitudes on even spin structures look similar to \eqref{LoopAmplEven} and on non-separating and separating degenerations they evaluate the same way. Unfortunately, for $g > 3$ there is one big issue in that there are not enough conditions available ($2g$ instead of $3(g-1) + \delta_{1g}$) to fix the moduli integrations.\\

It remains to look at loop amplitudes on odd spin structures. Here is a big difference because not all of the zero modes of the auxiliary $\rho^a_{1,2}$ spinors can get canceled by zero modes from other auxiliary bosonic spinors. Therefore, they will appear in the determinants of the matrices. Exactly the terms with saturated zero mode components survive the Grassmann odd integration. When trying to evaluate the amplitude on a non-separating divisor for $g \!=\! 1$, the outcome would not have any resemblance with tree amplitudes.\\

\section{Non-trivial Gauge of the Tiny Group}
\label{Gauge}
The quotient group G$_Q$ is isomorphic to SL(2,$\mathbb{C}$), but, unfortunately, just to gauge the tiny group G$_T$ by 'eliminating' it and to use a single pair of supertwistors to represent momenta and supermomenta like it was done for tree amplitudes at the end of section \ref{ExMassATStr} is not possible for loop amplitudes because then some PCOs would vanish and nullify the amplitude. But it is sufficient to only let two particles $i$ and $j$ have more than two momenta in such a way that $\lambda^\mathrm{a}_A$ has non-zero components for $\dot{a}$ and $\ddot{a}$ in index $\mathrm{a}$ without the matrices $(H^{[fg]}_{[kl]})^I_{DJ}$ and $(G^{[pr]}_{[st]})^I_{DJ}$ in \eqref{AV} to contain any dependence on the dotted indices:\\
   
   Using SO(4, $\mathbb{C}$) indices $\alpha \!=\! 0,\ldots,3$ instead of spinor indices for the tiny group G$_T$ , with contractions between these indices done with $D^{\alpha\beta} \!=\! \text{diag}(1,1,1,-1)$, define for two particles $i$ and $j$:
   \begin{align*}
   &u_{i\alpha}^\mathrm{a} \!=\! (u_i^a, 0, u_i^{\dot{a}}, u_i^{\ddot{a}})_\alpha, &&u_{j\alpha}^\mathrm{a} \!=\! (u_j^a, 0, u_j^{\dot{a}}, u_j^{\ddot{a}})_\alpha,\nonumber\\
   &v_{i\alpha}^\mathrm{a} \!=\! (v_i^a, 0, \frac{1}{\sqrt{2}}(\dot{1}, -\dot{i}), \frac{1}{\sqrt{2}}(\ddot{1}, -\ddot{i}))_\alpha, &&v_{j\alpha}^\mathrm{a} \!=\! (v_j^a, 0, \frac{1}{\sqrt{2}}(\dot{i}, \dot{1}), \frac{1}{\sqrt{2}}(\ddot{i}, \ddot{1}))_\alpha,\nonumber\\
   &\epsilon_{i\alpha}^\mathrm{a} \!=\! (\epsilon_i^a, 0, \frac{1}{\sqrt{2}}(-\dot{i}, \dot{1}), \frac{1}{\sqrt{2}}(-\ddot{i}, \ddot{1}))_\alpha, &&\epsilon_{j\alpha}^\mathrm{a} \!=\! (\epsilon_j^a, 0, -\frac{1}{\sqrt{2}}(\dot{1}, \dot{i}), -\frac{1}{\sqrt{2}}(\ddot{1}, \ddot{i}))_\alpha,\nonumber\\
   &\hat{u}_{x\alpha}^\mathrm{a} \!=\! U_x (\frac{\hat{u}^a}{U_x}, 0, (-\dot{1},\dot{0}), (-\ddot{1}, \ddot{0}))_\alpha,&&U_x \!=\! u_{x\alpha}^\mathrm{a} u_{x\mathrm{a}}^\alpha \!=\!\!\braket{u_x \hat{u}},\ x \!=\! i,j,\nonumber\\
   &\kappa_A^{\ddot{a}} \!=\! \kappa_A^{\dot{a}}, \;\kappa_A^{\dot{a}} \; \text{arbitrary with}&& k^A_A \!=\! \kappa_A^{\dot{a}} \kappa^A_{\dot{a}} \!\ne\! 0 \!\ne\! \kappa_A^{\ddot{a}} \kappa^A_{\ddot{a}} ,\nonumber\\
   &\kappa_{iA}^{\dot{a}} \!=\! \kappa_A^{\dot{a}}, \kappa_{iA}^{\ddot{a}} \!=\! \kappa_A^{\ddot{a}}, &&\kappa_{jA}^{\dot{0}} \!=\! \kappa_A^{\dot{0}}, \kappa_{jA}^{\dot{1}} \!=\! -\kappa_A^{\dot{0}}, \kappa_{jA}^{\ddot{0}} \!=\! \kappa_A^{\ddot{0}}, \kappa_{jA}^{\ddot{1}} \!=\! -\kappa_A^{\ddot{1}},\nonumber\\
   &u_i^{\dot{1}} \!=\! (\dot{\sigma_{ij}}, \!\dot{1}), u_i^{\ddot{a}} \!=\! (\ddot{0},\!\ddot{1}), u_j^{\dot{a}} \!=\! (\dot{0}, \!\dot{1}), u_j^{\ddot{a}} \!=\! (\ddot{\sigma_{ji}}, \!\ddot{1}), && \sigma_{ij} \!\equiv\! \frac{\sqrt{\ud \sigma_i}\sqrt{\ud \sigma_j}}{S_{\alpha}(\sigma_i, \sigma_j | \tau)}, \quad u_i^{\dot{a}} u_{j\dot{a}} \!=\! u_i^{\ddot{a}} u_{j\ddot{a}} \!=\! \sigma_{ij}, \nonumber
   \end{align*}
   where it is understood that in the right hand side of the equations in the first four rows an item like $u^{\dot{a}}$ stands for $(0,u^{\dot{a}},0)$ etc. For all the other particles $\kappa_{kA}^{\dot{a}} \!=\! 0 \!=\! \kappa_{kA}^{\ddot{a}}, u_{k\alpha}^\mathrm{a} \!=\! (u_k^a,0,0)\delta^0_\alpha, k \!\ne\! i,j$, but $v_{k\alpha}^\mathrm{a} \!=\! v_{i\alpha}^\mathrm{a}, \epsilon_{k\alpha}^\mathrm{a} \!=\! \epsilon_{i\alpha}^\mathrm{a}$ for $k \!\ne\! j$ such that $v_{k\alpha}^\mathrm{a} \epsilon_{k\mathrm{a}\beta} \!=\! n_\alpha \delta_{\alpha\beta} (n_\alpha \!=\! 1$ except for $n_1 \!=\! 0)$ and $v_{k\alpha \mathrm{a}} \epsilon^\alpha_{k\mathrm{b}} \!=\! \varepsilon_{ab} \!+\! \varepsilon_{\dot{a}\dot{b}} \!-\! \varepsilon_{\ddot{a}\ddot{b}}$ for all particles although the non-zero $\alpha-$components only contribute for particles $i$ and $j$.
   Then 
   \begin{align*}
    &k^{AB} \!=\! \kappa_i^{\ddot{a}[A} \kappa_{i\ddot{a}}^{B]} \!=\! \kappa_i^{\dot{a}[A} \kappa_{i\dot{a}}^{B]} \!=\! -\kappa_j^{\dot{a}[A} \kappa_{j\dot{a}}^{B]} \!=\! \kappa_j^{\ddot{a}[A} \kappa_{j\ddot{a}}^{B]},\nonumber\\
    &\epsilon_{i\alpha}^{[A} \epsilon_j^{B]\alpha} = \epsilon_i^{[A} \epsilon_j^{B]},\quad \dot{\epsilon}_{i2}^{[A} \dot{\epsilon}_{j2}^{B]} \!=\! - k^{AB} \!=\! \ddot{\epsilon}_{i3}^{[A} \ddot{\epsilon}_{j3}^{B]}&&\epsilon_{i\alpha}^{[A} \tilde{\epsilon}_i^{B]\alpha} \!=\! \epsilon_i^{[A} \tilde{\epsilon}_i^{B]},\epsilon_{j\alpha}^{[A} \tilde{\epsilon}_j^{B]\alpha} \!=\! \epsilon_j^{[A} \tilde{\epsilon}_j^{B]},\nonumber\\
    &u_{i\alpha}^\mathrm{a} u_{j\mathrm{a}}^\alpha \!=\!  u_i^a u_{ja} \!\equiv\!  \braket{u_i u_j},\nonumber\\
    &\epsilon^{A \alpha}_i u_{i\alpha}^\mathrm{a} \hat{u}_{\mathrm{a}\beta} \!=\! \epsilon^A_{i\beta} U_i, &&\epsilon^{A \alpha}_j u_{j\alpha}^\mathrm{a} \hat{u}_{\mathrm{a}\beta} \!=\! \epsilon^A_{j\beta} U_j,\nonumber\\
    &\frac{1}{U_i} \hat{u}_{\mathrm{a}\alpha} \lambda^\mathrm{a}_A(\sigma) \epsilon_i^{A\alpha} \!=\! \frac{1}{\braket{u_i \hat{u}}} \!\braket{\hat{u} \lambda_A(\sigma)}\!\epsilon_i^A,&&\frac{1}{U_j} \hat{u}_{\mathrm{a}\alpha} \lambda^\mathrm{a}_A(\sigma) \epsilon_j^{A\alpha} \!=\! \frac{1}{\braket{u_j \hat{u}}}\!\braket{\hat{u} \lambda_A(\sigma)}\!\epsilon_j^A,\nonumber
    \end{align*}
    and the polarized scattering equations with the additional momenta $\kappa_{iA}^{\dot{a}}, \kappa_{iA}^{\ddot{a}}, \kappa_{jA}^{\dot{a}}, \kappa_{jA}^{\ddot{a}}$ follow from the scattering equations without them, i.e. in the loop amplitudes the additional momenta only impact the fields $\lambda_A^{\dot{a}}$ and $\lambda_A^{\ddot{a}}$ in the Pfaffian while the dotted indices do not contribute to the elements in $(H^{[fg]}_{[kl]})^I_{DJ}$ and $(G^{[pr]}_{[st]})^I_{DJ}$. This can easily be verified directly or by choosing particles $i$ and $j$ to be the ones with only fixed vertex operators and using $u_{i\alpha}^\mathrm{a} u_{j\mathrm{a}}^\alpha \!=\!  \braket{u_i u_j}$. Requiring $k^A_A \!=\! \kappa_A^{\dot{a}} \kappa^A_{\dot{a}} \!\ne\! 0 \!\ne\! \kappa_A^{\ddot{a}} \kappa^A_{\ddot{a}}$ prevents the amplitudes from having vanishing Pfaffians in \eqref{AV}.\\
    
    The invariant measures $\ud \mu(u)$ and $\ud \mu(v)$ become:
    \begin{align*}
    &\ud \mu(u_i) \!=\! \ud^2 \!u_i^a \,\ud^2 \!u_i^{\dot{a}} \,\ud^2 \!u_i^{\ddot{a}} \,\delta(u_i^{\dot{a}} \!-\! (\sigma_{ij}, 1))\delta(u_i^{\ddot{a}} \!-\!  (0, 1)),\\
    &\ud \mu(u_j) \!=\! \ud^2 \!u_j^a \,\ud^2 \!u_j^{\dot{a}} \,\ud^2 \!u_j^{\ddot{a}} \,\delta(u_j^{\dot{a}} \!-\! (0, 1))\delta(u_j^{\ddot{a}} \!-\!  (\sigma_{ji}, 1)),\\
    &\ud \mu(v_i) \!=\! \ud^2 \!v_i^a \,\ud^2 \!v_i^{\dot{a}} \,\ud^2 \!v_i^{\ddot{a}} \,\delta(v_i^{\dot{a}} \!-\! \frac{1}{\sqrt{2}}(1, -i))\delta(v_i^{\ddot{a}} \!-\! \frac{1}{\sqrt{2}}(1, -i)),\\
    &\ud \mu(v_j) \!=\! \ud^2 \!v_j^a \,\ud^2 \!v_j^{\dot{a}} \,\ud^2 \!v_j^{\ddot{a}} \,\delta(v_j^{\dot{a}} \!-\! \frac{1}{\sqrt{2}}(i, 1))\delta(v_j^{\ddot{a}} \!-\! \frac{1}{\sqrt{2}}(i, 1)).
    \end{align*}
    For the supermomenta one can choose, without obstruction, $\zeta_{i\mathcal{I}}^{\dot{a}} \!=\! \zeta_{i\mathcal{I}}^{\ddot{a}} 
    \!=\! \zeta_{j\mathcal{I}}^{\dot{a}} \!=\! \zeta_{j\mathcal{I}}^{\ddot{a}} \!=\! 0$ which is equivalent to use as gauge the simple little group SL(2,$\mathbb{C}$) for the fermionic components of the first supertwistor. The factorization measure can then be simplified as well, taking advantage of $\!\braket{\epsilon_{n+1} \epsilon_{n+2}} \!=\! 1$:
    \begin{equation*}
    \ud m_{\text{fw}}  \,\delta(k_{n+2 A}^A) = \frac{\ud \mu(\kappa_{n+1\mathrm{a}}^A) \,\ud \mu(\zeta_{n+1\mathrm{a}}^{\mathcal{I}})}{\text{vol}(\text{G}_Q){\kappa}} \rightarrow \frac{\prod_{l= n\!+\!1}^{n\!+\!2} \ud^4 \epsilon_l^A \,\ud^{\mathcal{N}}\!q_l^{\mathcal{I}}}{\text{vol}(\text{SL}(2, \mathbb{C}))_{\epsilon}}.\\
    \end{equation*}

\appendix

\bibliography{TwistorString}

\begin{thebibliography}{10}

\bibitem{Kunz:2024wsx}
C.~Kunz, ``{Extended Massive Ambitwistor String},''
  {\href{https://arxiv.org/abs/2406.01907}{{\ttfamily arXiv:2406.01907
  [hep-th]}}}, 2024.

\bibitem{Albonico:2022}
G.~Albonico, Y.~Geyer, and L.~Mason, ``{From Twistor-Particle Models to Massive
  Amplitudes},'' {\em
  \href{https://doi.org/10.3842/SIGMA.2022.045}{\emph{SIGMA} {\bfseries 18}
  (2022) p.45}}, [\href{https://arxiv.org/abs/2203.08087}{{\ttfamily 2203.08087
  [hep-th]}}].

\bibitem{Albonico:2023}
G.~Albonico, Y.~Geyer, and L.~Mason, ``{Massive ambitwistor-strings; twistorial
  models},'' {\em \href{https://doi.org/10.1007/JHEP01(2024)127}{\emph{JHEP}
  {\bfseries 01} (2024) p.127}},
  [\href{https://arxiv.org/abs/2301.11227}{{\ttfamily 2301.11227 [hep-th]}}].

\bibitem{Okano:2016gya}
S.~Okano and S.~Deguchi, ``{A no-go theorem for the $n$-twistor description of
  a massive particle},'' {\em \href{https://doi.org/10.1063/1.4976961}{\emph{J.
  Math. Phys.} {\bfseries 58} (2017) p.031701}}, no.~3,
  [\href{https://arxiv.org/abs/1606.01339}{{\ttfamily 1606.01339 [hep-th]}}].

\bibitem{Geyer:2019ayz}
Y.~Geyer and L.~Mason, ``{Supersymmetric S-matrices from the worldsheet in 10
  \textbackslash{}\& 11d},'' {\em
  \href{https://doi.org/10.1016/j.physletb.2020.135361}{\emph{Phys. Lett. B}
  {\bfseries 804} (2020) p.135361}},
  [\href{https://arxiv.org/abs/1901.00134}{{\ttfamily 1901.00134 [hep-th]}}].

\bibitem{Geyer:2020}
Y.~Geyer, L.~Mason, and D.~Skinner, ``{Ambitwistor strings in six and five
  dimensions},'' {\em
  \href{https://doi.org/10.1007/JHEP08(2021)153}{\emph{JHEP} {\bfseries 08}
  (2021) p.153}}, [\href{https://arxiv.org/abs/2012.15172}{{\ttfamily
  2012.15172 [hep-th]}}].

\bibitem{Arkani-Hamed:2017jhn}
N.~Arkani-Hamed, T.-C. Huang, and Y.-t. Huang, ``{Scattering amplitudes for all
  masses and spins},'' {\em
  \href{https://doi.org/10.1007/JHEP11(2021)070}{\emph{JHEP} {\bfseries 11}
  (2021) p.70}}, [\href{https://arxiv.org/abs/1709.04891}{{\ttfamily 1709.04891
  [hep-th]}}].

\bibitem{Penrose:1986}
R.~Penrose and W.~Rindler, {\em {SPINORS AND SPACE-TIME. VOL. 2: SPINOR AND
  TWISTOR METHODS IN SPACE-TIME GEOMETRY}}.
\newblock Cambridge Monographs on Mathematical Physics, Cambridge University
  Press, 4 1988.

\bibitem{Geyer:2018}
Y.~Geyer and L.~Mason, ``{Polarized Scattering Equations for 6D
  Superamplitudes},'' {\em
  \href{https://doi.org/10.1103/PhysRevLett.122.101601}{\emph{Phys. Rev. Lett.}
  {\bfseries 122} (2019) p.101601}}, no.~10,
  [\href{https://arxiv.org/abs/1812.05548}{{\ttfamily 1812.05548 [hep-th]}}].

\bibitem{Verlinde:1986kw}
E.~P. Verlinde and H.~L. Verlinde, ``{Chiral Bosonization, Determinants and the
  String Partition Function},'' {\em
  \href{https://doi.org/10.1016/0550-3213(87)90219-7}{\emph{Nucl. Phys. B}
  {\bfseries 288} (1987) p.357}}.

\bibitem{Nelson:1986ab}
P.~C. Nelson, ``{Lectures on Strings and Moduli Space},'' {\em
  \href{https://doi.org/10.1016/0370-1573(87)90082-2}{\emph{Phys. Rept.}
  {\bfseries 149} (1987) p.337}}.

\bibitem{Eguchi:1986ui}
T.~Eguchi and H.~Ooguri, ``{Chiral Bosonization on Riemann Surface},'' {\em
  \href{https://doi.org/10.1016/0370-2693(87)90084-0}{\emph{Phys. Lett. B}
  {\bfseries 187} (1987) Iss.1-2 p.127--134}}.

\bibitem{DHoker:1988pdl}
E.~D'Hoker and D.~H. Phong, ``{The Geometry of String Perturbation Theory},''
  {\em \href{https://doi.org/10.1103/RevModPhys.60.917}{\emph{Rev. Mod. Phys.}
  {\bfseries 60} (1988) p.917}}.

\bibitem{Knizhnik:1986kf}
V.~G. Knizhnik, ``{ANALYTIC FIELDS ON RIEMANNIAN SURFACES},'' {\em
  \href{https://doi.org/10.1016/0370-2693(86)90304-7}{\emph{Phys. Lett. B}
  {\bfseries 180} (1986) Iss.3 p.247--254}}.

\bibitem{DHoker:1989cxq}
E.~D'Hoker and D.~H. Phong, ``{Conformal Scalar Fields and Chiral Splitting on
  Superriemann Surfaces},'' {\em
  \href{https://doi.org/10.1007/BF01218413}{\emph{Commun. Math. Phys.}
  {\bfseries 125} (1989) p.469}}.

\bibitem{Polchinski:1998rq}
J.~Polchinski, {\em {String theory. Vol. 1: An introduction to the bosonic
  string}}.
\newblock Cambridge Monographs on Mathematical Physics,
  \href{https://doi.org/10.1017/CBO9780511816079}{\emph{Cambridge University
  Press} (2007)}, 12.

\bibitem{fay1973theta}
J.~Fay, {\em Theta Functions on Riemann Surfaces}.
\newblock \href{https://doi.org/10.1007/BFb0060090}{\emph{Lecture Notes in
  Mathematics, Springer} (1973)}.

\bibitem{Geyer:2018xwu}
Y.~Geyer and R.~Monteiro, ``{Two-Loop Scattering Amplitudes from Ambitwistor
  Strings: from Genus Two to the Nodal Riemann Sphere},'' {\em
  \href{https://doi.org/10.1007/JHEP11(2018)008}{\emph{JHEP} {\bfseries 11}
  (2018) p.008}}, [\href{https://arxiv.org/abs/1805.05344}{{\ttfamily
  1805.05344 [hep-th]}}].

\bibitem{verlinde_rijksuniversiteit_utrecht_1988}
H.~Verlinde and R.~U. (Netherlands) {\em
  \href{https://inis.iaea.org/records/jj4tv-nhg80}{The path-integral
  formulation of supersymmetric string theory}}, Sep 1988.

\bibitem{Witten:1985cc}
E.~Witten, ``{Noncommutative Geometry and String Field Theory},'' {\em
  \href{https://doi.org/10.1016/0550-3213(86)90155-0}{\emph{Nucl. Phys. B}
  {\bfseries 268} (1986) p.253--294}}.

\bibitem{Erbin:2021smf}
H.~Erbin, {\em {String Field Theory: A Modern Introduction}}.
\newblock
  \href{https://link.springer.com/book/10.1007/978-3-030-65321-7}{\emph{Lecture
  Notes in Physics} {\bfseries 980} (2021) 3},
  [\href{https://arxiv.org/abs/2301.01686}{{\ttfamily 2301.01686 [hep-th]}}].

\bibitem{Verlinde:1987sd}
E.~P. Verlinde and H.~L. Verlinde, ``{Multiloop Calculations in Covariant
  Superstring Theory},'' {\em
  \href{https://doi.org/10.1016/0370-2693(87)91148-8}{\emph{Phys. Lett. B}
  {\bfseries 192} (1987) p.95--102}}.

\bibitem{Friedan:1985ge}
D.~Friedan, E.~J. Martinec, and S.~H. Shenker, ``{Conformal invariance,
  supersymmetry and string theory},'' {\em
  \href{https://doi.org/10.1016/S0550-3213(86)80006-2}{\emph{Nucl. Phys. B}
  {\bfseries 271} (1986) p.93--165}}.

\bibitem{Polchinski:1998rr}
J.~Polchinski, {\em {String theory. Vol. 2: Superstring theory and beyond}}.
\newblock Cambridge Monographs on Mathematical Physics,
  \href{https://doi.org/10.1017/CBO9780511618123}{\emph{Cambridge University
  Press} (2007)}, 12.

\bibitem{Livine:2011gp}
E.~R. Livine and J.~Tambornino, ``{Spinor Representation for Loop Quantum
  Gravity},'' {\em \href{https://doi.org/10.1063/1.3675465}{\emph{J. Math.
  Phys.} {\bfseries 53} (2012) p.012503}},
  [\href{https://arxiv.org/abs/1105.3385}{{\ttfamily 1105.3385 [gr-qc]}}].

\bibitem{DHoker:2001jaf}
E.~D'Hoker and D.~H. Phong, ``{Two loop superstrings 4: The Cosmological
  constant and modular forms},'' {\em
  \href{https://doi.org/10.1016/S0550-3213(02)00516-3}{\emph{Nucl. Phys. B}
  {\bfseries 639} (2002) p.129--181}},
  [\href{https://arxiv.org/abs/hep-th/0111040}{{\ttfamily hep-th/0111040}}].

\bibitem{Albonico:2020}
G.~Albonico, Y.~Geyer, and L.~Mason, ``{Recursion and worldsheet formulae for
  6d superamplitudes},'' {\em
  \href{https://doi.org/10.1007/JHEP08(2020)066}{\emph{JHEP} {\bfseries 08}
  (2020) p.66}}, [\href{https://arxiv.org/abs/2001.05928}{{\ttfamily 2001.05928
  [hep-th]}}].

\bibitem{Witten:2012bh}
E.~Witten, ``{Superstring Perturbation Theory Revisited},'' 9
  {\href{https://arxiv.org/abs/1209.5461}{{\ttfamily arXiv:1209.5461
  [hep-th]}}}, 2012.

\bibitem{Tuite:2010}
M.~P. Tuite and A.~Zuevsky, ``{The Szeg\"o Kernel on a Sewn Riemann Surface},''
  {\em \href{https://doi.org/10.1007/s00220-011-1310-1}{\emph{Commun. Math.
  Phys.} {\bfseries 306} (2011) p.617--645}},
  [\href{https://arxiv.org/abs/1002.4114}{{\ttfamily 1002.4114 [math.QA]}}].

\bibitem{Alvarez-Gaume:1986rcs}
L.~Alvarez-Gaume, G.~W. Moore, and C.~Vafa, ``{Theta Functions, Modular
  Invariance and Strings},'' {\em
  \href{https://doi.org/10.1007/BF01210925}{\emph{Commun. Math. Phys.}
  {\bfseries 106} (1986) p.1--40}}.

\bibitem{Naculich:2014naa}
S.~G. Naculich, ``{Scattering equations and BCJ relations for gauge and
  gravitational amplitudes with massive scalar particles},'' {\em
  \href{https://doi.org/10.1007/JHEP09(2014)029}{\emph{JHEP} {\bfseries 09}
  (2014) p.029}}, [\href{https://arxiv.org/abs/1407.7836}{{\ttfamily 1407.7836
  [hep-th]}}].

\bibitem{Geyer:2016}
Y.~Geyer, {\em {Ambitwistor Strings: Worldsheet Approaches to perturbative
  Quantum Field Theories}}.
\newblock PhD thesis, Oxford U., Inst. Math., 2016,
  [\href{https://arxiv.org/abs/1610.04525}{{\ttfamily 1610.04525 [hep-th]}}].

\bibitem{Bern:2018jmv}
Z.~Bern, J.~J. Carrasco, W.-M. Chen, A.~Edison, H.~Johansson,
  J.~Parra-Martinez, R.~Roiban, and M.~Zeng, ``{Ultraviolet Properties of
  $\mathcal N = 8$ Supergravity at Five Loops},'' {\em
  \href{https://doi.org/10.1103/PhysRevD.98.086021}{\emph{Phys. Rev. D}
  {\bfseries 98} (2018) p.086021}}, no.~8,
  [\href{https://arxiv.org/abs/1804.09311}{{\ttfamily 1804.09311 [hep-th]}}].

\bibitem{Feger:2019tvk}
R.~Feger, T.~W. Kephart, and R.~J. Saskowski, ``{LieART 2.0 \textendash{} A
  Mathematica application for Lie Algebras and Representation Theory},'' {\em
  \href{https://doi.org/10.1016/j.cpc.2020.107490}{\emph{Comput. Phys. Commun.}
  {\bfseries 257} (2020) p.107490}},
  [\href{https://arxiv.org/abs/1912.10969v2}{{\ttfamily 1912.10969v2
  [hep-th]}}].

\end{thebibliography}
\end{document}